\documentclass[reprint,
preprintnumbers,
nofootinbib,
amsmath,amssymb,
aps,
prd,
superscriptaddress,
]{revtex4-2}

\usepackage{graphicx}
\usepackage{dcolumn}
\usepackage{bm}
\usepackage[hidelinks]{hyperref}
\usepackage[all]{hypcap}
\usepackage{xcolor}
\usepackage{enumitem}
\usepackage{multirow}
\usepackage{orcidlink}

\hypersetup{
	colorlinks,
	linkcolor={red!50!black},
	citecolor={blue!50!black},
	urlcolor={blue!80!black}
}

\newcommand{\order}[1]{\mathcal{O}(#1)}

\DeclareMathOperator{\Tr}{Tr}
\renewcommand{\Re}{\operatorname{Re}}

\begin{document}

\preprint{ADP-24-20/T1259}

\title{Centre vortex evidence for a second finite-temperature QCD transition}

\author{Jackson A. Mickley\,\orcidlink{0000-0001-5294-2823}}
\affiliation{Centre for the Subatomic Structure of Matter, Department of Physics, The University of Adelaide, South Australia 5005, Australia}

\author{Chris Allton\,\orcidlink{0000-0003-0795-124X}}
\affiliation{Department of Physics, Swansea University, Swansea, SA2 8PP, United Kingdom}
\affiliation{School of Mathematics and Physics, The University of Queensland, St.\ Lucia, Brisbane, Queensland 4072, Australia}

\author{Ryan Bignell\,\orcidlink{0000-0001-8401-1345}}
\affiliation{School of Mathematics and Hamilton Mathematics Institute, Trinity College, Dublin 2, Ireland}

\author{Derek Leinweber\,\orcidlink{0000-0002-4745-6027}}
\affiliation{Centre for the Subatomic Structure of Matter, Department of Physics, The University of Adelaide, South Australia 5005, Australia}

\begin{abstract}
Evidence for the existence of a second finite-temperature transition in quantum chromodynamics (QCD) is obtained through the study of centre vortex geometry and its evolution with temperature. The dynamical anisotropic ensembles of the \textsc{Fastsum} Collaboration are utilised to conduct a comprehensive analysis at eight temperatures beyond the established chiral transition. Visualisations of the centre vortex structure in temporal and spatial slices of the lattice reveal that vortex percolation persists through the chiral transition and ceases at a temperature that is approximately twice the chiral transition temperature $T_c$. This implies that confinement is retained through temperatures up to $T \approx 2\,T_c$, pointing toward a second transition corresponding to deconfinement. The loss of percolation is quantified by the vortex cluster extent, providing a clear signal for the deconfinement transition. Additional vortex statistics, including temporal correlations, vortex and branching point densities, the number of secondary clusters and vortex chain lengths between branching points, are scrutinised as a function of temperature. All ten measures investigated herein show the characteristics of two transitions in QCD, encompassing the chiral transition at $T_c$ and the deconfinement transition at $T \approx 2\,T_c$. Performing an inflection point analysis on the vortex and branching point densities produces an estimate of $T_c$ that agrees with the known \textsc{Fastsum} value. By the same procedure, a precise estimate of the deconfinement point is extracted as $T_d = 321(6)\,$MeV.
\end{abstract}

\maketitle

\section{Introduction} \label{sec:intro}
QCD has a rich phase structure that has been the subject of many theoretical studies over the years. Along the temperature ($T$) axis, QCD is known to be confining at low-to-moderate temperatures and is presumed to be deconfining in the large $T$ limit due to asymptotic freedom arguments. The nature and position of the transition between these two limits has been studied extensively using the lattice approach \cite{Wilson:1974sk}. The earliest \cite{Rebbi:1979sg} of these simulations used the quenched approximation, and results have confirmed a first-order phase transition at a temperature $T_\text{quenched} \simeq 290\,$MeV \cite{Borsanyi:2022xml} where the theory becomes deconfined.

It had been assumed that this transition would persist as the quenched approximation was removed, i.e. as dynamical quarks were ``turned on''. This was understood schematically via the ``Columbia plot'' \cite{Brown:1990ev} which shows the nature of the transition in the $m_l$-$m_s$ plane, where $m_l$ represents the two light quark flavour masses and $m_s$ the strange quark mass. On this plot, the quenched theory is at the $m_l=m_s=\infty$ limit. Studies found that the first-order transition does not extend very far into finite values of $m_{l,s}$, and at the physical point the transition is a ``pseudocritical'' or crossover transition. The value of this temperature has been accurately determined at the physical quark masses as $T_\text{c}^\text{phys} \simeq 158\,$MeV \cite{HotQCD:2018pds, Borsanyi:2020fev, Gavai:2024mcj}. It is generally accepted that $T_c$ corresponds to the restoration of chiral symmetry.

The conventional assumption is that this crossover transition at $T_c$ is simply the continuation of the transition found earlier in the quenched theory. However, this would imply that quark deconfinement occurs at $T_c$, whereas there is some evidence that the string tension remains nonzero \cite{Bazavov:2023dci, Spriggs:2023ccb} and hadrons are still bound for $T\gtrsim T_{c}$ \cite{Burnier:2015tda, Rothkopf:2019ipj, Aarts:2023nax}. This calls into question what the deconfinement temperature $T_d$ is for physical quark masses.

To answer this question, we study centre vortices in QCD for a range of temperatures from 47 to 760\,MeV. This is the first such study in full QCD at finite temperature. Centre vortices are obtained by a two-step process that first fixes the gauge to maximal centre gauge. The link degrees of freedom are then factorised into centre elements, which give rise to the centre vortices, and a remainder with perturbative properties. By analysing the structure of these vortices using ten measurements of seven different properties, we uncover unambiguous evidence of two transition temperatures. We confirm that the first of these occurs at $T_c$, with the second transition corresponding to deconfinement occurring at $T_d \simeq 1.9\, T_c$.

This work uses the $2+1$ dynamical flavour anisotropic ensembles from the {\sc Fastsum} Collaboration which have light quarks heavier than nature corresponding to $m_\pi = 239(1)\,$MeV, and consequently $T_c = 167(3)\,$MeV \cite{Aarts:2022krz}.

This paper is structured as follows. We commence with a brief review of other approaches in lattice QCD that allude to a second transition in Sec.~\ref{sec:rev}. This is followed by a summary of our simulation details in Sec.~\ref{sec:simdetails} and an overview of centre vortices in Sec.~\ref{sec:centrevortices}. Section~\ref{sec:anisotropy} covers how to include the effects of anisotropy in the gauge-fixing procedure. Thereafter, visualisations of centre vortex structures are shown in Sec.~\ref{sec:visualisations}. Our main analysis is presented in Secs.~\ref{sec:statistics} and \ref{sec:branching}. Temporal correlation functions, vortex cluster extents, vortex densities and the prevalence of secondary clusters are studied in Sec.~\ref{sec:statistics}. In Sec.~\ref{sec:branching}, vortex branching points are analysed, including calculations of their densities and the lengths of the vortex ``chains'' between branches. Finally, we provide a discussion on our results in Sec.~\ref{sec:discussion} and conclude our primary findings in Sec.~\ref{sec:conclusion}.

\section{Two transitions in the literature} \label{sec:rev}
There have been several recent lattice QCD calculations~\cite{Aarts:2023vsf} with dynamical fermions that have hinted at the presence of a second phase transition in full QCD. These studies suggest that this may occur at $T/T_{c} \approx 2$, but there is considerable uncertainty in its location with proposals covering the wide range of $T\approx 200$--$500\,$MeV. References \cite{Kotov:2021rah, Kotov:2021ujj, Kotov:2021mpi, Kotov:2022inz} confirm that the first transition at $T_c$ restores chiral symmetry, while some of the studies \cite{Cardinali:2021mfh, Hanada:2023krw, Hanada:2023rlk, Glozman:2024ded} speculate a connection between the second transition and the deconfinement temperature, $T_d$, at which confinement gives way to a quark-gluon plasma.

Reference \cite{Cardinali:2021mfh} studies the emergence of this deconfinement temperature through the medium of monopole condensation in $N_f=2+1$ QCD. In pure gauge SU(2) and SU(3), monopole condensation occurs at a temperature that agrees with the standard deconfinement temperature~\cite{DAlessandro:2010jdd, Bonati:2013bga}. This motivates its use in full QCD~\cite{Cardinali:2021mfh}, wherein evidence is found for the emergence of a different phase which they call $T_{BEC}$ at $T \sim 275\,$MeV.

Other work~\cite{Glozman:2016swy, Rohrhofer:2019qal, Rohrhofer:2019qwq, Glozman:2022zpy, Philipsen:2022wjj, Chiu:2023hnm, Glozman:2024ded}, reviewed in Refs.~\cite{Glozman:2022zpy, Glozman:2024ded}, investigates meson correlators to examine the existence of a chiral-spin symmetric phase above $T_c$. The chiral-spin symmetry is an emergent symmetry where the quarks are bound into hadronlike states predominantly by the chromoelectric interaction~\cite{Chiu:2023hnm}. This causes the formation of a number of different meson multiplets dependent upon which symmetry is dominant at each temperature~\cite{Chiu:2023hnm}. Defining at which temperature these symmetries manifest is difficult as they are expected only to be approximate. Thus, the temperature determined depends upon how approximate the multiplet formation is allowed to be. 

Related work includes Ref.~\cite{Cohen:2024ffx}, which examines expectations for the renormalised Polyakov loop~\cite{Petreczky:2015yta} and finds good agreement with a model for the smooth disappearance of chiral-spin symmetry.

There are also a number of studies concerning the gauge topology~\cite{Shuryak:2017fkh, Hanada:2023krw, Hanada:2023rlk} that utilise Polyakov loop observables and the topological charge to hypothesise the presence of an intermediate phase they call ``partial deconfinement" between $T/T_{c} \approx 0.9$--$1.9$. This corresponds to $T \approx 174$--$348\,$MeV. Motivated by axion physics, Refs.~\cite{Petreczky:2016vrs, Athenodorou:2022aay} examine the topological susceptibility at a wide range of temperatures. They find a change such that the topological susceptibility can be well described by a dilute-instanton gas approximation for temperatures $T \gtrsim 250\,$MeV and specifically note the coincidence of this approach with quenched studies.

Further considerations~\cite{Kotov:2021rah, Kotov:2021ujj, Kotov:2021mpi, Kotov:2022inz} include modified topological observables that show the end of the ``usual" $\order{4}$ scaling behaviour at around $T \simeq 300\,$MeV. The $\order{4}$ scaling behaviour correctly reproduces the chiral transition temperature $T_c$. A new phase of thermal QCD is also proposed by Ref.~\cite{Alexandru:2019gdm} based upon arguments of scale invariance. They examine the volume and temperature dependence exhibited by the spectral density of the overlap Dirac operator and observe a qualitative change in behaviour that occurs in the region $T\approx 200$--$250\,$MeV, as expected from scale invariance. A newer study~\cite{Meng:2023nxf} uses two sets of simulations at $T=187$ and $234\,$MeV to place the upper bound at $T=234\,$MeV.

We stress that while these studies propose a second thermal transition in QCD, none are able to obtain a precise value for its position. In addition to presenting strong evidence for the existence of a second transition that corresponds to deconfinement, this paper crucially gives an accurate determination of the transition temperature.

\section{Simulation details} \label{sec:simdetails}
\begin{table*}
	\centering
	\caption{\label{tab:ensembles} \textsc{Fastsum} Generation 2L ensembles used in this work. The lattice size is $32^3 \times N_\tau$, with temperature $T = 1/(a_\tau N_\tau)$. The spatial lattice spacing is $a_s = 0.11208(31)$\,fm with renormalised anisotropy $\xi = a_s/a_\tau = 3.453(6)$, and the pion mass is $m_\pi = 239(1)$\,MeV \cite{Wilson:2019wfr}. The estimate for $T_c$ comes from an analysis of the renormalised chiral condensate, which provides $T_c = 167(3)\,$MeV \cite{Aarts:2020vyb, Aarts:2022krz}. Full details of these ensembles may be found in Refs.~\cite{Aarts:2020vyb,Aarts:2022krz}.}
	\begin{ruledtabular}
		\begin{tabular}{r|rrrrrrrrrrrrr}
			$N_\tau$ & 128 & 64 & 56 & 48 & 40 & 36 & 32 & 28 & 24 & 20 & 16 & 12 & 8 \\
			\colrule \\[-0.9em]
			$T$ (MeV) & 47 & 95 & 109 & 127 & 152 & 169 & 190 & 217 & 253 & 304 & 380 & 507 & 760 \\
			$T/T_c$ & 0.284 & 0.569 & 0.650 & 0.758 & 0.910 & 1.01 & 1.14 & 1.30 & 1.52 & 1.82 & 2.28 & 3.03 & 4.55
		\end{tabular}
	\end{ruledtabular}
\end{table*}

The thermal ensembles of the \textsc{Fastsum} Collaboration~\cite{Aarts:2020vyb} are used in this study. These have $2+1$ flavours of $\order{a}$-improved Wilson fermions on  anisotropic lattices. The renormalised anisotropy is $\xi \equiv a_s/a_\tau = 3.453(6)$~\cite{Dudek:2012gj, Aarts:2020vyb}. The lattice action follows that of the Hadron Spectrum Collaboration~\cite{Edwards:2008ja}. The gauge action is a Symanzik-improved~\cite{Symanzik:1983dc, Symanzik:1983gh} anisotropic gauge action with tree-level mean-field-improved coefficients and the fermion action employs mean-field-improved Wilson-clover~\cite{Sheikholeslami:1985ij, Zanotti:2004qn} fermions with stout-smeared spatial links~\cite{Morningstar:2003gk}. Two sweeps of smearing at $\rho=0.14$ are employed. Full details of the action and parameter values can be found in Ref.~\cite{Aarts:2020vyb}.

The ``Generation 2L" ensembles which have a pion mass of $m_\pi = 239(1)\,$MeV \cite{Wilson:2019wfr,Aarts:2022krz} are used. While this is still heavier than physical, it represents an important step toward the physical regime [in the previous ``Generation 2" ensembles the pion mass was $m_\pi = 384(4)\,$MeV~\cite{HadronSpectrum:2008xlg, Aarts:2014nba, Aarts:2014cda}]. The strange quark has been approximately tuned to its physical value via the tuning of the light and strange pseudoscalar masses~\cite{HadronSpectrum:2008xlg, HadronSpectrum:2012gic, Cheung:2016bym}. 

The ensembles are generated using a fixed-scale approach, such that the temperature is varied by changing $N_\tau$, as $T=1/(a_\tau N_\tau)$. A summary of the ensembles is given in Table~\ref{tab:ensembles}. There are five ensembles below the pseudocritical chiral transition temperature $T_c = 167(3)\,$MeV, one close to $T_c$ and seven above $T_c$. The estimate for $T_c$ comes from an analysis of the renormalised chiral condensate~\cite{Aarts:2020vyb}. This wide range and large number of temperatures is crucial to the success of this study. The ensembles were generated using \textsc{openQCD-Fastsum}~\cite{glesaaen_jonas_rylund_2018_2216356}, an extension of \textsc{openQCD-1.6}.

\section{Centre vortices} \label{sec:centrevortices}
Centre vortices~\cite{tHooft:1977nqb, tHooft:1979rtg, Nielsen:1979xu, Greensite:2003bk} are regions of the gauge field that carry magnetic flux quantised according to the centre of SU(3),
\begin{equation}
	\mathbb{Z}_3 = \left\{ \exp\left(\frac{2\pi i}{3}\, n \right) \mathbb{I} \;\middle|\; n = -1,0,1 \right\} \,.
\end{equation}
Physical vortices in the QCD ground-state fields have a finite thickness. Any Wilson loop that encircles a vortex has the effect of multiplying the loop by an element of $\mathbb{Z}_3$,
\begin{equation}
	W(C) \longrightarrow z\,W(C) \,.
\end{equation}

In contrast, on the lattice ``thin" centre vortices are extracted through a well-known gauge-fixing procedure that seeks to bring each link variable $U_\mu(x)$ as close as possible to an element of $\mathbb{Z}_3$, known as maximal centre gauge (MCG). These thin vortices form closed surfaces in four-dimensional Euclidean spacetime and, thus, one-dimensional structures in a three-dimensional slice of the four-dimensional spacetime.

Fixing to MCG is typically performed by finding the gauge transformation $\Omega(x)$ to maximise the ``mesonic" functional~\cite{Montero:1999by}
\begin{equation} \label{eq:mesonicMCG}
	R_\mathrm{mes} = \sum_{x,\,\mu} \,\left| \Tr U_\mu^{\Omega}(x) \right|^2 \,.
\end{equation}
The links are subsequently projected onto the centre:
\begin{equation}
	U_\mu^{\Omega}(x) \longrightarrow Z_\mu(x) = \exp\left(\frac{2\pi i}{3} \, n_\mu(x) \right) \mathbb{I} \in \mathbb{Z}_3 \,,
\end{equation}
with $n_\mu(x) \in \{-1,0,1\}$ identified as the centre phase nearest to $\arg \Tr U_\mu(x)$ for each link. Finally, the locations of vortices are identified by nontrivial plaquettes in the centre-projected field:
\begin{equation} \label{eq:centreprojplaq}
	P_{\mu\nu}(x) = \prod_\square Z_\mu(x) = \exp\left(\frac{2\pi i}{3} \, m_{\mu\nu}(x) \right)\mathbb{I}
\end{equation}
with $m_{\mu\nu}(x) = \pm 1$. The value of $m_{\mu\nu}(x)$ is referred to as the \textit{centre charge} of the vortex, and we say the plaquette is pierced by a vortex.

Due to a Bianchi identity satisfied by the vortex fields~\cite{Engelhardt:1999wr, Spengler:2018dxt}, the centre charge is conserved such that the vortex topology constitutes closed sheets in four dimensions, or closed lines in three-dimensional slices of the lattice. Although gauge dependent, numerical evidence strongly suggests the projected vortex locations correspond to the physical vortices of the original fields~\cite{DelDebbio:1998luz, Langfeld:2003ev, Montero:1999by, Faber:1999gu}. This allows one to investigate the significance of centre vortices through the vortex-only field $Z_\mu(x)$.

At low temperatures, the centre vortex flux is understood to primarily form a single connected cluster that spreads through the entire volume, known as the ``percolating cluster".  Centre vortex percolation naturally generates an area-law falloff for large Wilson loops~\cite{Bertle:1999tw, Engelhardt:1999fd, Engelhardt:1999wr, Engelhardt:1998wu, Greensite:2016pfc}. This allows one to extract a linear confining potential by considering space-time Wilson loops, which asymptotically give access to the static quark-antiquark potential $V(r)$:
\begin{equation} \label{eq:staticquarkpotential}
	\langle W(r,t) \rangle \sim \exp\left(-V(r) \, t\right) \,.
\end{equation}

In the pure $\mathrm{SU}(3)$ gauge theory, there is a transparent connection between centre vortices and confinement. Above the critical temperature, the vortex sheet rearranges to primarily pierce space-space plaquettes, representing an alignment of the sheet with the temporal dimension~\cite{Engelhardt:2003wm, Spengler:2018dxt, Mickley:2024zyg}. Even though it remains percolating in spatial dimensions, this results specifically in an absence of vortices piercing space-time Wilson loops. A trivial expectation value in Eq.~(\ref{eq:staticquarkpotential}) follows, signifying a vanishing static quark potential.

These findings will be useful to guide the present work with dynamical fermions, providing a baseline to compare against and motivate specific methods. Initial work exploring the effect of centre vortices in full QCD at zero temperatures illustrates the substantial differences in vacuum structure within the vortex framework compared to pure gauge~\cite{Biddle:2022zgw, Biddle:2022acd, Biddle:2023lod}. It is therefore prescient to explore how this translates into the finite-temperature regime and whether the relationship between vortices and confinement is as clear cut as discovered in the pure gauge sector. This will be a vital step toward understanding the finite-temperature phase structure of full QCD.

\section{Anisotropic considerations} \label{sec:anisotropy}
\begin{table*}
	\caption{\label{tab:anisotropytest} The physical space-space and space-time vortex densities in $\text{fm}^{-2}$ obtained with a variety of MCG functionals on our lowest temperature ($N_\tau = 128$) ensemble. Without any anisotropy correction, the space-time density is considerably larger. This can be resolved by including a factor of $\xi^2$ on the temporal term of the functional ($R^{\xi}$), though to ensure the resulting gauge is ``smooth" a two-step procedure must be performed in which the ``isotropic" functional is first applied as a preconditioner ($R\to R^{\xi}$). Mesonic and baryonic MCG functionals perform equally well in matching space-space and space-time densities with this procedure, though the latter is lower for the mesonic functional yet higher for the baryonic functional.}
	\begin{ruledtabular}
		\begin{tabular}{cD{.}{.}{3.6}D{.}{.}{3.6}D{.}{.}{3.6}                              D{.}{.}{3.6}D{.}{.}{3.6}D{.}{.}{3.6}}
			Plaquette orientation & \multicolumn{1}{c}{$R_\mathrm{mes}$} & \multicolumn{1}{c}{$R_\mathrm{mes}^{\xi}$} & \multicolumn{1}{c}{$R_\mathrm{mes} \to R_\mathrm{mes}^{\xi}$} & \multicolumn{1}{c}{$R_\mathrm{bar}$} & \multicolumn{1}{c}{$R_\mathrm{bar}^{\xi}$} & \multicolumn{1}{c}{$R_\mathrm{bar} \to R_\mathrm{bar}^{\xi}$} \\
			\colrule \\[-1em]
			Space-space & 9.94(1) & 18.47(1) & 11.75(1) & 10.61(1) & 21.32(1) & 13.78(2) \\
			Space-time  & 13.21(1) & 18.38(1) & 11.38(1) & 14.37(1) & 22.17(1) & 14.25(2) \\[0.1em]
			\colrule \\[-1em]
			\% difference & 28.20(15) & 0.49(5) & 3.24(12) & 30.08(11) & 3.89(6) & 3.35(15) \\[-0.25em]
		\end{tabular}
	\end{ruledtabular}
\end{table*}
Before proceeding to a comprehensive analysis of vortex geometry, it is crucial to consider whether any modification to the MCG-fixing procedure outlined in Sec.~\ref{sec:centrevortices} is needed to account for the anisotropic lattice spacing. A simple test for this is to carry out the gauge fixing and centre projection, and compare the proportion of space-space and space-time plaquettes pierced. More precisely, \textit{physical} vortex densities are computed in which these proportions are divided by the area of a single space-space or space-time plaquette; these are $a_s^2$ and $a_s a_\tau$, respectively. With a finer resolution in the temporal dimension, there are additional opportunities for space-time plaquettes to be pierced. If vortex identification were unaffected by the anisotropy, then we would expect these two densities to be equal at near-zero temperatures. Their values using the standard MCG functional of Eq.~(\ref{eq:mesonicMCG}) are provided in Table~\ref{tab:anisotropytest}.

For later use, a relative difference between the two values is also supplied. This is calculated as
\begin{equation} \label{eq:percenterror}
	\% \text{ difference} = \frac{\left| \rho_\mathrm{ss} - \rho_\mathrm{st} \right|}{\frac{1}{2} \left( \rho_\mathrm{ss} + \rho_\mathrm{st} \right)} \,,
\end{equation}
where the subscripts ``s" and ``t" denote space and time respectively.

We see that the vortex matter obtained with $R_\mathrm{mes}$ has a considerably higher density of space-time plaquettes pierced. This suggests the identification of centre vortices is sensitive to an anisotropic lattice spacing. Although the discrepancy is small compared to the anisotropy factor $\xi = 3.453$, we nonetheless desire to match the space-space and space-time densities to remove this apparent anisotropy dependence of the projected vortices.

\subsection{Continuum limit} \label{subsec:continuumMCG}
To improve the gauge-fixing functional, consider first the simple case of Landau gauge, defined on the lattice by the gauge that maximises
\begin{equation} \label{eq:landau}
	R_\mathrm{Landau} = \sum_{x,\,\mu} \Re \Tr U_\mu^{\Omega}(x) \,.
\end{equation}
Following the process outlined in Ref.~\cite{Bonnet:1999mj}, the continuum version of the lattice Landau gauge with a direction-dependent spacing $a_\mu$ can be constructed as
\begin{equation}
	\sum_\mu a_\mu^2 \, \partial_\mu A_\mu(x) = 0 \,.
\end{equation}
To recover the true continuum Landau gauge condition, $\sum \partial_\mu A_\mu = 0$, on an anisotropic lattice, the temporal term in Eq.~(\ref{eq:landau}) must be multiplied by a factor of $\xi^2 = (a_s/a_\tau)^2$. That is, the modified functional
\begin{gather}
	R_\mathrm{Landau}^\xi = \sum_{x,\,\mu} \xi_\mu^2 \, \Re \Tr U_\mu^\Omega(x) \,, \\
	\xi_\mu = \begin{cases}
		1, & \mu = 1, \, 2, \, 3 \\
		\xi, & \mu = 4
	\end{cases}
\end{gather}
should be maximised.

Returning now to $R_\mathrm{mes}$ of Eq.~(\ref{eq:mesonicMCG}), we write $\Omega(x) = \exp\left(-i \, \omega_a(x) \, t_a\right)$, where the $t_a$ are the $\mathrm{SU}(3)$ generators. By taking the functional derivative with respect to the $\omega_a$ on $\left|\Tr U_\mu^\Omega(x)\right|^2 = \left(\Tr U_\mu^\Omega(x)\right) \left(\Tr U_\mu^\Omega(x)\right)^*$, one arrives at
\begin{multline} \label{eq:funcderivative}
	\frac{\delta R_\mathrm{mes}}{\delta \omega_a(x)} = i \, \sum_\mu \Tr \bigg\{ \Big[\! \left(\Tr U_\mu^\Omega(x-\hat{\mu})\right)^{\!*} U_\mu^\Omega(x-\hat{\mu}) \\ - \left(\Tr U_\mu^\Omega(x) \right)^{\!*} U_\mu^\Omega(x) - \mathrm{h.c.} \, \Big] t_a \bigg\} \,.
\end{multline}
At this stage, one might be tempted to perform a naive perturbative expansion of the links about the identity to connect to the continuum limit. This is valid in Landau gauge, for which maximising the functional tends to bring the links near to the identity. For MCG, they can be near any one of the three centre elements.

To overcome this, we write $U_\mu(x) = Z_\mu(x) \, R_\mu(x)$, where $Z_\mu(x)$ is the centre-projected field and $R_\mu(x)$ is the vortex-removed field, the latter of which will overwhelmingly lie near the identity at a maximum of $R_\mathrm{mes}$. As explained in Ref.~\cite{Engelhardt:1999xw}, it is possible to construct a continuum gauge field $A_\mu'(x)$ that reproduces $R_\mu(x)$ on the lattice:
\begin{equation}
	R_\mu(x) = e^{iga_\mu A_\mu'(x)} \,.
\end{equation}
After substituting $U_\mu(x) = Z_\mu(x) \, R_\mu(x)$ into Eq.~(\ref{eq:funcderivative}), this allows one to expand $R_\mu(x)$ about the identity, ultimately leading to a Landau gaugelike condition on the vortex-removed field:
\begin{equation}
	\frac{\delta R_\mathrm{mes}}{\delta \omega_a(x)} = 0 \implies \sum_\mu a_\mu^2 \, \partial_\mu A_\mu'(x) = 0 \,.
\end{equation}
From this, it is clear that an analogous correction to Landau gauge must be applied on an anisotropic lattice to ensure an isotropic continuum limit. In other words, we consider instead the functional
\begin{equation} \label{eq:mesonic_anisotropic}
	R_\mathrm{mes}^\xi = \sum_{x,\,\mu} \xi_\mu^2 \, \left|\Tr U_\mu^\Omega(x) \right|^2 \,.
\end{equation}

Looking again in Table~\ref{tab:anisotropytest}, we see that the space-space and space-time vortex densities obtained with direct fixing to this modified functional now strongly agree with each other. However, this brings about another problem. One might have expected in accounting for the anisotropy that the space-time density would lower to match the space-space density. Instead, we see both densities are substantially larger. This has a simple explanation. We fix directly to Eq.~(\ref{eq:mesonic_anisotropic}) starting from a random gauge. With the factor of $\xi^2$ on the temporal links, which in our case is $\xi^2 \simeq 11.9$, the temporal links are heavily favoured in the maximisation. As a result, the spatial links are barely optimised and the realised gauge is extremely ``rough".

Instead, we employ a two-step procedure in which one first fixes to the standard ``isotropic functional" of Eq.~(\ref{eq:mesonicMCG}) and uses this as a launching point for fixing to the improved functional of Eq.~(\ref{eq:mesonic_anisotropic}). In doing so, one successfully locates superior local maxima of the anisotropy-corrected functional with an improved balance between spatial and temporal links. Table~\ref{tab:anisotropytest} reveals vortex densities from this two-step gauge fixing still within a few percent of each other and, additionally, comparable in magnitude to the original values obtained with the isotropic functional. We advocate for this preconditioning also in Landau and other similar gauges that would suffer an identical deficiency.

\subsection{Alternate functionals} \label{subsec:alternatefunctionals}
Another avenue we explore in accounting for the anisotropy is to consider alternate MCG functionals. In particular, we examine the ``baryonic" functional~\cite{DelDebbio:1998luz, Faber:1999sq, Langfeld:2003ev}:
\begin{equation} \label{eq:baryonicMCG}
	R_\mathrm{bar} = \sum_{x, \, \mu} \Re \left(\Tr U_\mu^\Omega(x) \right)^{\!3} \,,
\end{equation}
proposed as an alternative to the mesonic functional. Both Eqs.~(\ref{eq:mesonicMCG}) and (\ref{eq:baryonicMCG}) attempt to maximise the overlap between the links and elements of $\mathbb{Z}_3$. We are interested as to whether either functional is preferred for matching space-space and space-time densities. It is not difficult to see that the baryonic functional possesses an equivalent anisotropy correction:
\begin{equation} \label{eq:baryonic_anisotropic}
	R_\mathrm{bar}^\xi = \sum_{x, \, \mu} \xi_\mu^2 \, \Re \left(\Tr U_\mu^\Omega(x) \right)^{\!3} \,.
\end{equation}
In Table~\ref{tab:anisotropytest}, we further provide the vortex densities obtained with purely $R_\mathrm{bar}$, direct fixing to $R_\mathrm{bar}^\xi$ and $R_\mathrm{bar}^\xi$ with the isotropic preconditioning (i.e. $R_\mathrm{bar} \to R_\mathrm{bar}^\xi$). These follow the same pattern as with the mesonic functional. Namely, the two-step gauge fixing is necessary to reconcile the disparity between space-space and space-time densities and simultaneously ensure a smooth gauge.

Based on the relative error defined in Eq.~(\ref{eq:percenterror}), there is no strong inclination toward either the mesonic or baryonic functional for matching the vortex densities under the two-step anisotropy-corrected gauge fixing; they are indistinguishable within statistical uncertainty. One key difference is that with the baryonic functional, it is the space-time density that is still marginally higher, while this is inverted for the mesonic functional. If the anisotropy were to have an effect, one would intuitively expect the space-time density to be larger since a smaller temporal lattice spacing creates increased chances for fine vortex structure to be resolved. In other words, we have overcorrected with the mesonic functional. For this reason, we select the baryonic functional for providing the more probable arrangement of vortex densities.

We stress that the calculations in this paper have been carried out with the various combinations of gauge-fixing procedures (mesonic/baryonic and isotropic/anisotropy-corrected with isotropic preconditioning) and the qualitative aspects of the results, in their entirety, are independent of the chosen method. The purpose of this section, then, is to justify a procedure for presenting the precise quantitative values. To summarise, our choice is the anisotropy-corrected baryonic functional with isotropic preconditioning, $R_\mathrm{bar} \to R_\mathrm{bar}^\xi$.

\section{Visualisations} \label{sec:visualisations}
The first course of action is to visualise the centre vortex structure at a selection of temperatures across the pseudocritical chiral transition in full QCD, drawing a qualitative comparison to the equivalent pure gauge results~\cite{Mickley:2024zyg} and prior zero temperature work with dynamical fermions~\cite{Biddle:2023lod}. This is achieved utilising techniques previously established in Ref.~\cite{Biddle:2019gke}. To construct a three-dimensional visualisation, we slice through a given dimension of the full four-dimensional lattice by holding the selected coordinate fixed. For each plaquette in the three-dimensional slice, this leaves one orthogonal direction that is used to identify the plaquette. A vortex is then rendered as a jet existing on the dual lattice and piercing the associated nontrivial plaquette.

\begin{figure}
	\centering
	\includegraphics[width=\linewidth]{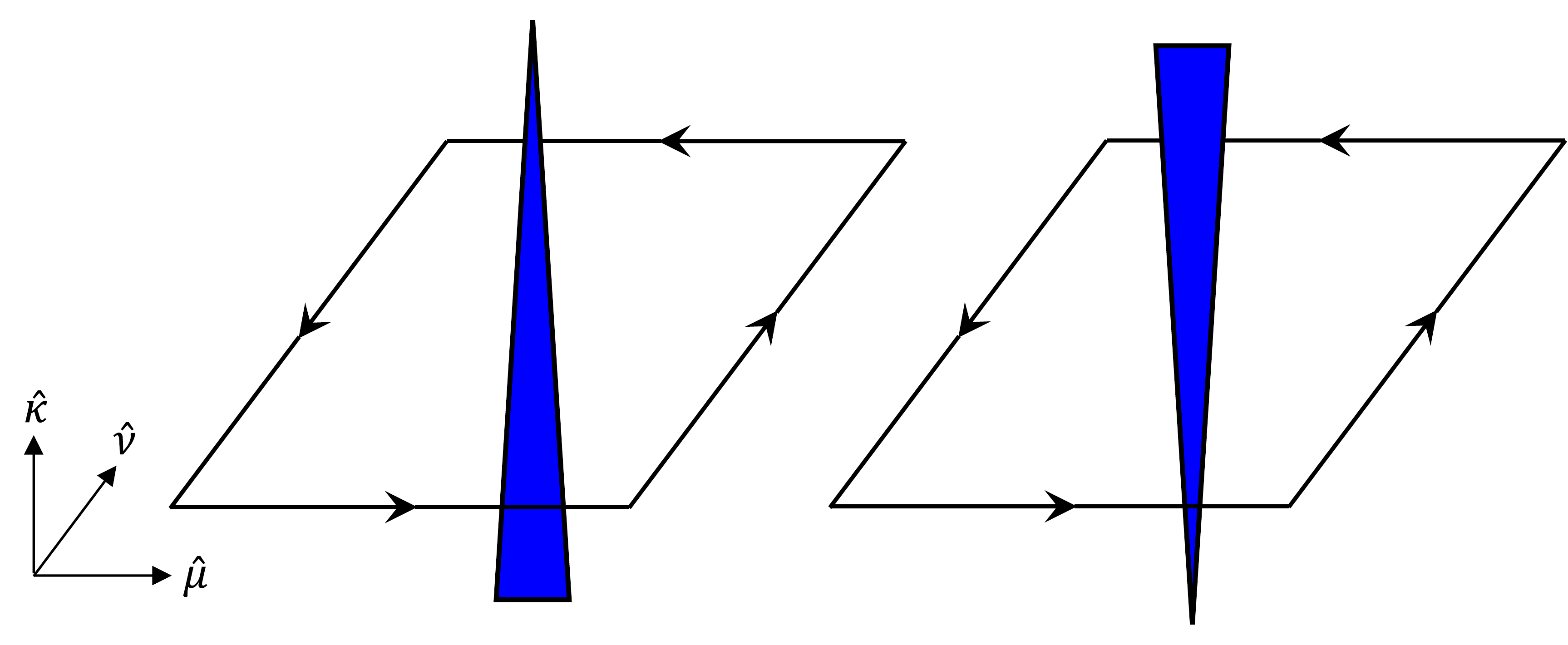}
	\caption{\label{fig:visconvention} The visualisation convention for centre vortices. An $m=+1$ vortex (\textbf{left}) is represented by a jet in the available orthogonal dimension, with the direction given by the right-hand rule. An $m=-1$ vortex (\textbf{right}) is rendered by a jet in the opposite direction, such that we always show the flow of $m=+1$ centre charge.}
\end{figure}

The orientation of an $m = +1$ vortex is determined by applying a right-hand rule. Moreover, since the flow of $m = -1$ centre charge is indistinguishable from an opposite flow of $m = +1$ centre charge, we display an $m = -1$ vortex as a jet pointing in the opposite direction from the right-hand rule. In other words, the visualisations exclusively show the flow of $m = +1$ centre charge. This convention is demonstrated in Fig.~\ref{fig:visconvention}.

With our understanding of the phase transition in the pure gauge theory, it is essential to produce separate visualisations from slicing along the temporal and spatial dimensions of the lattice, which we refer to as ``temporal" and ``spatial" slices, respectively. The former of these hold the temporal coordinate fixed, and the three-dimensional volume displayed comprises the three spatial dimensions. For the latter, we fix the $x$ coordinate such that we visualise a $y$-$z$-$\tau$ slice encompassing the remaining two space dimensions and the temporal dimension itself. In spatial slices of the lattice, we must accordingly be careful to properly accommodate the anisotropy present in the temporal dimension, achieved through a simple scaling of the grid points and vortex jets along the temporal axis by the anisotropy factor $\xi$.

\begin{figure*}
	\centering
	\includegraphics[width=0.48\linewidth]{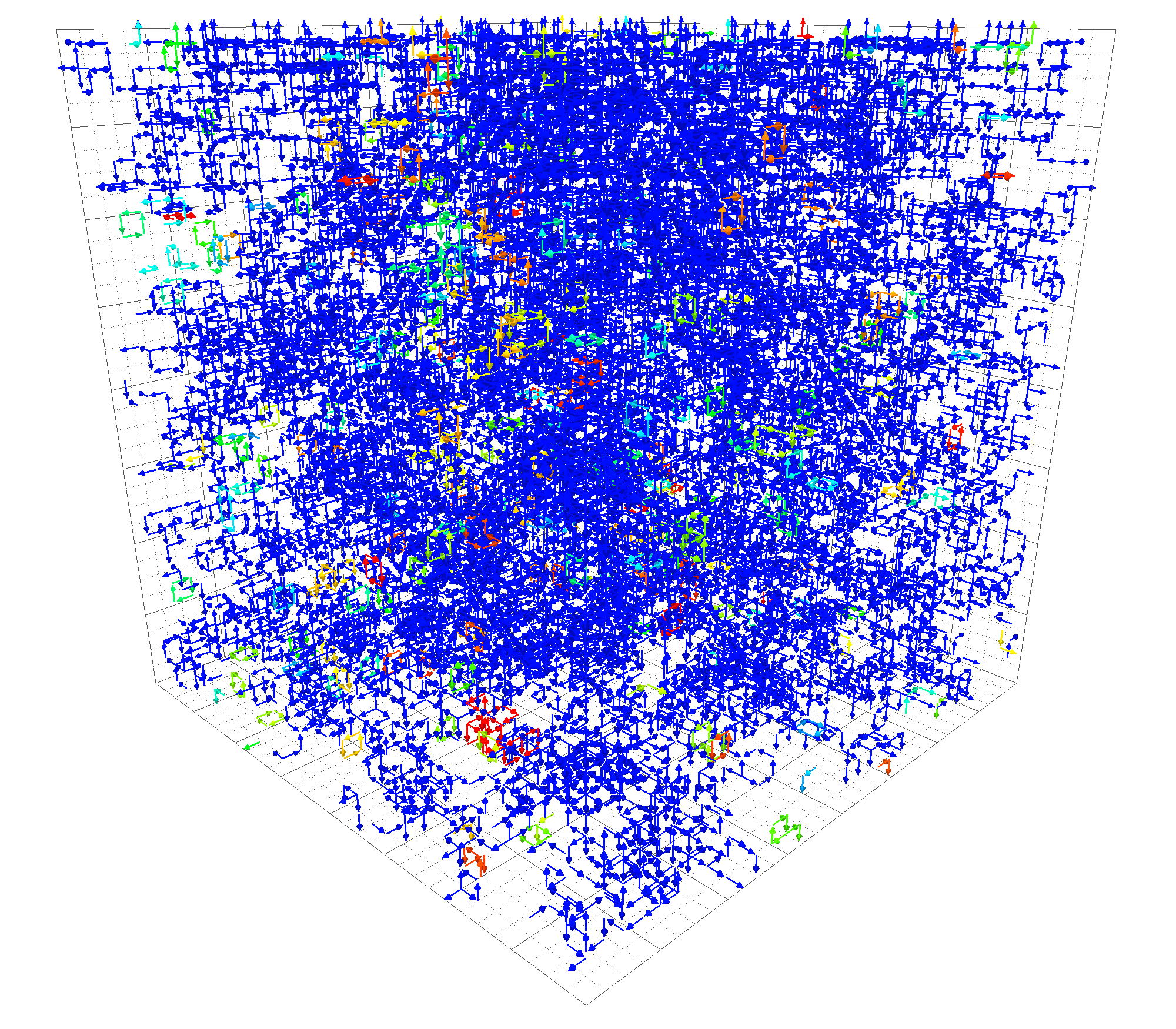}
	\includegraphics[width=0.48\linewidth]{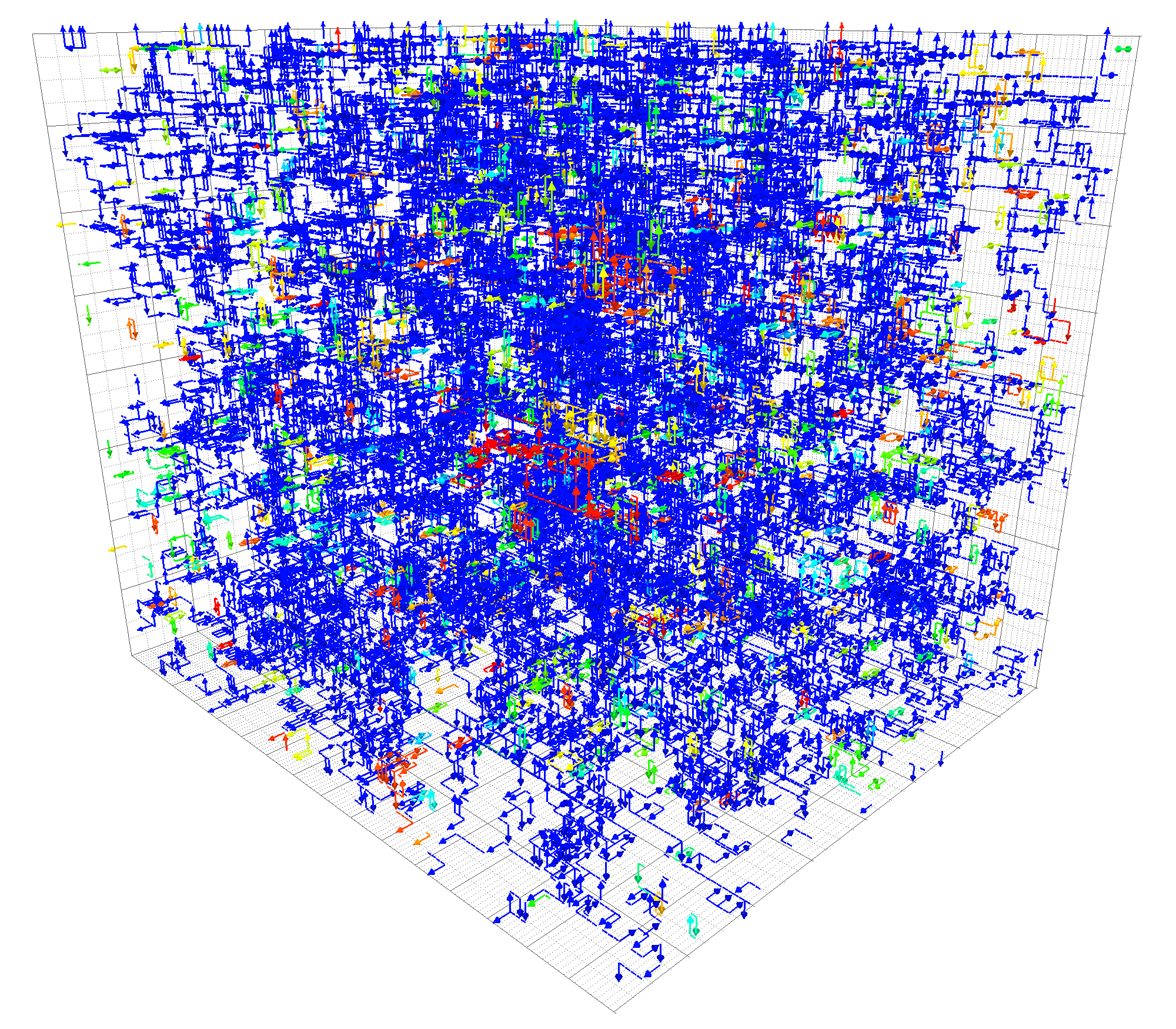}
	\vspace{-1em}
	\caption{\label{fig:Nt128vis} Centre vortex structure in temporal (\textbf{left}) and spatial (\textbf{right}) slices below the chiral transition temperature at $T/T_c \simeq 0.28$, corresponding to $N_\tau = 128$. In this and the following four figures, the compressed direction on the right-hand side is the temporal dimension. The significant reduction in jet length obscures the directional information of the arrowhead. The largest vortex cluster is coloured dark blue.}
\end{figure*}

\begin{figure*}
	\centering
	\includegraphics[width=0.48\linewidth]{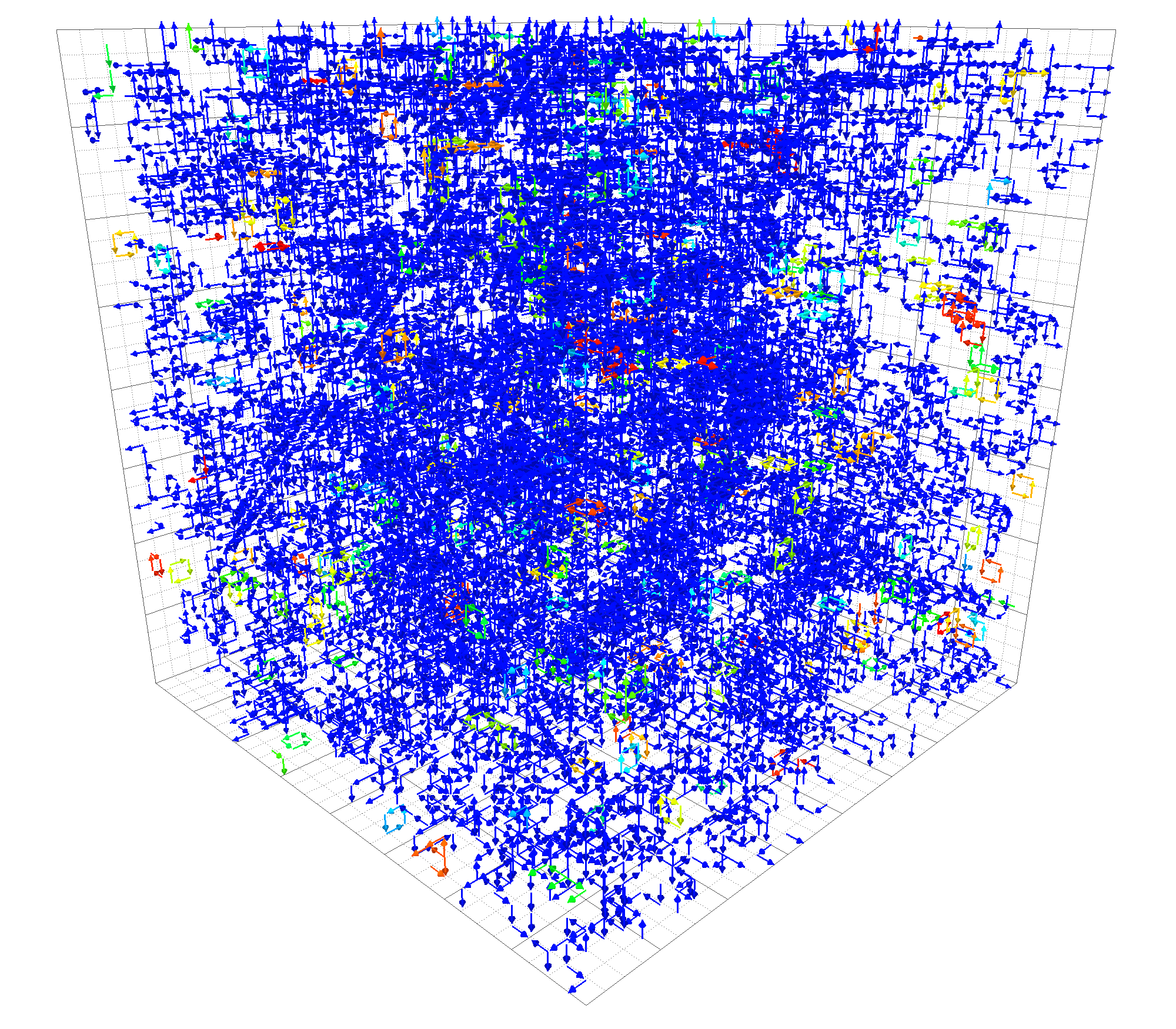}
	\includegraphics[width=0.48\linewidth]{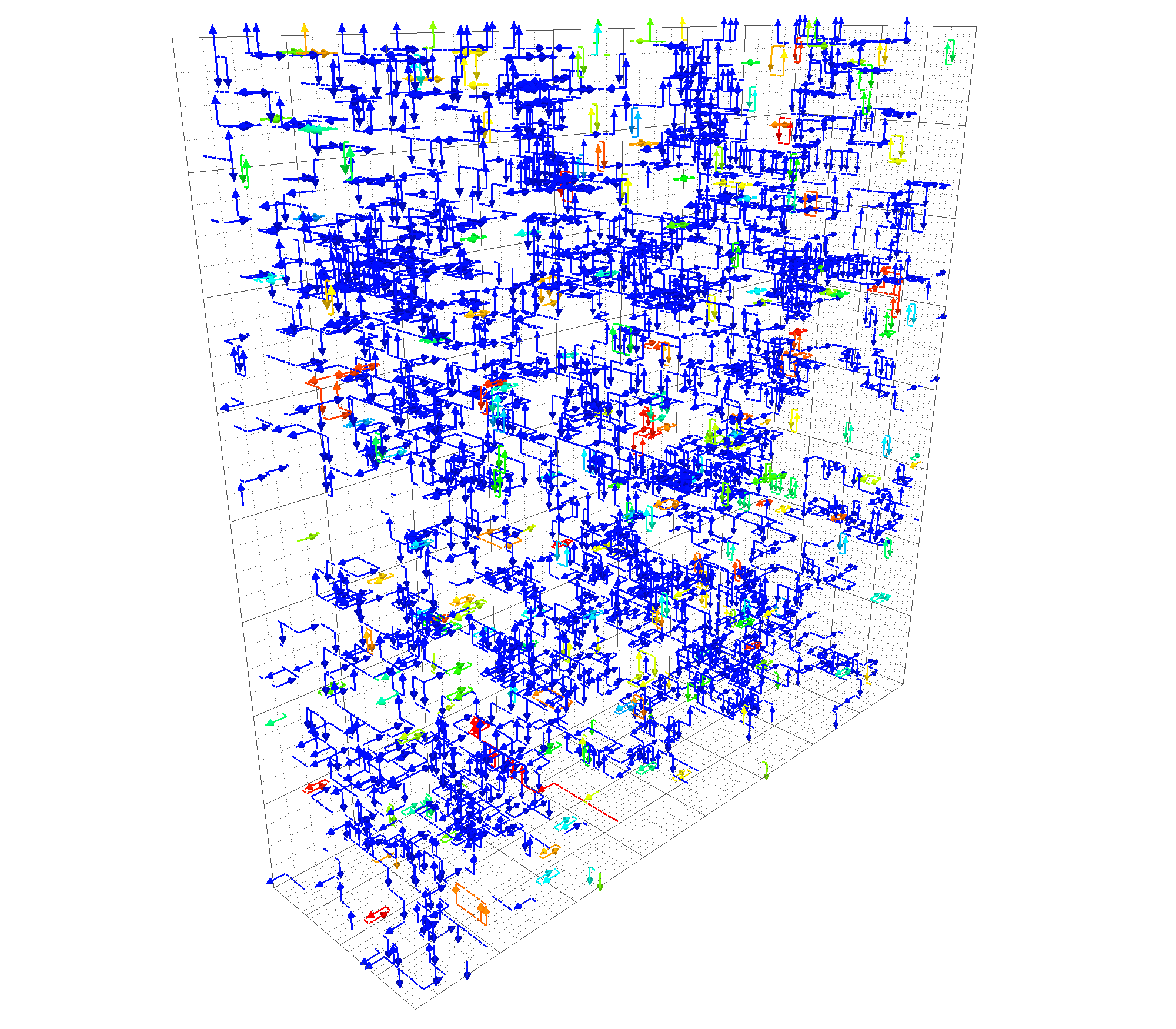}
	\vspace{-1em}
	\caption{\label{fig:Nt32vis} Centre vortex structure in temporal (\textbf{left}) and spatial (\textbf{right}) slices above the chiral transition at $T/T_c \simeq 1.14$, corresponding to $N_\tau = 32$}
\end{figure*}

\begin{figure*}
	\centering
	\includegraphics[width=0.48\linewidth]{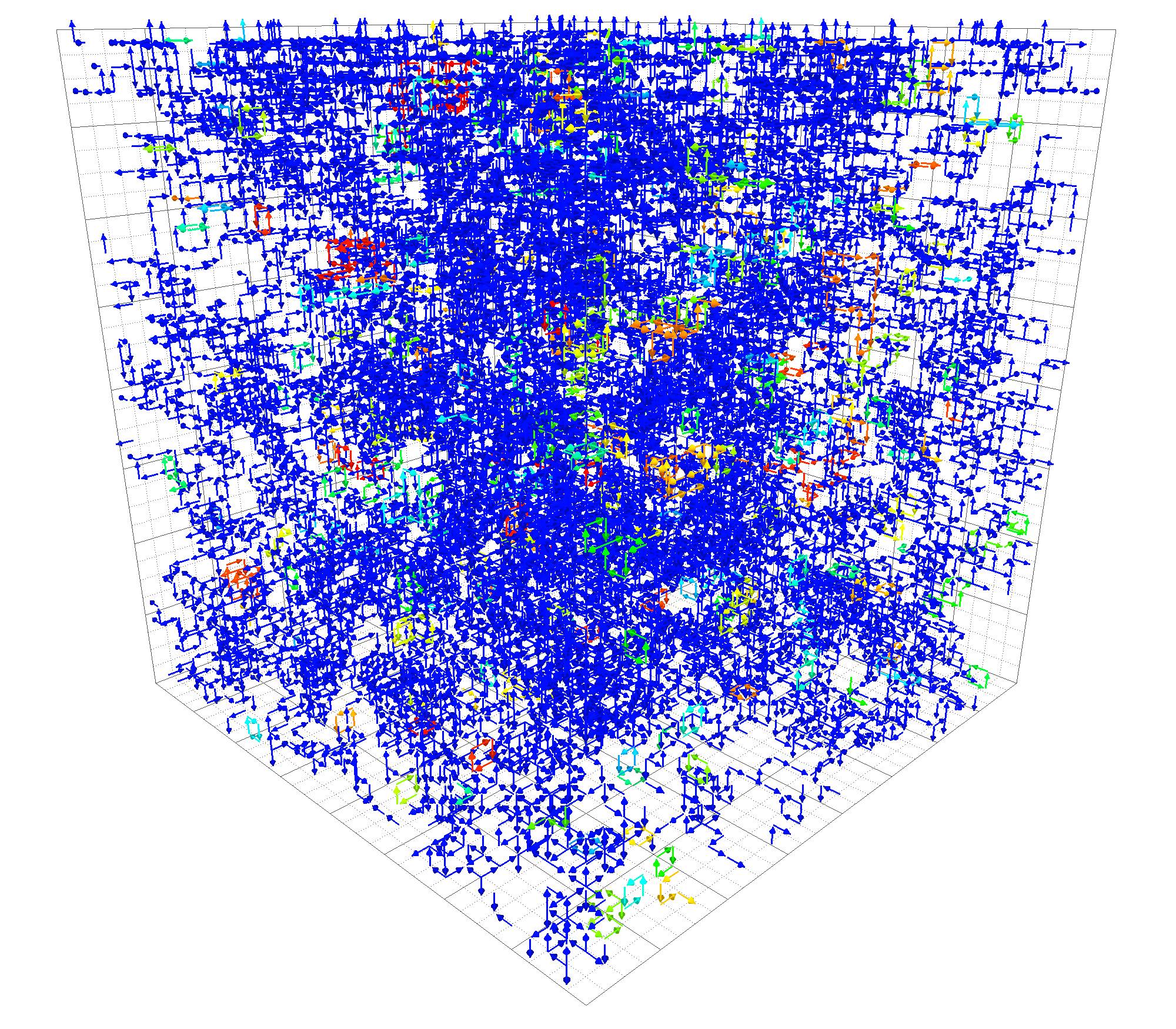}
	\includegraphics[width=0.48\linewidth]{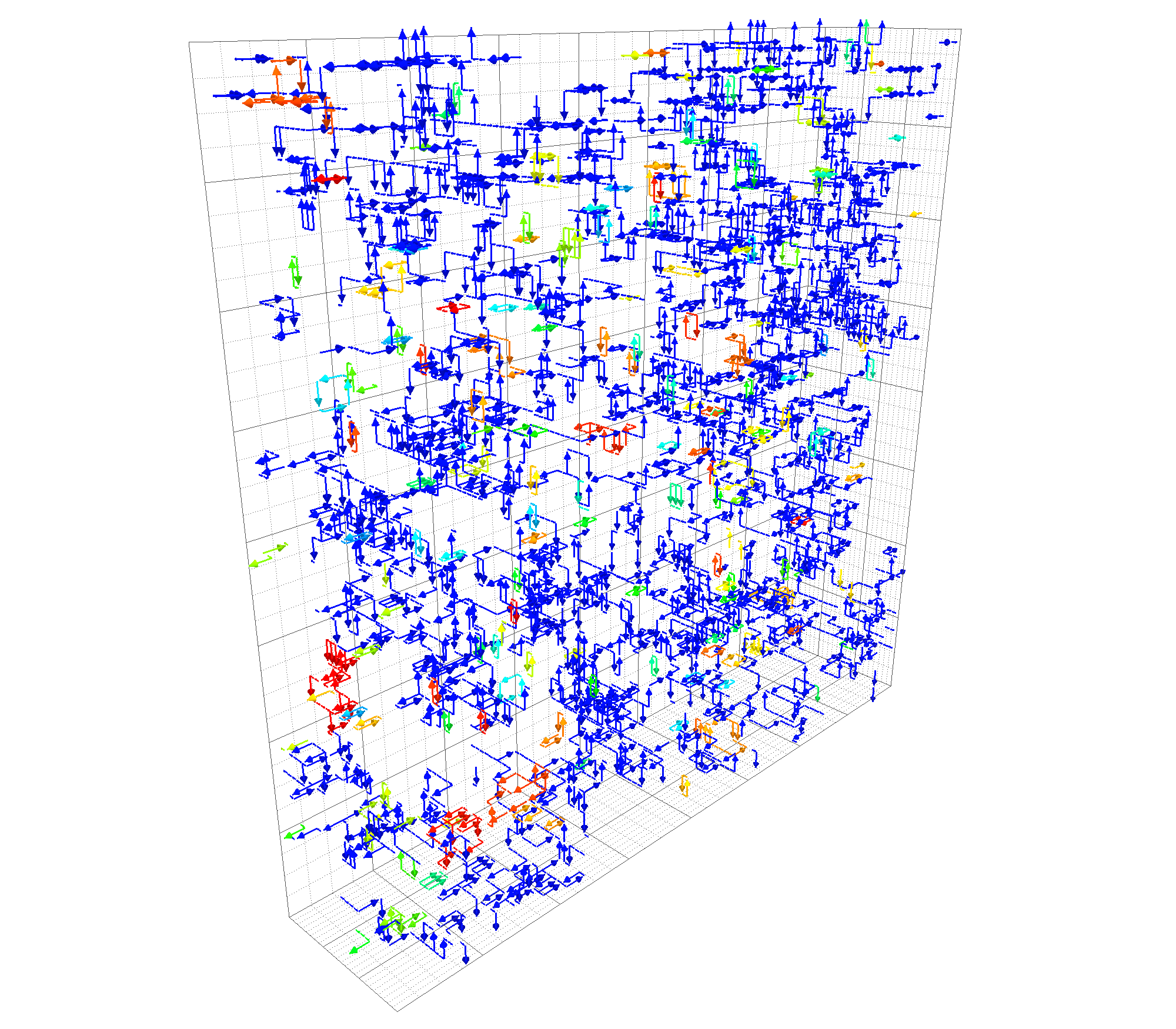}
	\vspace{-1em}
	\caption{\label{fig:Nt24vis} Centre vortex structure in temporal (\textbf{left}) and spatial (\textbf{right}) slices above the chiral transition at $T/T_c \simeq 1.52$, corresponding to $N_\tau = 24$.}
\end{figure*}

\begin{figure*}
	\centering
	\includegraphics[width=0.48\linewidth]{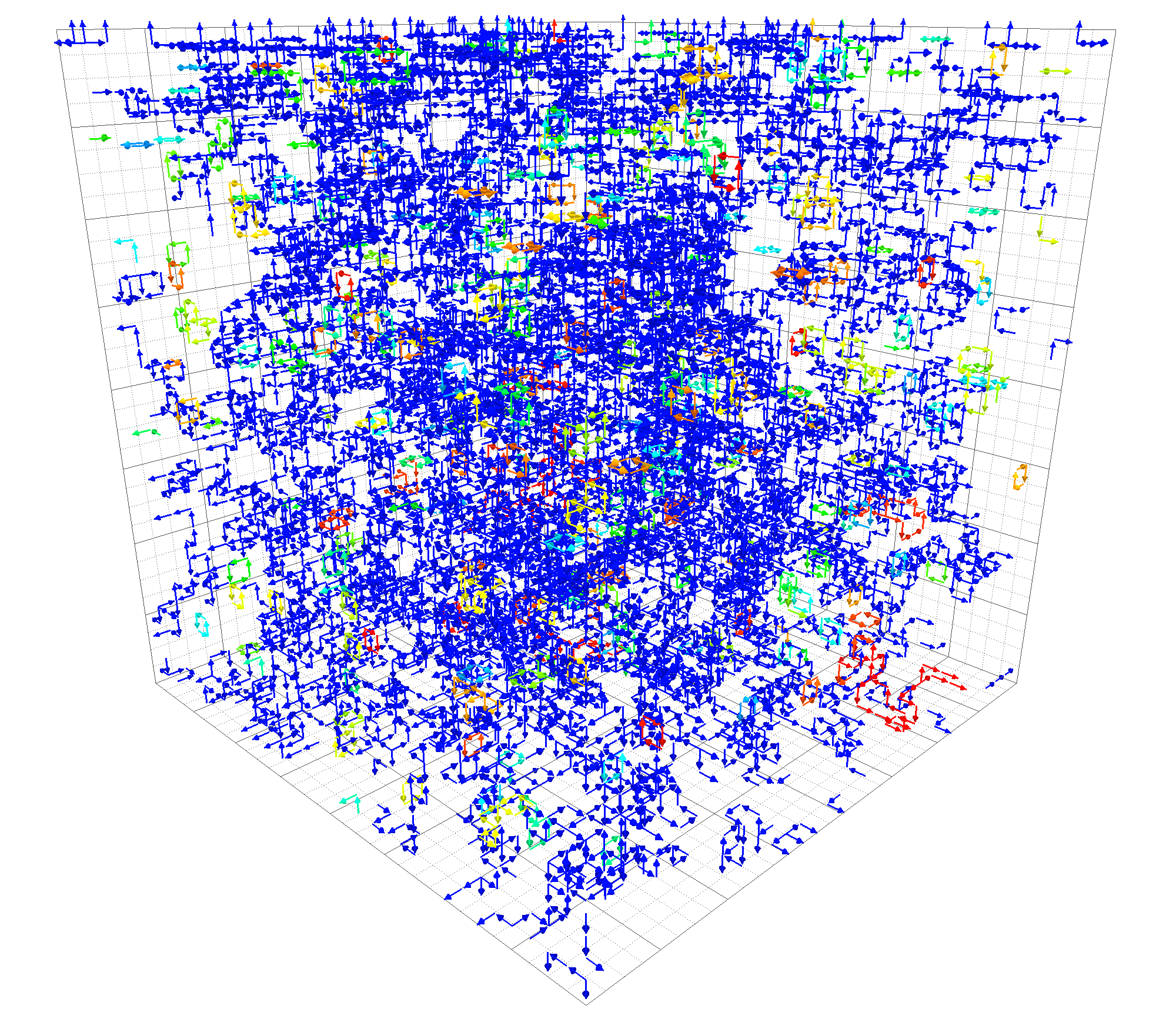}
	\includegraphics[width=0.48\linewidth]{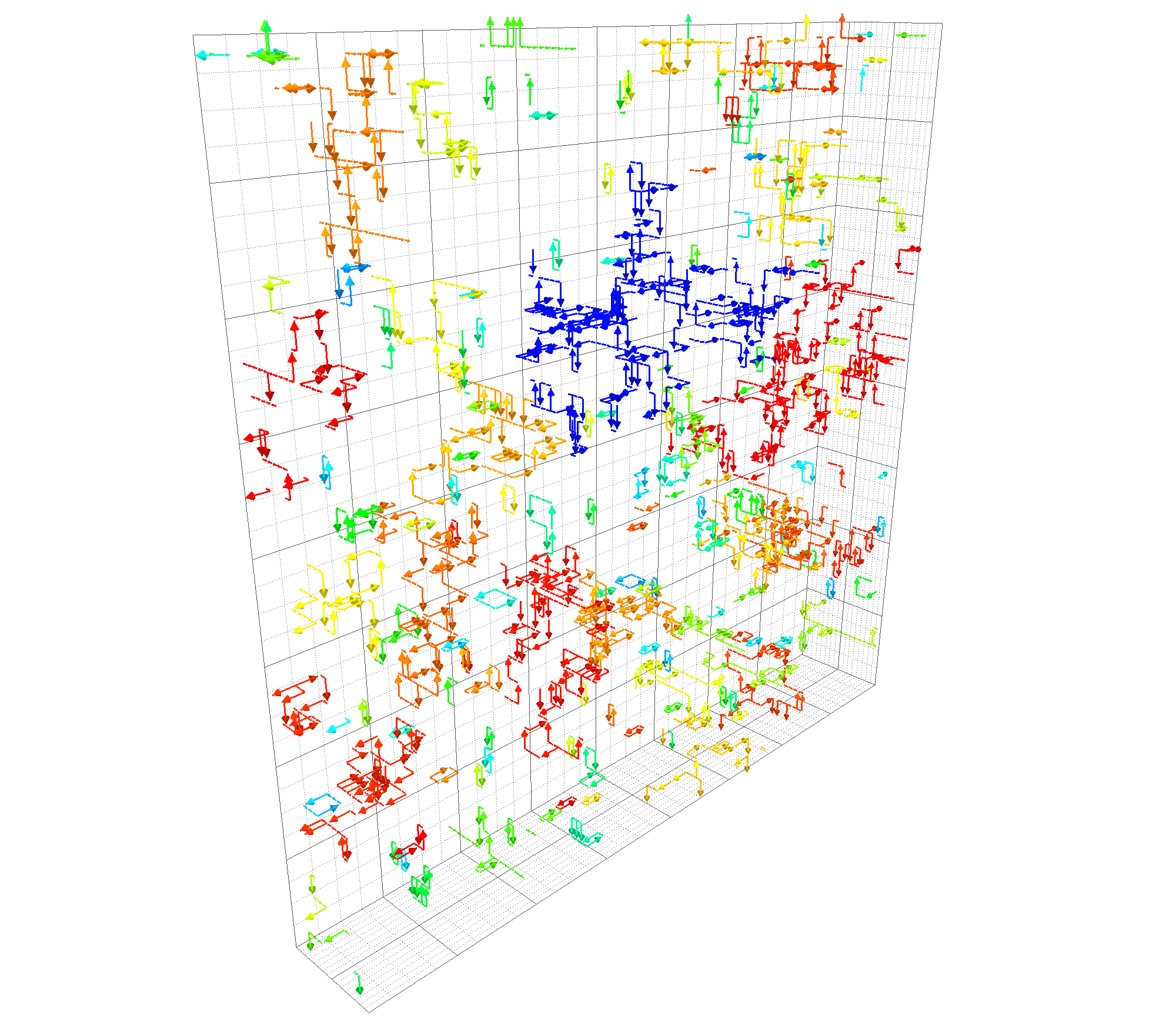}
	\vspace{-1em}
	\caption{\label{fig:Nt16vis} Centre vortex structure in temporal (\textbf{left}) and spatial (\textbf{right}) slices above the chiral transition at $T/T_c \simeq 2.28$, corresponding to $N_\tau = 16$.}
\end{figure*}

\begin{figure*}
	\centering
	\includegraphics[width=0.48\linewidth]{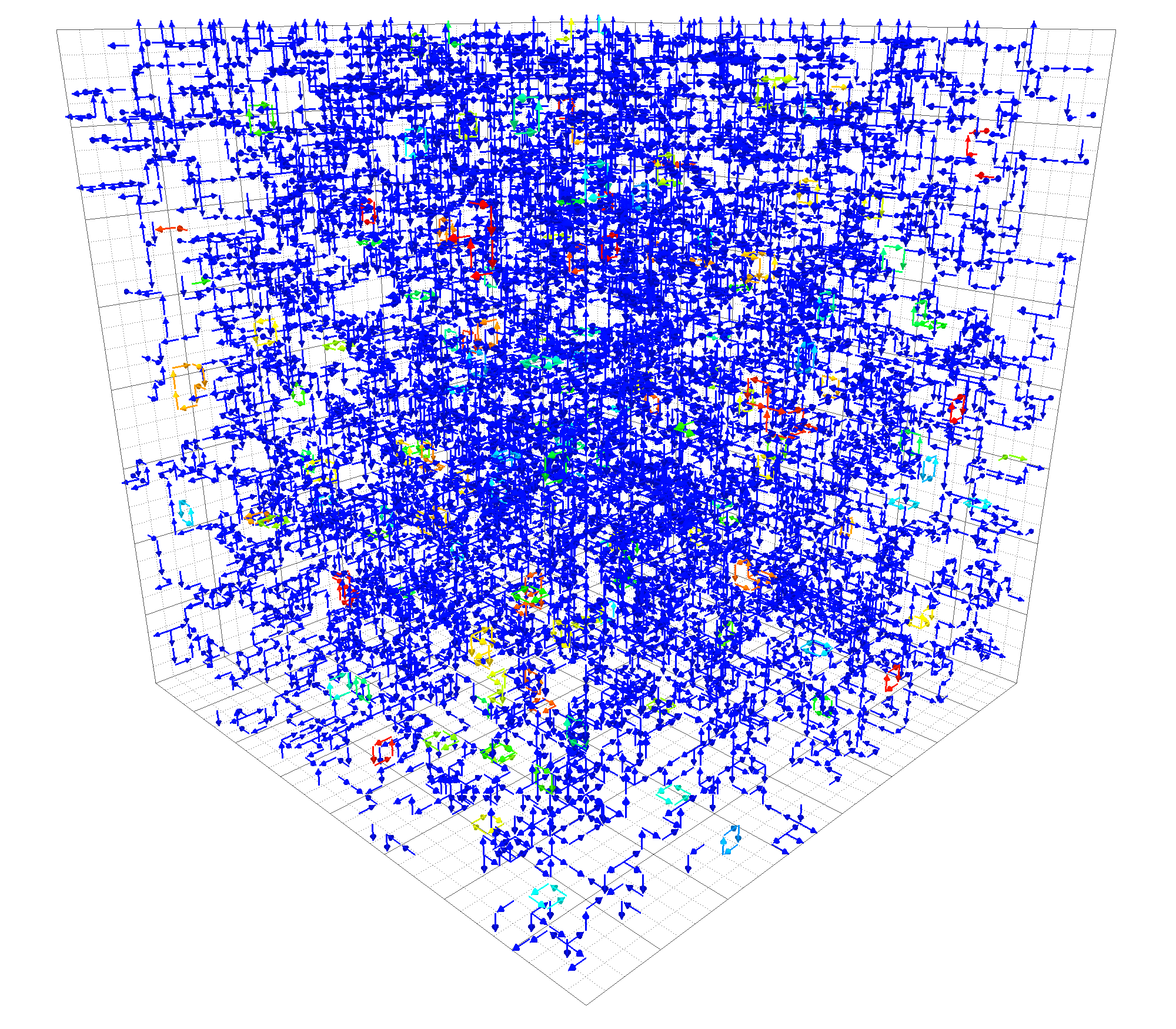}
	\includegraphics[width=0.48\linewidth]{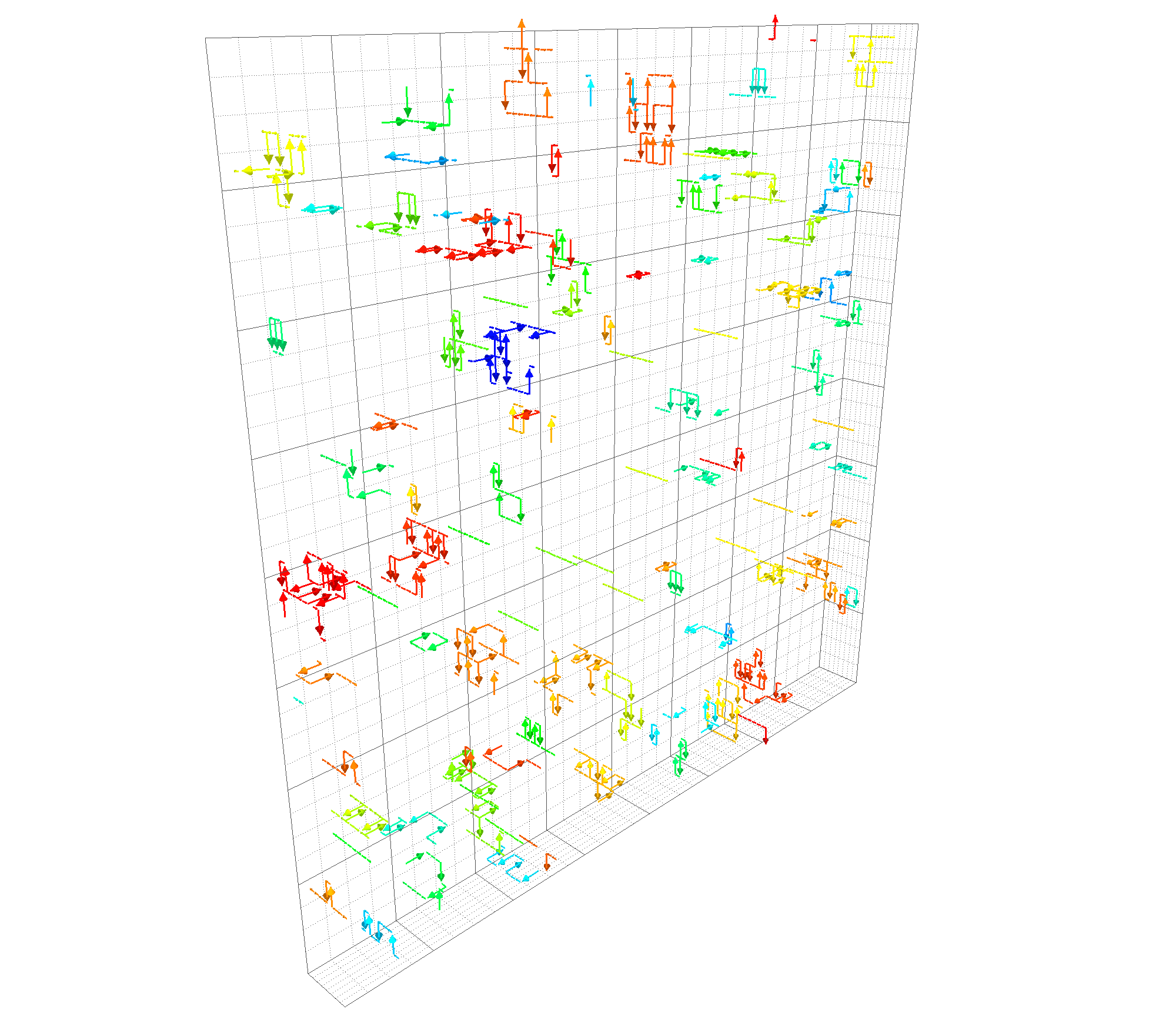}
	\vspace{-1em}
	\caption{\label{fig:Nt8vis} Centre vortex structure in temporal (\textbf{left}) and spatial (\textbf{right}) slices above the chiral transition at $T/T_c \simeq 4.55$, corresponding to $N_\tau = 8$.}
\end{figure*}

We produce these visualisations on our ensembles with $N_\tau = 128$, $32$, $24$, $16$ and $8$. For reference, $T_c$ corresponds to $N_\tau^c = 36.4$. Only one temperature below $T_c$ ($N_\tau = 128$) is displayed since we observe the same external structure everywhere in this regime. The remaining ensembles are chosen to include a temperature slightly above $T_c$, our highest temperature and two others to interpolate between these extremes. Typical vortex structures in temporal and spatial slices are presented for these ensembles throughout Figs.~\ref{fig:Nt128vis}--\ref{fig:Nt8vis}.

Below $T_c$ (Fig.~\ref{fig:Nt128vis}), we find equivalent structures in both temporal and spatial slices. These comprise a single percolating cluster (coloured in dark blue) that dominates the volume, with many small secondary clusters scattered throughout the lattice. Compared against centre vortices at finite temperature in pure gauge~\cite{Mickley:2024zyg}, there is a clear significant increase both in the amount of vortex matter and number of secondary clusters over the full temperature range. This is consistent with the previous findings including dynamical fermions at zero temperature~\cite{Biddle:2023lod}.

Moving above $T_c$, we continue to observe comparable structures in temporal slices across all temperatures. One qualitative change which can be elucidated from the visualisations is a reduction of vortex density in Fig.~\ref{fig:Nt16vis}, at $T/T_c \simeq 2.28$, compared to the temperatures below it. Along this line, there also appear to be fewer secondary clusters in the highest-temperature visualisation (Fig.~\ref{fig:Nt8vis}). At this point in our presentation, it is unclear whether this is a consequence of the specific slice displayed. This will be investigated quantitatively in Sec.~\ref{subsec:secondary}. If verified, this would broadly match an equivalent finding in pure $\mathrm{SU}(3)$ which saw a drop in amount of secondary clusters at $T_c$.

Turning to spatial slices, there is surprisingly no clear divergence in vortex structure compared to temporal slices as the chiral transition is crossed. This is apparent from Figs.~\ref{fig:Nt32vis} and \ref{fig:Nt24vis}, which still exhibit vortex percolation in both temporal and spatial slices despite having $T > T_c$. This is a stark contrast from the evolution in the pure gauge sector, which features a sudden strong alignment of the vortex sheet with the temporal dimension at the critical temperature. Continuing on to the two highest-temperature visualisations, however, does finally reveal a shift in the vortex geometry similar to the pure gauge theory. Indeed, a percolating cluster is notably absent from the spatial slices of Figs.~\ref{fig:Nt16vis} and \ref{fig:Nt8vis}. Even though the vortex sheet remains percolating in temporal slices, this indicates a loss of percolation in the full four-dimensional volume. Instead, the three-dimensional spatial slices consist exclusively of many small vortex clusters with a finite size; this is especially apparent in Fig.~\ref{fig:Nt8vis}. Consequently, any sufficiently large space-time Wilson loop will have a trivial expectation value in the vortex field, with an equal amount of $m = +1$ and $m = -1$ vortices piercing the enclosed area. This indicates confinement in the vortex-only fields has been lost.

As described in Sec.~\ref{sec:centrevortices}, the connection between centre vortices and confinement in the pure gauge theory is made abundantly clear in the deconfined phase, where the vortex sheet principally aligns with the temporal dimension. In spatial slices of the lattice, this manifests as short disconnected vortex lines winding around the temporal axis~\cite{Mickley:2024zyg}. With dynamical fermions, this is observed to a mild extent in Fig.~\ref{fig:Nt8vis}. However, it is certainly not the predominant feature as most clusters still involve many instances of pierced space-time plaquettes. The key property in connection to deconfinement, as described, is instead that percolation is lost and the vortex clusters are bounded.

\section{Vortex statistics} \label{sec:statistics}
With the evidently subdued changes in vortex structure with temperature in full QCD compared to pure gauge, it is essential to conduct a thorough investigation into the evolution of intrinsic vortex statistics. We will commence by pursuing a comparison to the pure gauge behaviour through the correlation introduced in Ref.~\cite{Mickley:2024zyg} that quantifies the temporal alignment of the vortex sheet. Subsequently, we proceed to utilise a more direct measure of percolation and the loss thereof, the cluster extent. Following on, we probe the vortex density as a function of temperature, including an explicit comparison between the proportion of space-time and space-space plaquettes pierced. This will be accompanied by an examination of the secondary clusters present in the three-dimensional slices. Finally, we conclude this analysis by considering the temperature dependence of branching point geometry.

Throughout this section, all statistics are obtained using 100 bootstrap ensembles, with errors calculated through the standard deviation of the bootstrap estimates.

\subsection{Vortex structure correlation} \label{subsec:correlation}
\begin{figure}
	\centering
	\includegraphics[width=\linewidth]{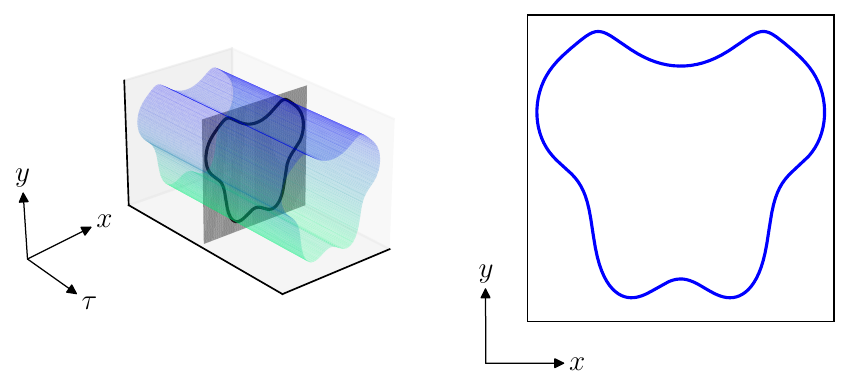}
	
	\vspace{-0.5em}
	
	\includegraphics[width=\linewidth]{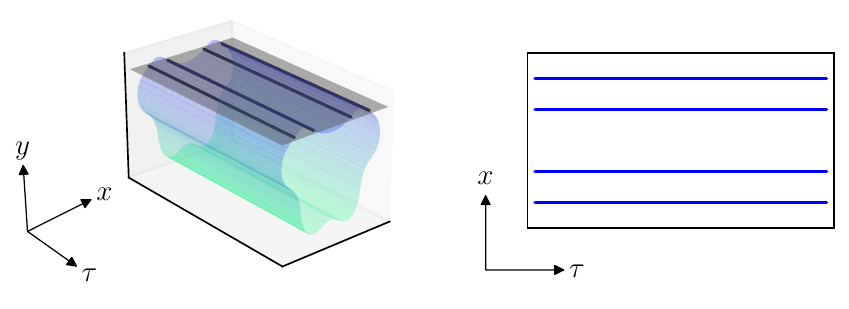}
	
	\vspace{-0.5em}
	
	\includegraphics[width=\linewidth]{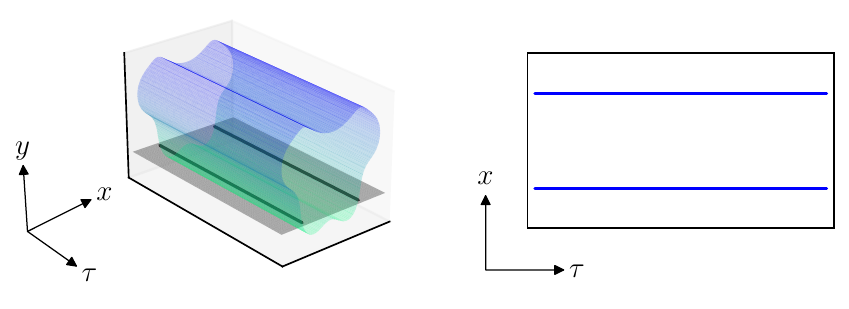}
	\caption{\label{fig:alignment_illustration} A cylindrical surface aligned with the temporal axis in three dimensions. In taking a temporal slice of the surface by fixing the $\tau$-coordinate (\textbf{top}), it is clear that an identical closed curve is obtained regardless of the coordinate of the slice (one can imagine ``sliding" the $x$-$y$ cross section along the temporal dimension). This allows the alignment to be quantified by comparing the structure across temporal slices. Alternatively, taking a spatial slice (\textbf{middle}/\textbf{bottom}) results in disconneccted lines parallel to the temporal axis, as identified with the vortex structure in Fig.~\ref{fig:Nt8vis}. Evidently, the position and number of these lines can vary with the slice coordinate.}
\end{figure}

As identified through the visualisations, there appears to be a subtle preference for the vortex sheet to align with the temporal dimension at very high temperatures. Since it is softer than the alignment in the pure gauge theory, we consider it beneficial to calculate a correlation of the vortex structure across temporal slices. In Sec.~\ref{sec:visualisations}, it was explained how any alignment is reflected in spatial slices through vortex lines parallel to the temporal dimension. An equivalent manifestation is that the vortex structure is frozen in temporal slices, experiencing minimal change from slice to slice. This can be understood through Fig.~\ref{fig:alignment_illustration}, which illustrates a cylinderlike surface oriented with the temporal axis in three dimensions.

\begin{figure}
	\centering
	\includegraphics[width=\linewidth]{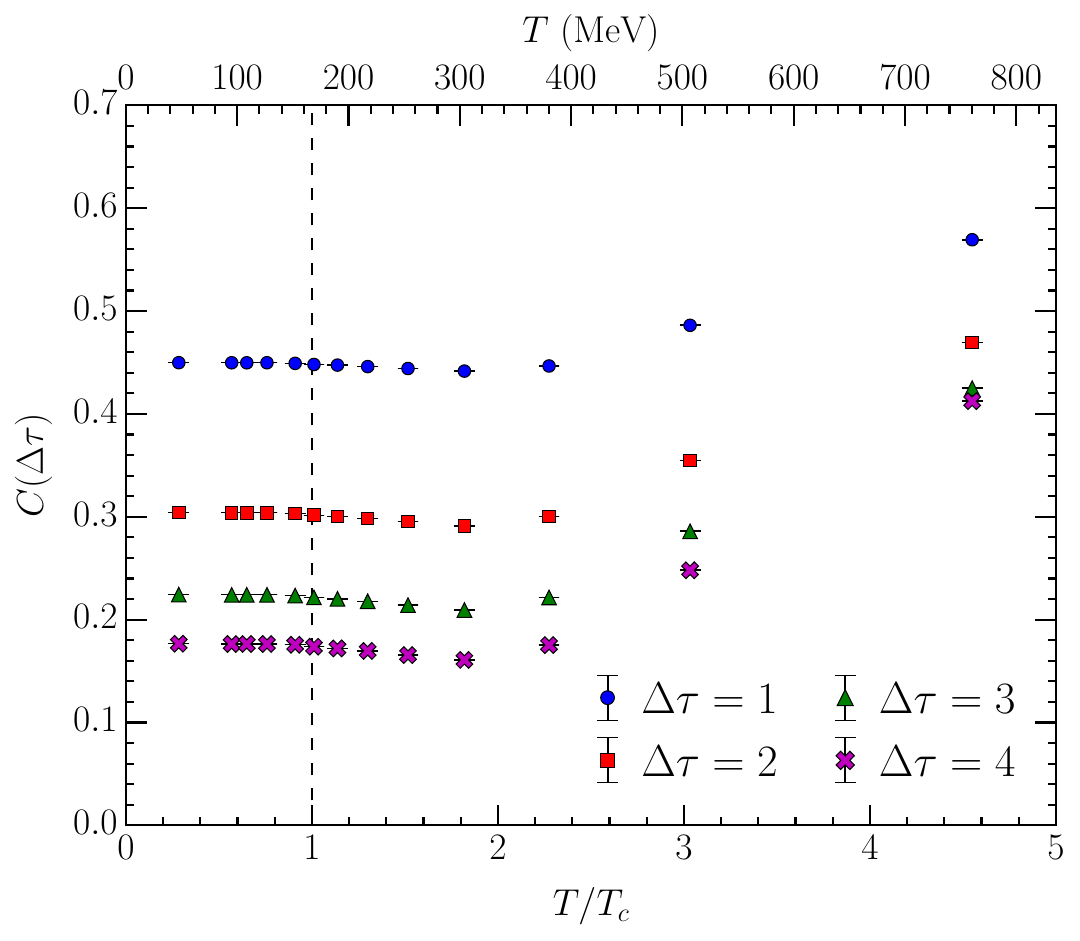}
	\vspace{-2em}
	\caption{\label{fig:temporalcorrelation} The correlation $C(\Delta\tau)$ defined in Eq.~(\ref{eq:correlation}) as a function of temperature for $\Delta\tau\in\{1,\,2,\,3,\,4\}$. It is approximately constant below $T_c$ and features a very slight decrease over $1\leq T/T_c \lesssim 2$. There is subsequently a turning point, beyond which the correlation gradually increases.}
		
	\vspace{1.5em}
		
	\includegraphics[width=\linewidth]{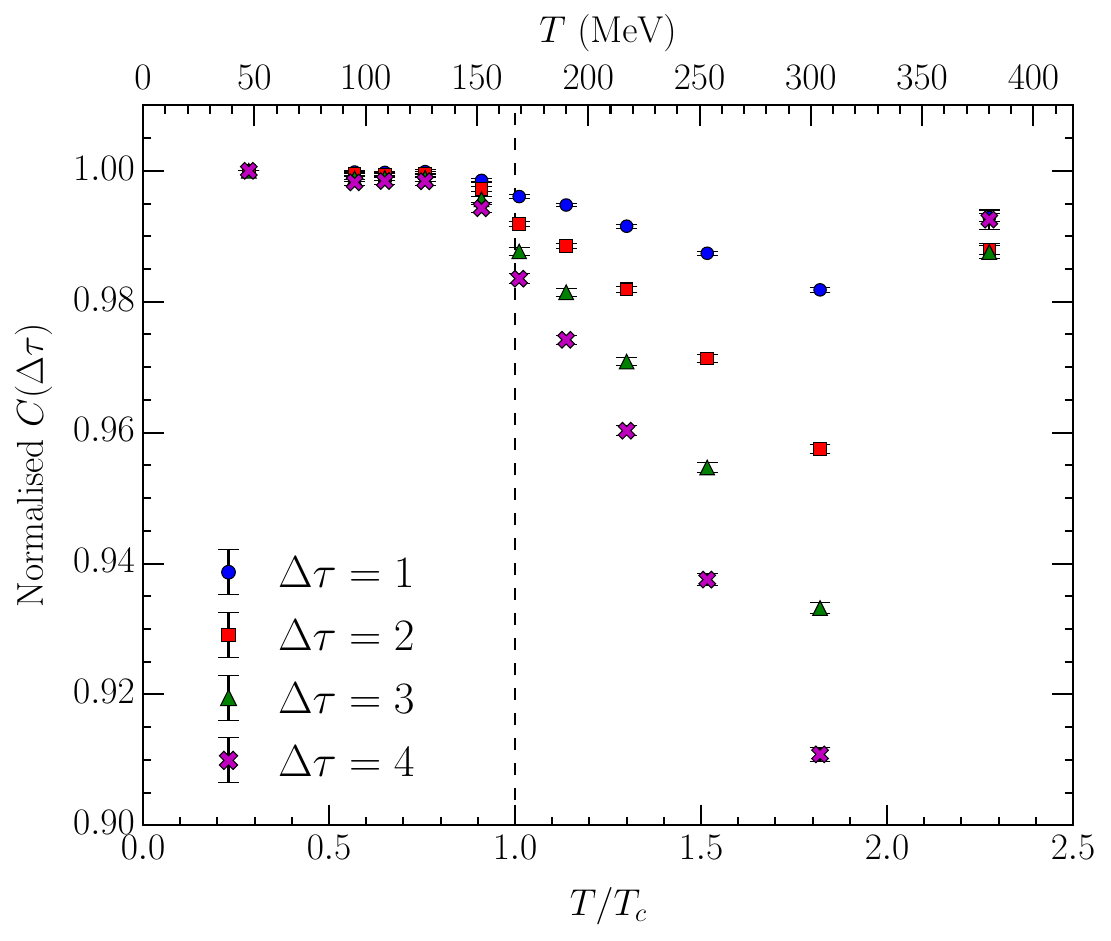}
	\vspace{-2em}
	\caption{\label{fig:normalisedcorrelation} The correlation $C(\Delta\tau)$ normalised by its value on our lowest-temperature ensemble separately for each $\Delta\tau$. This reveals a statistically significant decrease that occurs between $1 \leq T/T_c \lesssim 2$, before a turning point at $T/T_c \approx 2$ leads to the increase in Fig.~\ref{fig:temporalcorrelation}. The decrease becomes more pronounced as $\Delta\tau$ increases.}
\end{figure}

We start by recalling how the correlation measure is constructed~\cite{Mickley:2024zyg}. To quantify the similarity of the vortex structure between various temporal slices, an indicator function is defined:
\begin{equation}
	\chi_{ij}(\mathbf{x}, \tau; \Delta\tau) = \begin{cases}
		1, & \!\!\! m_{ij}(\mathbf{x},\tau) \, m_{ij}(\mathbf{x},\tau\!+\!\Delta\tau) > 0 \,, \\
		0, & \!\!\! \text{otherwise} \,, \\
	\end{cases}
\end{equation}
where $m_{ij}(\mathbf{x}, \tau)$ is the centre charge of the plaquette $P_{ij}(\mathbf{x}, \tau)$, as defined in Eq.~(\ref{eq:centreprojplaq}). It takes the value $1$ if a nontrivial plaquette at spatial position $\mathbf{x}$ in temporal slice $\tau$ is also pierced by a vortex in the same direction in the later temporal slice $\tau+\Delta\tau$. This naturally leads to the correlation
\begin{equation} \label{eq:correlation}
	C(\Delta\tau) = \frac{1}{N_\mathrm{vor} \, N_\tau} \,\sum_{\substack{\mathbf{x},\,\tau,\\i,\,j}} \chi_{ij}(\mathbf{x}, \tau; \Delta\tau) \,,
\end{equation}
where $N_\mathrm{vor}$ is the average number of vortices per temporal slice. It represents the probability that a given pierced plaquette remains invariant between slices $\tau$ and $\tau+\Delta\tau$, such that $C(\Delta\tau) = 1$ for every $\Delta\tau$ signals perfect alignment [note that $C(\Delta\tau = 0) \equiv 1$ by default].  Due to periodic boundary conditions, $\Delta\tau$ is effectively restricted to $\Delta\tau \leq \lfloor N_\tau/2 \rfloor$. Consequently, we compare $\Delta\tau = 1$--$4$ across all our ensembles. This evolution is presented in Fig.~\ref{fig:temporalcorrelation}.

Noting that a temporal alignment of the vortex sheet is embodied by an increase in the correlation, we find no evidence for an alignment until close to twice the chiral transition temperature. There is a slight upturn in value of $C(\Delta\tau)$ on our ensemble corresponding to $T/T_c \simeq 2.28$, which then continues to grow over the remaining temperature range. This verifies our claim that there is still a soft alignment of the centre vortex sheet with the temporal dimension at high temperatures in the presence of dynamical fermions, though this does not onset until well above $T_c$. It appears the change in vortex geometry that occurs at $T_c$ in the pure gauge theory does not match with the chiral transition in full QCD but rather with a higher temperature close to $2\,T_c$.

To further substantiate this last point, a close inspection of Fig.~\ref{fig:temporalcorrelation} reveals that the correlation \textit{decreases} between $1\leq T/T_c \lesssim 2$, following a minute drop in value at $T_c$ itself. Independently for each $\Delta\tau$, we normalise $C(\Delta\tau)$ by its value on our lowest-temperature ensemble and plot the result up to and including the point at $T/T_c \simeq 2.28$. The end product is presented in Fig.~\ref{fig:normalisedcorrelation}.

This highlights that the vortex correlation is sensitive to the chiral transition. There is a crossover that takes place through $T_c$, succeeded by a statistically significant decrease for intermediate temperatures $1\leq T_c \lesssim 2$. The turning point near $T/T_c \approx 2$ is also brought front and centre in Fig.~\ref{fig:normalisedcorrelation}, coinciding with the onset of the temporal alignment of the vortex sheet. The crossover at $T_c$ and subsequent decrease is stronger for larger $\Delta\tau$. If the correlation weakens, this signifies that the vortex structure undergoes greater change between consecutive slices. One could imagine that these changes ``stack up" over multiple slices, leading to this observation.

It is also interesting to note that the value of the correlation on consecutive slices $C(\Delta\tau=1)$ at low temperatures is substantially greater here than observed on the pure gauge ensembles in Ref.~\cite{Mickley:2024zyg}. This can be directly attributed to the anisotropy. By reducing the physical separation between temporal slices, without changing the distance a vortex must move to pierce an adjacent space-space plaquette (controlled by the spatial lattice spacing), the probability that said vortex remains invariant between consecutive slices will be considerably larger.

Additionally, the distance between $C(\Delta\tau)$ for the various choices of $\Delta\tau$ diminishes as the temperature rises above $T/T_c \gtrsim 2$. This is especially apparent when comparing $\Delta\tau=3$ and $4$ at the highest temperature, which lie virtually on top of each other. Indeed, $C(\Delta\tau)$ would necessarily be constant with $\Delta\tau$ (and equal to $1$) if the vortex sheet were perfectly aligned with the temporal dimension. It follows that one expects the separation between the $C(\Delta\tau)$ to lessen as the alignment strengthens.

\subsection{Cluster extent} \label{subsec:extents}
Following on from the correlation, we now seek an explicit assessment of the key change in vortex geometry crossing the transition---the loss of percolation in spatial slices. In particular, it is curious to note that the increase in correlation at $T/T_c \simeq 2.28$ coincides with the absence of percolation observed in the corresponding visualisation, Fig.~\ref{fig:Nt16vis}. This is achieved with the cluster extent. For this, we follow the procedure in Refs.~\cite{Engelhardt:1999fd, Biddle:2023lod}, though modified to account for the anisotropy present in our ensembles.

\begin{figure}
	\centering
	\includegraphics[width=\linewidth]{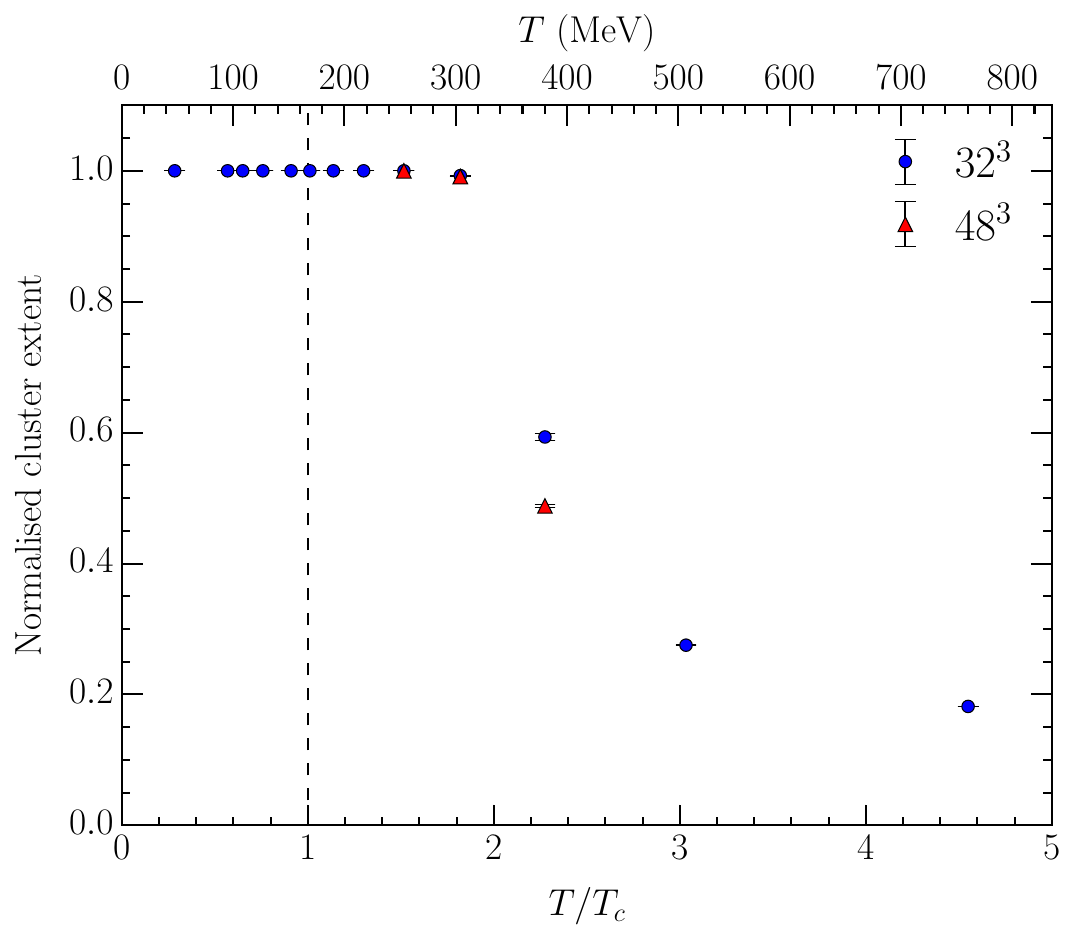}
	\vspace{-2em}
	\caption{\label{fig:clusterextent} The normalised cluster extent for spatial slices defined in Eq.~(\ref{eq:clusterextent}). It attains a value of $\approx\! 1$ for all $T/T_c \lesssim 2$, indicating the vortex structure remains percolating until twice the chiral transition temperature. It thereafter rapidly decreases as percolation is lost.}
\end{figure}

For each spatial slice, we ascertain the largest physical pairwise distance $d_\mathrm{cluster}$ between any two vortices within the same cluster (accounting for periodic boundary conditions). This is then normalised by the maximum possible such distance,
\begin{gather} \label{eq:clusterextent}
	d_\mathrm{norm} = \frac{d_\mathrm{cluster}}{d_\mathrm{max}} \,, \\
	d_\mathrm{max} = \sqrt{2 \left(\! \frac{a_s N_s}{2} \!\right)^{\! 2} \! + \left(\! \frac{a_\tau N_\tau}{2} \!\right)^{\! 2}} \,,
\end{gather}
where $N_s$ and $N_\tau$ are the number of sites in the spatial and temporal dimensions, respectively. This is averaged over all spatial slices for each configuration. The normalised cluster extent is presented with temperature in Fig.~\ref{fig:clusterextent}.

If the vortex sheet percolates all spacetime, we expect $d_\mathrm{norm} = 1$ in every slice within a very small margin of error, since any given slice will contain a primary cluster that fills the entire three-dimensional volume. Unsurprisingly, this is the observed behaviour everywhere below $T_c$. Remarkably, however, it is in fact unchanged all the way up to $T/T_c \approx 2$. A value marginally (but clearly) less than $1$ is attained initially on our ensemble with $T/T_c \simeq 1.82$, before a sharp decline in value subsequently takes hold. This indicates values of the cluster extent consistently less than $1$ across spatial slices, thereby signifying the loss of percolation in the vortex structure. Figure \ref{fig:clusterextent} therefore confirms our presumption from the visualisations that no discernible change to the vortex sheet's percolating nature occurs at $T_c$ itself.

This is an extremely interesting finding. Given vortex percolation implies a confining static quark potential, Fig.~\ref{fig:clusterextent} entails that the vortex-only fields retain confinement through temperatures as high as $T \approx 1.8\,T_c$. Such behaviour opens the possibility for a ``transitional" period between confinement and deconfinement, with two transition temperatures. The established $T_c$ describes the pseudocritical chiral transition, and a second transition temperature $T_d$ would describe deconfinement.

As recounted in Sec.~\ref{sec:rev}, subtle hints for the existence of three distinct phases in full QCD have been previously noted~\cite{Aarts:2023vsf, Cardinali:2021mfh, Glozman:2016swy, Rohrhofer:2019qal, Rohrhofer:2019qwq, Glozman:2022zpy, Philipsen:2022wjj, Chiu:2023hnm, Glozman:2024ded, Cohen:2024ffx, Petreczky:2015yta, Shuryak:2017fkh, Hanada:2023krw, Hanada:2023rlk, Kotov:2021rah, Kotov:2021ujj, Kotov:2021mpi, Kotov:2022inz, Alexandru:2019gdm}. These place the second transition temperature within the ballpark of $T_d/T_c \approx 2$. However, its precise value is not well known, with proposals ranging from $T_d \approx 200$--$500$\,MeV.

The middle of this range is in agreement with the geometrical changes to the centre vortex sheet that transpire at high temperatures. As the subtle preference for the vortex sheet to align with the temporal dimension emerges at roughly the same temperature percolation ceases, this indicates both of these properties could be used to characterise the second transition temperature.

Still, in the context of this work one must consider the impact of a finite volume on the cluster extent. One could imagine that for a sufficiently small physical volume, the vortex sheet would artificially appear percolating with the finite cluster size unable to be resolved. Hence, we must account for the possibility that increasing the volume could shift the perceived loss of percolation to lower temperatures. We investigate this by generating three new ensembles with a spatial volume of $48^3$ at the temperatures $T/T_c \simeq 1.52$, $1.82$ and $2.28$. These surround the loss of percolation currently observed in Fig.~\ref{fig:clusterextent}.

Focusing initially on $T/T_c \simeq 2.28$ in Fig.~\ref{fig:clusterextent}, we find the normalised cluster extent has been reduced by increasing the volume. This is the expected behaviour if percolation has been lost, for the finite cluster size occupies a smaller fraction of the total volume. It follows that the ratio in Eq.~(\ref{eq:clusterextent}) decreases. Conversely, if the vortex sheet is percolating the cluster size should grow to fill the larger volume. This is our finding at $T/T_c \simeq 1.52$, with no difference between the normalised cluster extents on the two volumes.

It is curious that the value of the normalised cluster extent at $T/T_c \simeq 1.82$, which is marginally less than 1, is also unchanged. Certainly, this point cannot be clearly above the percolation transition for otherwise the normalised cluster extent would decrease, as previously justified. Furthermore, it cannot be significantly below the transition, since then we imagine it would be exactly 1 in accordance with all lower temperatures, or at least tend to 1 in the infinite volume limit. It seems probable then that this temperature of $T \simeq 304$\,MeV lies in very close vicinity to the transition point.

We also compare the \textit{physical} (i.e. unnormalised) cluster extent $d_\mathrm{cluster}$. These are provided in Table~\ref{tab:clusterextents}, along with the normalised extents for reference.
\begin{table}
	\caption{\label{tab:clusterextents} The normalised ($d_\mathrm{norm}$) and physical ($d_\mathrm{cluster}$) cluster extents compared between spatial volumes of $32^3$ and $48^3$, for each considered temperature. Of particular note is $T/T_c \simeq 2.28$, at which percolation has definitively been lost. Here, in moving to the larger volume, the normalised extent decreases, yet its physical counterpart increases.}
	\begin{ruledtabular}
		\begin{tabular}{cD{.}{.}{2.8}D{.}{.}{2.8}D{.}{.}{2.7}D{.}{.}{2.7}}
			\multirow{2}{*}{$T/T_c$} & \multicolumn{2}{c}{$d_\mathrm{norm}$} & \multicolumn{2}{c}{$d_\mathrm{cluster}$ (fm)} \\
			& \multicolumn{1}{c}{$32^3$} & \multicolumn{1}{c}{$48^3$} & \multicolumn{1}{c}{$32^3$} & \multicolumn{1}{c}{$48^3$} \\
			\colrule \\[-0.9em]
			1.52 & 1.0000(0)  & 1.0000(0)  & 2.564(0)  & 3.821(0) \\
			1.82 & 0.9925(7)  & 0.9912(6)  & 2.536(2)  & 3.782(2) \\
			2.28 & 0.5939(44) & 0.4879(23) & 1.513(11) & 1.859(9) \\[-0.15em]
		\end{tabular}
	\end{ruledtabular}
\end{table}
We find that although the normalised extent diminishes at $T/T_c \simeq 2.28$, the physical extent grows. Increasing the lattice dimensions provides an opportunity to reveal larger clusters than contained within the smaller volume, thus increasing the average. Crucially, the extent has not simply scaled up with the volume.

\subsection{Vortex density} \label{subsec:vortexdensity}
\begin{figure}
	\centering
	\includegraphics[width=\linewidth]{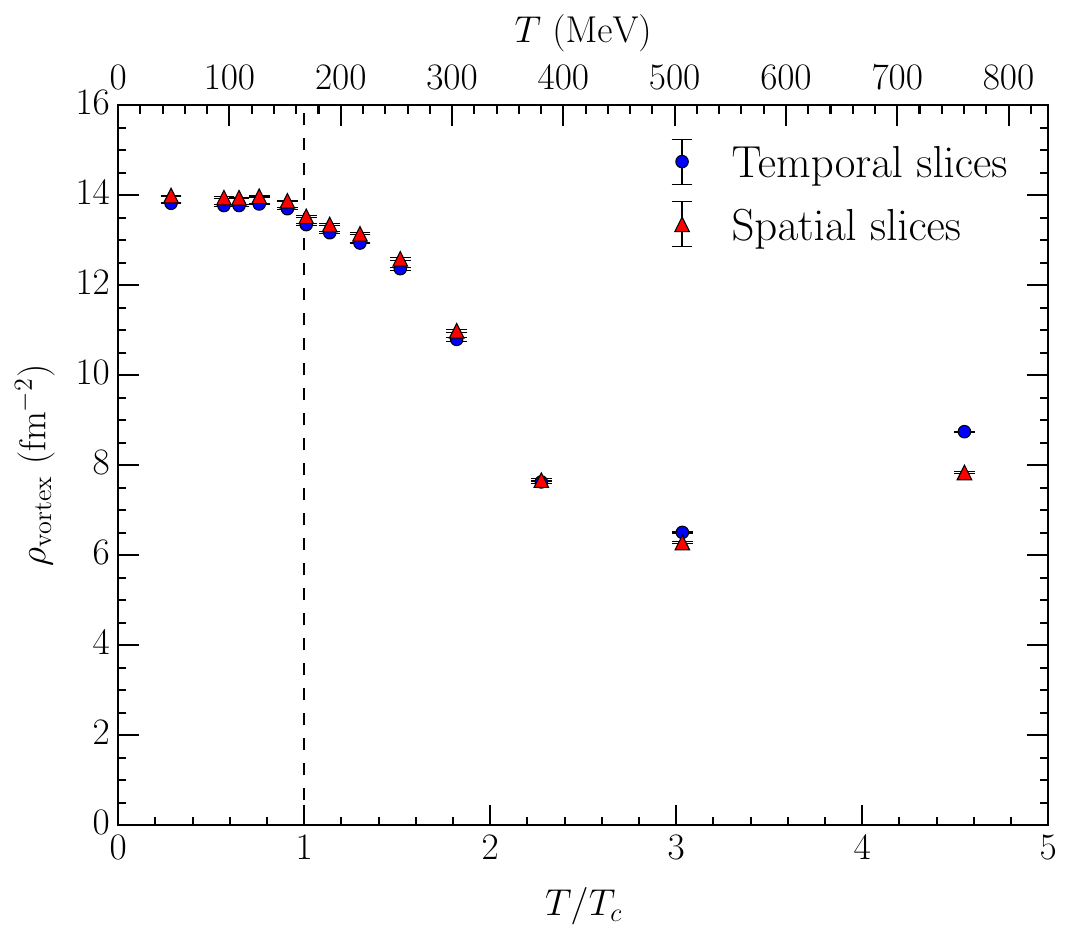}
	\vspace{-2em}
	\caption{\label{fig:vortexdensity} The vortex density in temporal and spatial slices of the lattice. Although approximately constant below $T_c$, there is a small drop in value as the chiral transition is crossed. This continues to decrease above $T_c$ except for a sudden increase at the highest temperature.}
	
	\vspace{1.5em}
	
	\includegraphics[width=\linewidth]{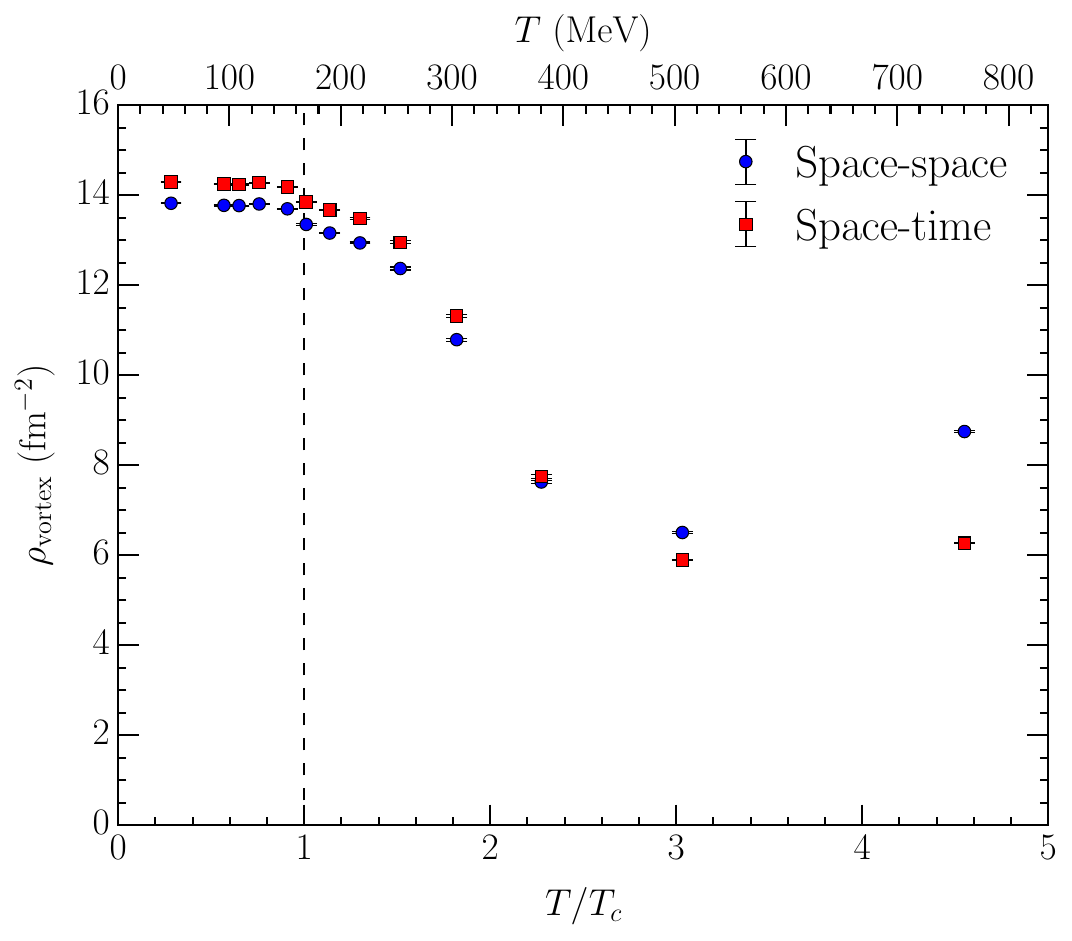}
	\vspace{-2em}
	\caption{\label{fig:ss_st} The vortex density decomposed into space-space and space-time areas. As discussed prior, the anisotropy effects a higher density of space-time plaquettes at low temperatures, though this undergoes an inversion for $T/T_c \gtrsim 2$. This can be attributed to the slight alignment of the vortex sheet with the temporal dimension, which results in a greater relative proportion of space-space plaquettes being pierced.}
\end{figure}

We now move to investigate a more intrinsic aspect of centre vortices, the vortex density. This is defined simply as the number of plaquette piercings per unit area. In this section, we consider decomposing the vortex density in two ways. First, to connect to the visualisations we calculate a density in temporal and spatial slices of the lattice. Denoting by $N_\mathrm{slice}$ the number of sites in a given three-dimensional slice, the areas spanned by the plaquettes in temporal and spatial slices are
\begin{align}
	A_\mathrm{t} = 3 \, N_\mathrm{slice} \, a_s^2 \,, && A_\mathrm{s} = N_\mathrm{slice} \left( a_s^2 + 2 \, a_s a_\tau \right) \,,
\end{align}
respectively. Thus, the required definitions of the vortex density are
\begin{equation}
	\rho_\mathrm{vortex} = \frac{N_\mathrm{pierced}}{A_{\mathrm{t,s}}} \,,
\end{equation}
where $N_\mathrm{pierced}$ is the number of plaquettes in that slice pierced by a vortex. These are shown in Fig.~\ref{fig:vortexdensity}.

Like the correlation calculated earlier, the vortex density is seen to be sensitive to the chiral crossover at $T_c$. In a similar vein, it is  roughly constant below $T_c$ in both temporal and spatial slices, with a small but sudden drop in value crossing the chiral transition. There is an ensuing steady decline in the densities up to $T/T_c \approx 3$ before they increase in moving to our highest temperature. This is distinct from the correlation, which featured its turning point closer to $T/T_c \approx 2$. The late upturn matches equivalent behaviour in the pure gauge sector that saw an increase in vortex density following the initial drop at $T_c$~\cite{Mickley:2024zyg}.

The reduction in density following $T_c$ was originally hypothesised from the visualisations, in particular via Fig.~\ref{fig:Nt16vis} at $T/T_c \simeq 2.28$. A lower vortex density for the percolating cluster implies a weaker area-law falloff for Wilson loops, indicating a lessening of the string tension over this temperature range even though the static quark potential remains linear. In this sense, one could argue that confinement ``weakens" in the transitional phase between the chiral and deconfinement transitions, before it is lost at $T_d$.

Another prominent feature of Fig.~\ref{fig:vortexdensity} is the crossing of spatial- and temporal-slice densities that takes place at $T/T_c \approx 2$. Given that spatial slices comprise both space-space and space-time plaquettes, this motivates exploring the temperature evolution separately for space-space and space-time vortex densities. These were originally calculated at zero temperature in Sec.~\ref{sec:anisotropy} and are displayed now for all temperatures in Fig.~\ref{fig:ss_st}. Note that in averaging over all slices, the temporal-slice density shown in Fig.~\ref{fig:vortexdensity} is identical to a direct calculation of the total proportion of space-space plaquettes pierced. The purpose of Fig.~\ref{fig:ss_st}, then, is to replace the density in spatial slices (which reflects a mixture of plaquette orientations) to exclusively the proportion of space-time plaquettes pierced.

Predictably, there is stronger tension between the space-space and space-time densities than among temporal and spatial slices. This is also true at high temperatures. In particular, there is only a very soft upturn in the density of space-time plaquettes pierced moving to our highest temperature in comparison to space-space plaquettes. This emphasises that the increase in vortex density is linked primarily to the percolating cluster that persists in spatial dimensions following deconfinement, approaching a state of disorder after an initial thinning out at intermediate temperatures.

As a result, the crossing of densities for $T/T_c \gtrsim 2$ is also more pronounced for the space-space and space-time density analysis. With the deeper understanding offered by Fig.~\ref{fig:ss_st}, it is clear that the inversion of the densities is a direct by-product of the gentle alignment of the vortex sheet with the temporal dimension for $T/T_c \gtrsim 2$. As the sheet aligns with the temporal dimension, there are fewer instances of space-time plaquettes pierced relative to space-space piercings. The trend in Fig.~\ref{fig:ss_st} follows.

\begin{figure}
	\centering
	\includegraphics[width=\linewidth]{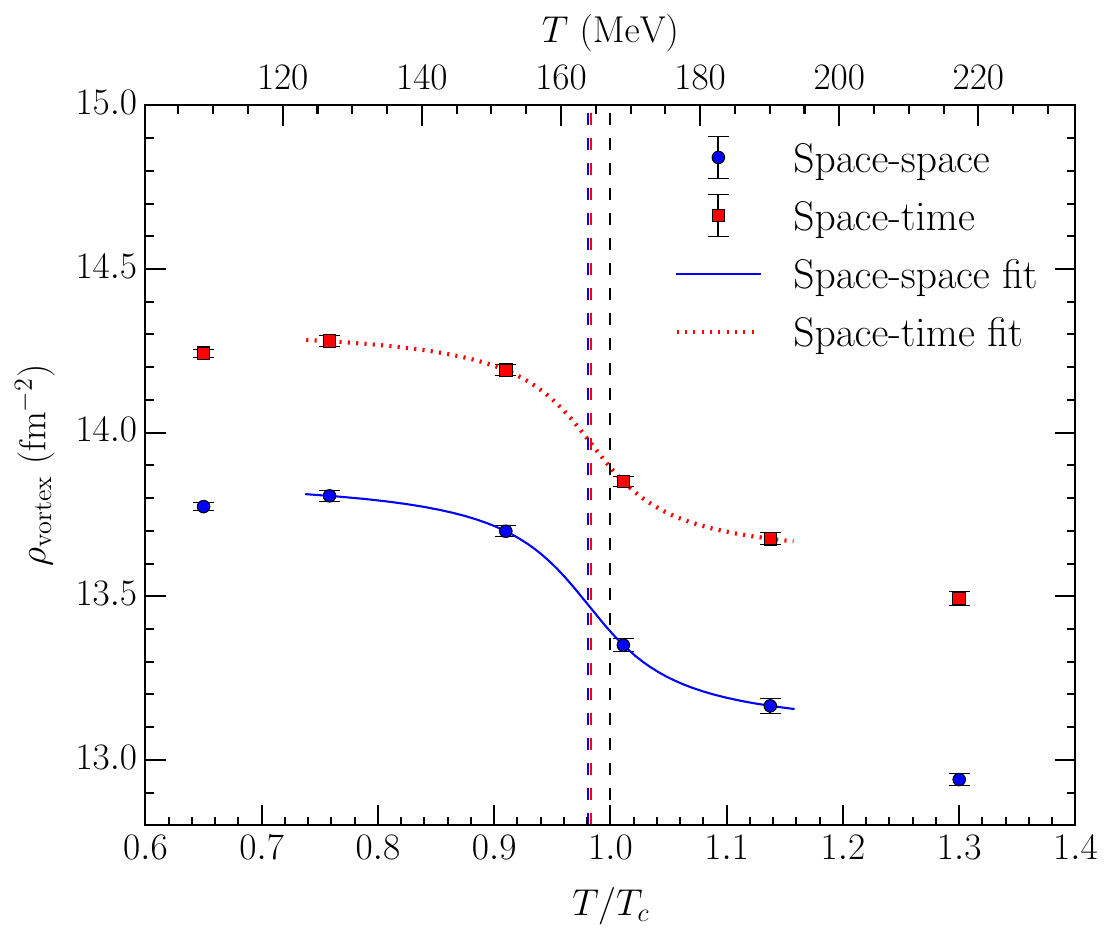}
	\vspace{-2em}
	\caption{\label{fig:vortex_density_near_Tc} The space-space and space-time vortex densities enlarged on $T_c$. There is visible crossover behaviour through the transition point, described by the ansatz in Eq.~(\ref{eq:crossover_ansatz}). The resulting inflection points are strongly coincident with each other and in reasonable agreement with $T_c$.}
		
	\vspace{1.5em}
		
	\includegraphics[width=\linewidth]{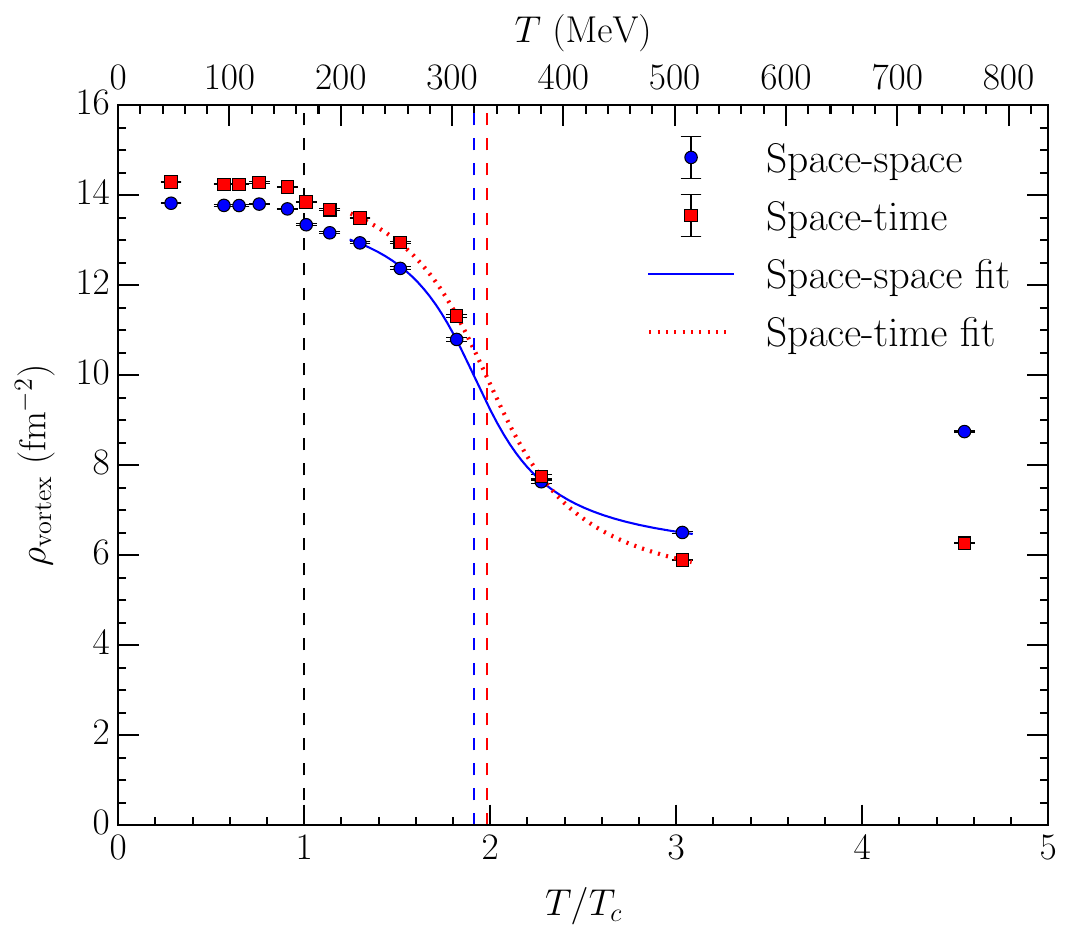}
	\vspace{-2em}
	\caption{\label{fig:vortex_density_fits} The space-space and space-time vortex densities with the ansatz of Eq.~(\ref{eq:crossover_ansatz}) fit to a range of temperatures around $T/T_c \approx 2$. We perform the fits starting at the first temperature above the range used near $T_c$ in Fig.~\ref{fig:vortex_density_near_Tc}. The two inflection points are again similar to each other, and provide an improved estimate of the proposed second transition temperature.}
	
	\vspace{-0.4em}
\end{figure}

Given that the vortex density is susceptible to the chiral transition, as with the correlation measure we proceed to explore its behaviour around $T_c$ in greater detail. For this, we proceed referring exclusively to the space-space and space-time densities as these provide a more natural division of the vortex density. In Fig.~\ref{fig:vortex_density_near_Tc} we provide an enlargement of the densities in a narrow range around $T_c$.

In fact, this reveals that a crossover takes place through the chiral transition, with the value nearest to but below $T_c$ attaining a visibly lower density than the point to its left. This encourages performing fits to extract estimates of the transition point which can be compared against $T_c$. The crossover behaviour is well described by the sigmoid functional form~\cite{Burger:2018fvb, Aarts:2019hrg, Aarts:2020vyb}
\begin{equation} \label{eq:crossover_ansatz}
	f(x) = c_0 + c_1 \arctan\left(c_2 \, (x - x_c)\right) \,.
\end{equation}
Its inflection point occurs at the critical value $x = x_c$. We note that as Eq.~(\ref{eq:crossover_ansatz}) features four parameters, and there are only four temperatures in Fig.~\ref{fig:vortex_density_near_Tc} that conform to this description, the resulting fits are ``perfect". Nevertheless, we still believe it is interesting to see how close the inflection points land to $T_c$. The fits are overlaid in Fig.~\ref{fig:vortex_density_near_Tc}, along with vertical dashed lines corresponding to the inflection points; their precise values are provided in Table~\ref{tab:inflectionpoints}.
\begin{table*}
	\caption{\label{tab:inflectionpoints} The inflection points $T_1$ and $T_2$ obtained from fitting Eq.~(\ref{eq:crossover_ansatz}) to various vortex quantities, provided both in terms of $T_c$ and in MeV. These translate to estimates of the chiral transition temperature $T_c$ and the proposed deconfinement point $T_d$ relevant to this work. The statistical uncertainty is represented by the number in the first set of the brackets, while the second error for the ratios $T_{1,2}/T_c$ reflects the systematic uncertainty from $T_c = 167(3)\,$MeV.}
	\begin{ruledtabular}
		\begin{tabular}{clcccc}
			Quantity & \multicolumn{1}{c}{Fit} & $T_1/T_c$ & $T_1$ (MeV) & $T_2/T_c$ & $T_2$ (MeV) \\
			\colrule \\[-1em]
			\multirow{2}{*}{$\rho_\mathrm{vortex}$} 
			& Space-space & 0.981(8)(18) & 163.9(1.3) & 1.913(6)(34) & 319.7(1.0) \\
			& Space-time  & 0.983(8)(18) & 164.1(1.4) & 1.983(6)(36) & 331.4(1.0) \\
			\colrule \\[-1em]
			\multirow{2}{*}{$\rho_\mathrm{branch}$} 
			& Temporal slices & 0.982(8)(18) & 164.1(1.3) & 1.881(5)(34) & 314.3(0.9) \\
			& Spatial slices  & 0.983(8)(18) & 164.2(1.3) & 1.915(6)(34) & 320.0(0.9) \\
			\colrule \\[-1em]
			\multirow{2}{*}{$\lambda_\mathrm{branch}$} 
			& Temporal slices & 0.984(9)(18) & 164.5(1.6) & 1.921(7)(35) & 321.0(1.2) \\
			& Spatial slices  & 0.983(9)(18) & 164.3(1.6) & 1.935(8)(35) & 323.4(1.4) \\[-0.25em]
		\end{tabular}
	\end{ruledtabular}
\end{table*}

The first error is the statistical uncertainty on the inflection points, while the second error for the ratios is the contribution from the \textsc{Fastsum} uncertainty on $T_c = 167(3)\,$MeV. It is reassuring that the inflection points obtained separately from the space-space and space-time vortex densities are in strong agreement with each other. In addition, the values of $T_1/T_c \simeq 0.98$, where $T_1$ denotes the inflection point near $T_c$, are certainly comparable to $1$, and impressively agree with $1$ within uncertainty. The dominant source of error is provided by $T_c$ rather than the inflection points. Similarly, it can be seen that the inflection points $T_1$ in MeV are consistent with $T_c$ within statistical uncertainty. In this regard, one could even propose performing these fits to the vortex density as an alternative method of estimating the chiral transition point. $T_1$ is therefore interpreted as an estimate of $T_c$. It will be interesting to consider a finer temperature resolution near $T_c$ to constrain the fits, and to investigate the approach to the continuum limit.

Having performed this analysis, we now return to the vortex density over the full temperature range, Fig.~\ref{fig:ss_st}, and observe similar crossover behaviour in the vicinity of $T/T_c \approx 2$ near the proposed deconfinement transition point. Accordingly, we perform the fits of Eq.~(\ref{eq:crossover_ansatz}) to a range of temperatures around $T/T_c = 2$. We reproduce Fig.~\ref{fig:ss_st} with these fits displayed in Fig.~\ref{fig:vortex_density_fits}.

The two newly obtained inflection points $T_2$ are evidently very similar, landing within a few percent of each other, though certainly with a wider gap than those associated with the fits around $T_c$. This is to be expected given the sharper decrease in density exhibited by space-time plaquettes and the considerably wider temperature range over which the fits are performed. Similar to the interpretation of $T_1$ as an estimate of $T_c$, we now interpret $T_2$ as an estimate of the proposed deconfinement point $T_d$ at which vortex percolation is lost. Taking an average of the two inflection points places our current best estimate for the hypothesised second transition at $T_d/T_c \simeq 1.95$, or equivalently $T_d \simeq 326\,$MeV. This is a little larger than the cluster-extent estimate of $T_d \simeq 304\,$MeV.

\subsection{Secondary clusters} \label{subsec:secondary}
As an extension to our discussion on the amount of vortex matter, we additionally consider the number of secondary clusters found in three-dimensional slices of the vortex sheet. In the pure gauge theory, this number underwent a sharp drop at the critical temperature~\cite{Mickley:2024zyg}. With the abundance of secondary clusters observed with dynamical fermions compared to the pure gauge theory at near-zero temperatures~\cite{Biddle:2023lod}, it is therefore of interest to extend this comparison to high temperatures. In order to include both temporal and spatial slices in this discussion, we limit the temperature range for spatial slices to those that comprise a percolating cluster, such that the notion of ``primary" and ``secondary" clusters is well defined. For temporal slices, we perform a comparison across the full temperature range.

\begin{figure}
	\centering
	\includegraphics[width=\linewidth]{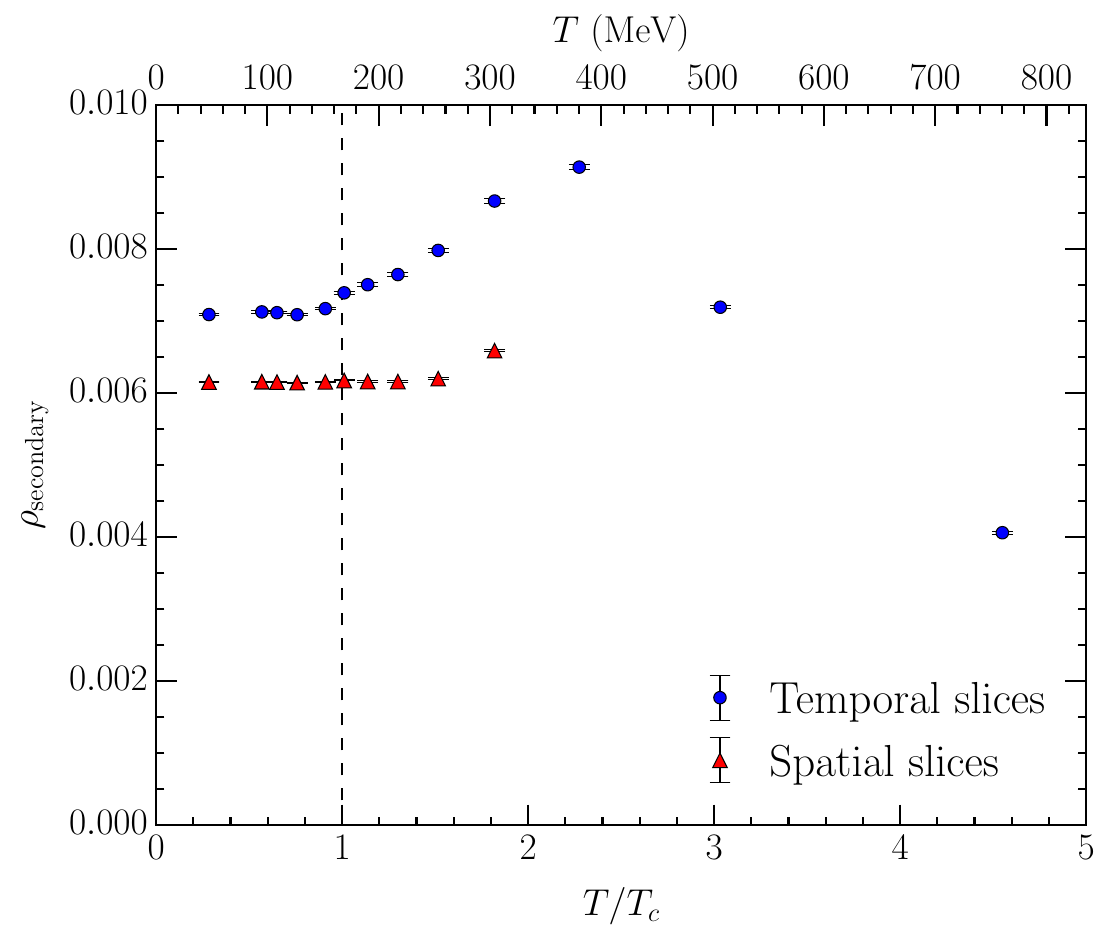}
	\vspace{-2em}
	\caption{\label{fig:secondaryclusters} The temperature dependence of the normalised ``density" of secondary clusters in temporal and spatial slices, as described in the text. Its value in the spatial slices is approximately constant. In temporal slices, however, it is sensitive to the chiral transition and increases over the range $1 \leq T/T_c \leq 2$ despite the diminishing overall vortex density. Subsequently, the amount of secondary clusters begins to decline for $T/T_c \gtrsim 2$.}
\end{figure}

To account for the variation in volume, we compute a normalised ``cluster density" $\rho_\mathrm{secondary}$ through dividing the number of secondary clusters by $N_\mathrm{slice}$. The result is presented in Fig.~\ref{fig:secondaryclusters}.
The reason for computing a dimensionless density is apparent. In doing so, the values in temporal and spatial slices at low temperatures are comparable to each other. If the physical volume were used in calculating $\rho_\mathrm{secondary}$, the latter would be scaled by almost a factor of $\xi$ relative to the former.

Figure \ref{fig:secondaryclusters} reveals that the number of secondary clusters in spatial slices is approximately constant relative to the slice volume, except for an increase at the final point displayed. This temperature of $T/T_c \simeq 1.82$ is the same at which the initial hint of percolation loss was observed in Fig.~\ref{fig:clusterextent}. The percolating cluster breaks apart into many smaller clusters, causing the increase in cluster density. If extending the spatial slices data to the full temperature range, the number rapidly rises to far exceed the scale of Fig.~\ref{fig:secondaryclusters}. This solidifies that a fair comparison between temporal and spatial slices is only valid when the vortex sheet percolates all four dimension.

On the other hand, the behaviour in temporal slices is more fascinating. Like the vortex density, the number of secondary clusters in temporal slices is sensitive to the chiral transition, this time experiencing an increase at $T_c$ that develops appreciably for $1 \leq T/T_c \leq 2$. This cannot be attributed to the overall amount of vortex matter, which actually decreases in temporal slices over this range. Thereafter, $\rho_\mathrm{secondary}$ features a turning point and decreases from $T/T_c \gtrsim 2$ onward. This adds to the growing collection of vortex statistics that point to a second transition temperature important in full QCD. This is the equivalent observation to the falloff in secondary clusters at the deconfinement transition with pure gauge. Since first noting this change in Ref.~\cite{Mickley:2024zyg}, we have developed a deeper understanding of its underlying cause in connection to the temporal alignment of the vortex sheet.

\begin{figure}
	\centering
	\includegraphics[width=\linewidth]{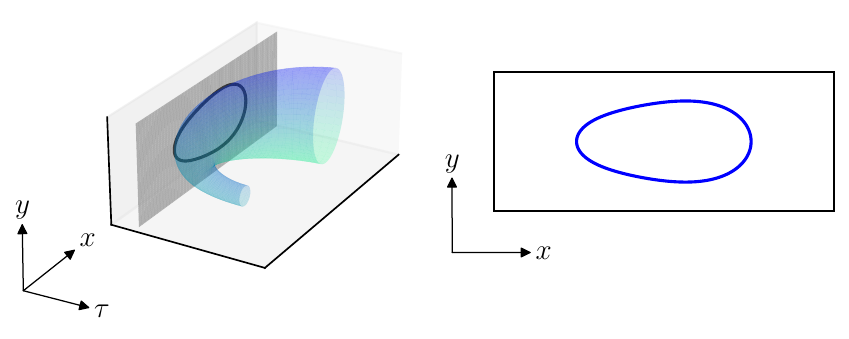}
	
	\vspace{-0.5em}
	
	\includegraphics[width=\linewidth]{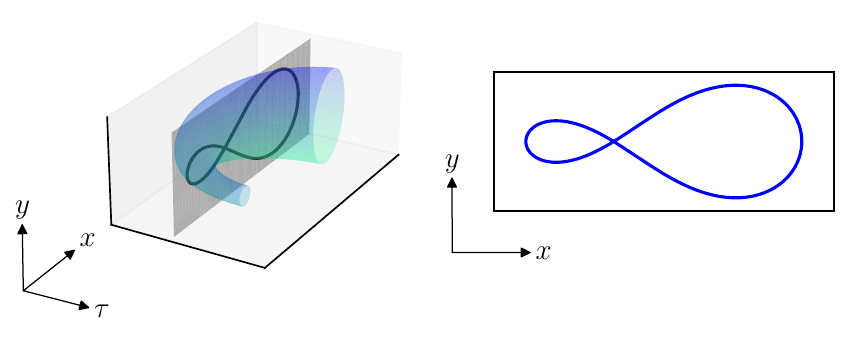}
	
	\vspace{-0.5em}
	
	\includegraphics[width=\linewidth]{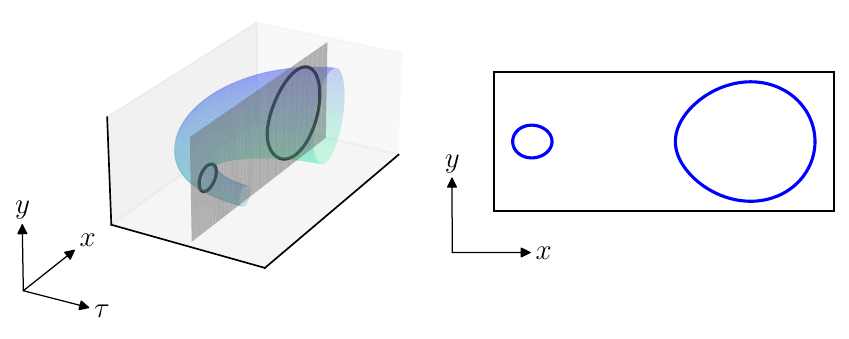}
	\caption{\label{fig:disconnected_clusters_illustration} An illustration of how disconnected clusters can arise by taking a cross section of a curved connected surface. Here, we are taking the equivalent of a temporal slice by fixing the $\tau$ coordinate and looking at an $x$-$y$ cross section. Since the surface curves back on itself in the $\tau$ dimension, depending on the coordinate of the slice we can see either a single closed curve (\textbf{top}), a curve with a self-intersection point just before it splits in two (\textbf{middle}), or two disconnected closed curves that lie in the same connected surface (\textbf{bottom}).}
\end{figure}

Even though secondary clusters appear disconnected from the percolating structure in a three-dimensional slice, it is possible that they lie within the same connected surface in four dimensions. Indeed, a single vortex sheet can curve back on itself in such a way that taking a slice of the surface results in disconnected curves in three dimensions. A basic illustration of how this can arise by considering a two-dimensional cross section of a surface in three dimensions is shown in Fig.~\ref{fig:disconnected_clusters_illustration}.

However, if the vortex sheet shifts to align with the temporal dimension, such instances where this arises become increasingly rare (cf.\ Fig.~\ref{fig:alignment_illustration}). Instead, the production of secondary clusters would occur predominantly when there are small, genuinely disconnected vortex sheets in the full four dimensions. This explains the sheer drop in the number of secondary clusters in the pure gauge sector, coinciding with the sudden strong alignment with the temporal dimension. With dynamical fermions, the steady decline in secondary clusters for $T/T_c \gtrsim 2$ matches the gentle shift toward alignment.

With this knowledge, we can also account for the increase over intermediate temperatures $1 \leq T/T_c \leq 2$ that is exclusive to temporal slices. Over this range it is understood that the vortex sheet \textit{prepares} for alignment, and Fig.~\ref{fig:secondaryclusters} reveals this manifests as ``wiggles" in the sheet along the temporal dimension. This idea is supported by the correlation in Fig.~\ref{fig:temporalcorrelation} that first decreased slightly in this intermediate phase. These frequent but small protrusions give rise to additional disconnected clusters in taking temporal slices of the full vortex sheet. This emphasises that the upward trend is not related to the quantity of vortex matter, but rather to the shape of the sheet in four dimensions. Subsequently, as percolation is lost and the alignment finally sets in, the amount of secondary clusters starts to decrease as described. Under this explanation, we would not expect to see an equivalent increase in spatial slices as the alignment singles out the temporal dimension.

\section{Branching points} \label{sec:branching}
\begin{figure}
	\centering
	\includegraphics[width=0.49\linewidth]{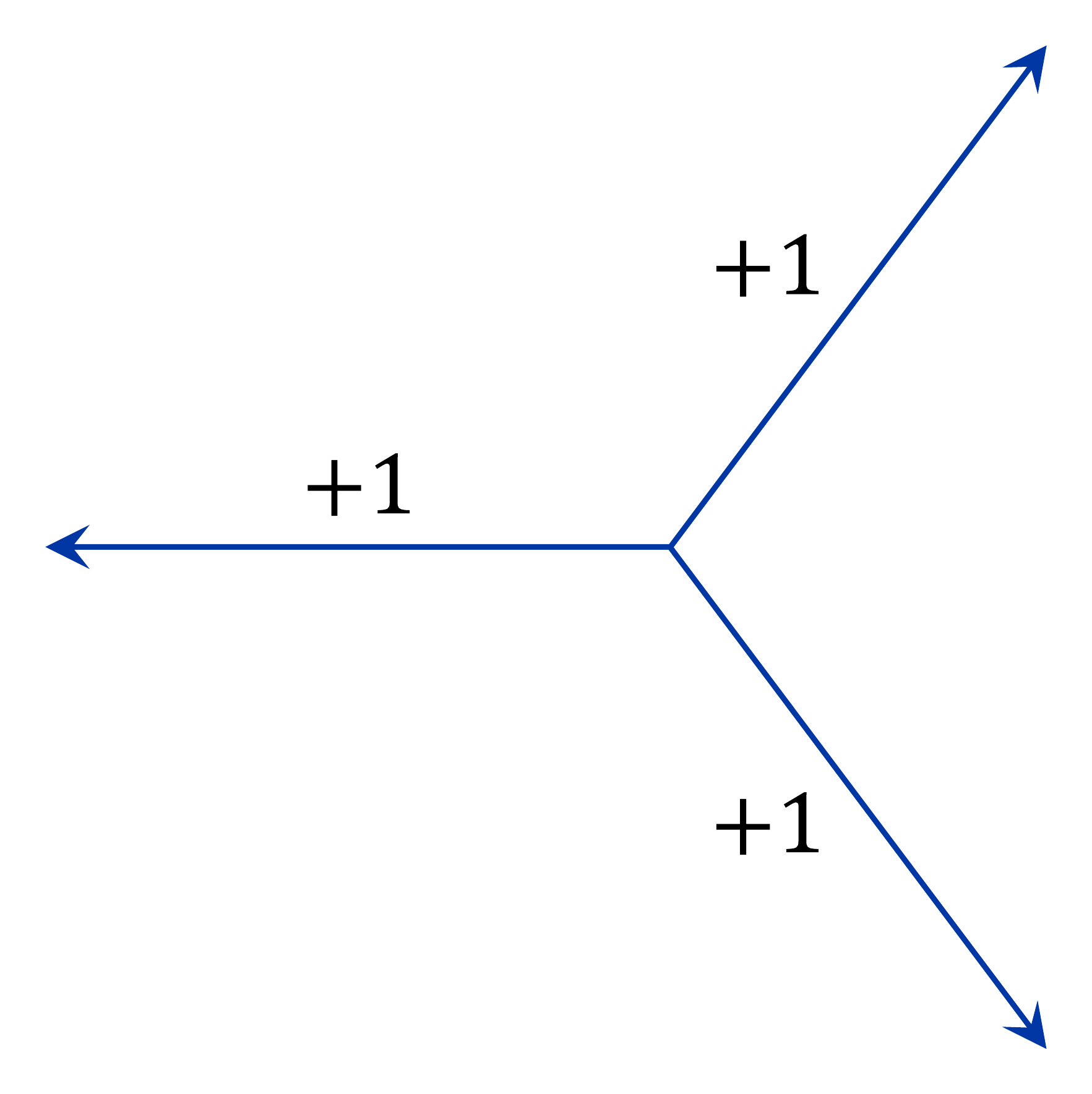}
	\includegraphics[width=0.49\linewidth]{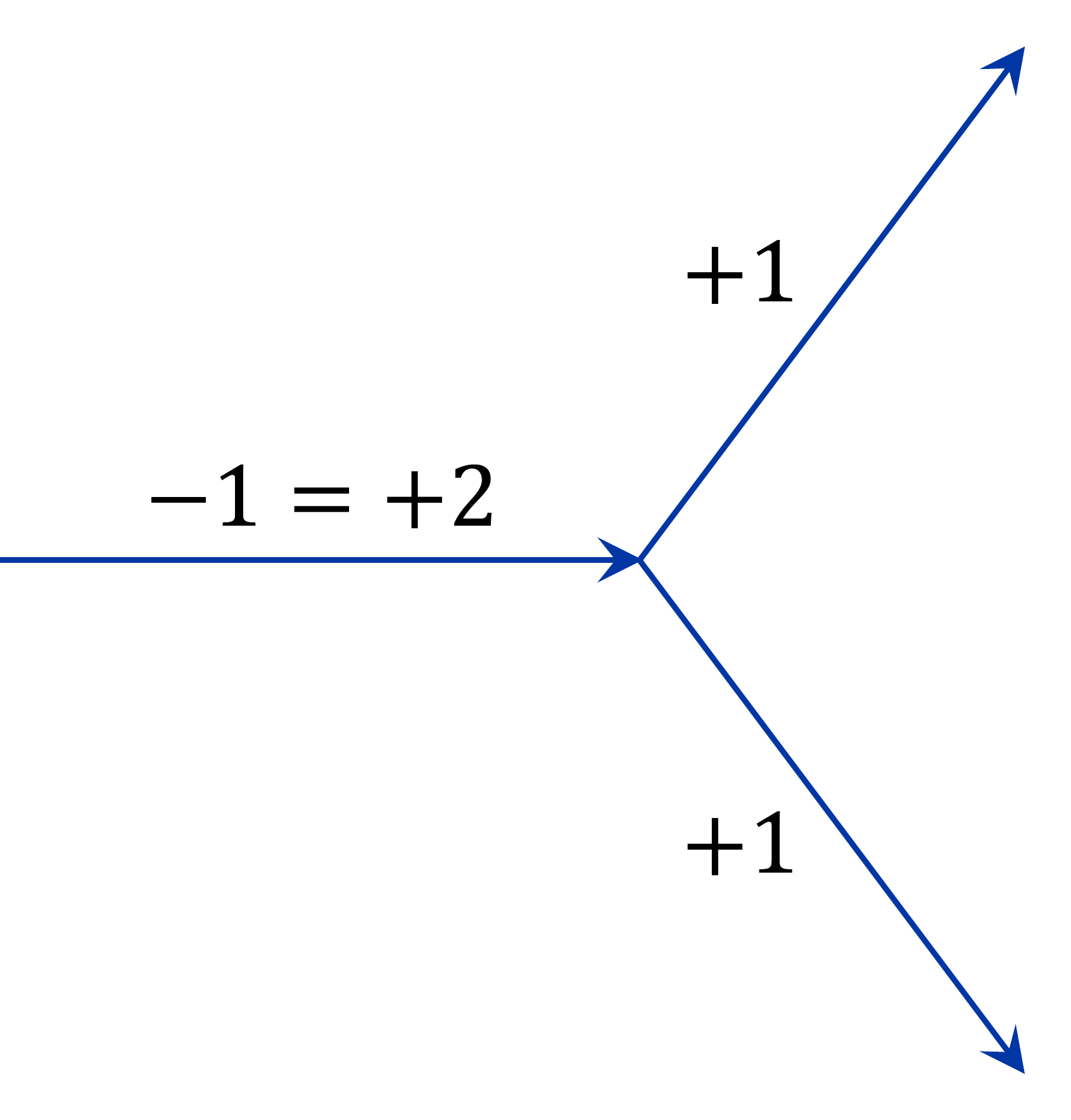}
	\caption{\label{fig:branching} Schematic of a monopole vertex (\textbf{left}) versus a branching point (\textbf{right}). The monopole follows our convention to show the directed flow of $m = +1$ centre charge. Reversal of the left-hand arrow indicates the flow of $m = -1$ charge, as seen on the right. Due to periodicity in the centre charge, $m = -1$ is equivalent to $m = +2$. Thus, the right-hand diagram depicts the branching of centre charge.}
\end{figure}

Due to the existence of two distinct nontrivial centre phases, $\mathrm{SU}(3)$ vortices are allowed to branch in which an $m=\pm 1$ vortex splits into two $m=\mp 1$ vortices. This is allowed due to the conservation of centre charge modulo $N$ in $\mathrm{SU}(N)$ gauge theory. Since reversing the orientation of a jet indicates the flow of the opposite centre charge, branching points can equivalently be interpreted as monopoles in which three vortices of the same centre charge emerge from, or converge to, a single point. This is how they manifest in the visualisations of Figs.~\ref{fig:Nt128vis}--\ref{fig:Nt8vis}. The equivalence between branching and monopole points is illustrated in Fig.~\ref{fig:branching}.

\subsection{Branching point densities} \label{subsec:branchingdensities}
\begin{figure}
	\centering
	\includegraphics[width=\linewidth]{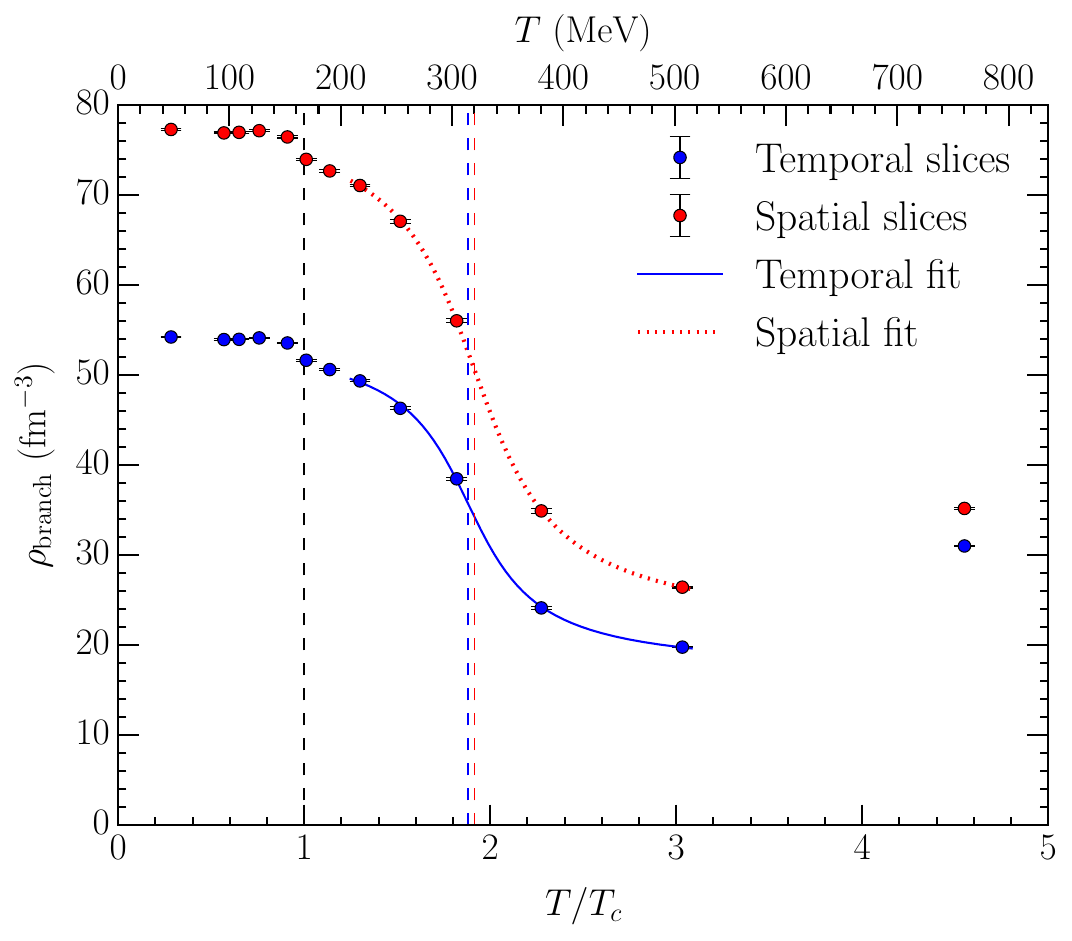}
	\vspace{-2em}
	\caption{\label{fig:branchingpointdensity} The branching point density in temporal and spatial slices of the lattice. It exhibits the same general trends as the vortex density, though is considerably more afflicted by anisotropic effects. Fits using Eq.~(\ref{eq:crossover_ansatz}) are again used to extract distinct estimates of the second transition temperature.}
	
	\vspace{1.5em}
	
	\includegraphics[width=\linewidth]{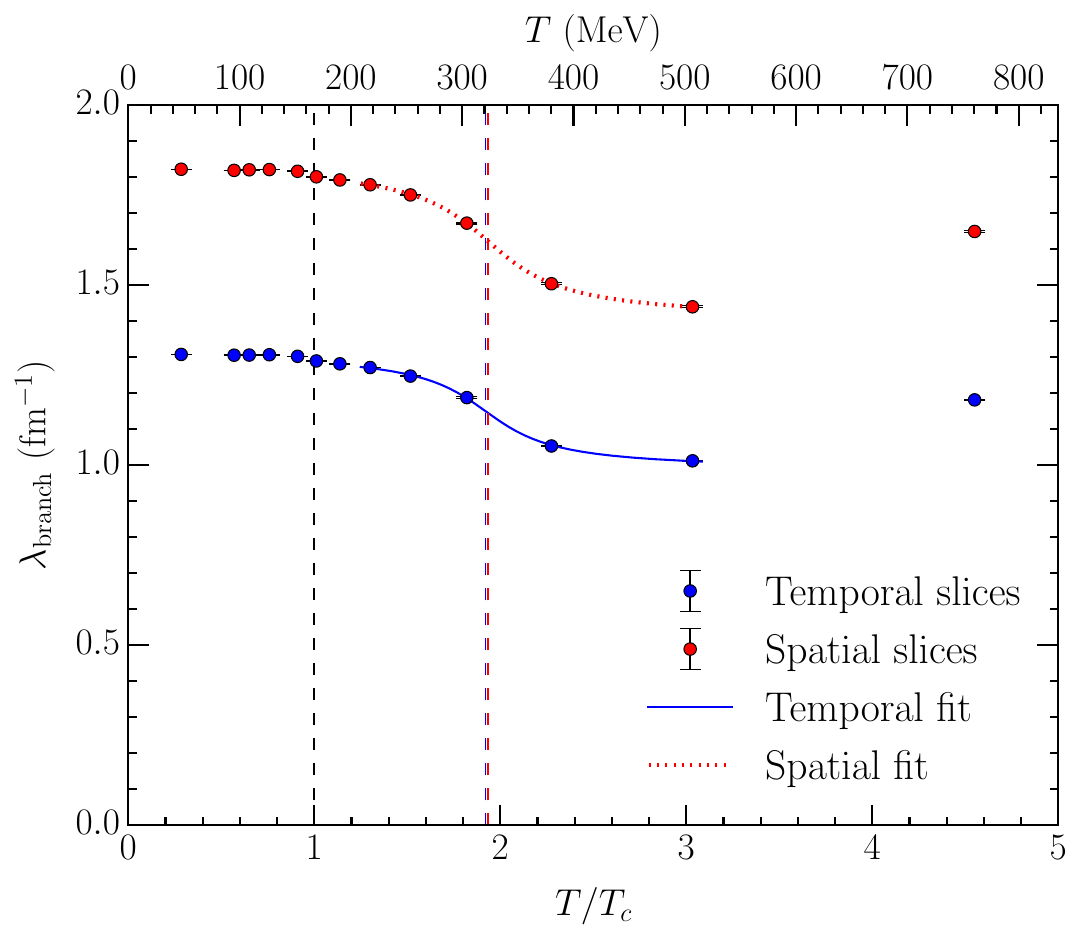}
	\vspace{-2em}
	\caption{\label{fig:linearbranchingdensity} The linear branching point density in temporal and spatial slices of the lattice. The same overarching trends are found as with the vortex and volume branching point densities, though visibly subdued. This quantity further reflects the increase in branching instances arising from the anisotropy.}
\end{figure}

The first quantity of interest is the branching point density $\rho_\mathrm{branch}$, defined as the number of branching points per unit volume for a given three-dimensional slice of the lattice. In temporal slices this volume is $N_\mathrm{slice} \, a_s^3$, while in spatial slices it is $N_\mathrm{slice} \, a_s^2 \, a_\tau$. The evolution of $\rho_\mathrm{branch}$ with temperature for these two cases is given in Fig.~\ref{fig:branchingpointdensity}.

The overarching trends obeyed by $\rho_\mathrm{branch}$ follow those of $\rho_\mathrm{vortex}$. This includes the small-yet-discernible cut at $T_c$, the continuing decline above $T_c$ and the increase in moving to our highest temperature. Notwithstanding, the falloff in branching for intermediate temperatures is more rapid than with the vortex density, especially in spatial slices. Most notable, though, is the discrepancy in branching point densities realised in temporal and spatial slices. This is in spite of the modified functional justified in Sec.~\ref{sec:anisotropy} to account for the anisotropy, and that succeeded in balancing the vortex densities. This suggests that branching point geometry is inherently sensitive to the presence of an anisotropic lattice spacing. By maintaining a finer resolution purely in the temporal dimension, there are more opportunities for a vortex to branch and pierce the additional space-time plaquettes in spatial slices. There is scope for further investigation in this regard to understand the sensitivity of branching points to the anisotropy compared to the standard vortex density. It will be interesting to learn if a more sophisticated modification to the vortex identification procedure could bring them to agreement.

Given the similar qualitative behaviour to the vortex density, we can also consider fitting Eq.~(\ref{eq:crossover_ansatz}) to the branching point densities to extract two further estimates of the deconfinement transition temperature. These fits and their associated inflection points are overlaid on Fig.~\ref{fig:branchingpointdensity}, with the values again provided in Table~\ref{tab:inflectionpoints}. We find the location of these inflection points to be marginally lower than those obtained using the vortex density; the average in this case is $T_d/T_c \simeq 1.90 \implies T_d \simeq 317\,$MeV. This is certainly within the vicinity of the estimate provided by the vortex density, and taking the lowest and highest of the four inflection points places the second transition point within $T_d \approx 314$--$331$\,MeV.

Another adjacent quantity is the \textit{linear} branching point density $\lambda_\mathrm{branch}$, i.e. the number of branching points per unit length of the vortex chain. This is calculated by dividing the number of branching points by the total length of the vortex path (in fm) within a given three-dimensional slice. Under the assumption that a vortex line features a constant rate of branching as it propagates through the slice, $\lambda_\mathrm{branch}$ provides an estimate of this rate. We will see in Sec.~\ref{subsec:chainlengths} that the internal branching point geometry is more complicated than this simple model, though $\lambda_\mathrm{branch}$ is still relevant in assessing the bulk branching properties of the vortex sheet. Its evolution with temperature in temporal and spatial slices is given in Fig.~\ref{fig:linearbranchingdensity}.

The same tendencies that we have come to expect are also reflected in $\lambda_\mathrm{branch}$, though certainly subdued in comparison to $\rho_\mathrm{vortex}$ and $\rho_\mathrm{branch}$. Importantly, this linear density also captures the enhanced branching in spatial slices that likely arises from the anisotropy. Indeed, Fig.~\ref{fig:linearbranchingdensity} underpins the idea that there is a greater rate of branching along the temporal dimension as allowed due to the extra space-time plaquettes. If one were to succeed in matching these branching rates, this would simultaneously realise equal (volume) branching point densities in temporal and spatial slices.

As before, we perform fits using Eq.~(\ref{eq:crossover_ansatz}) around $T/T_c \approx 2$ to describe the crossover behaviour. These are shown in Fig.~\ref{fig:linearbranchingdensity}, and the values given in Table~\ref{tab:inflectionpoints}. The two inflection points from temporal and spatial slices are found in this instance to be strongly coincident, with the estimate of the second transition temperature landing at $T_d \simeq 322$\,MeV. This is well within the bounds put forth from the vortex and volume branching point densities.

\subsection{Chain lengths} \label{subsec:chainlengths}
In addition to the bulk properties of branching points, we now turn to a detailed analysis of their intrinsic distribution throughout the vortex structure. Following Ref.~\cite{Biddle:2023lod}, we count the number of vortices between successive branching points along the vortex path in a three-dimensional slice. It has previously been proposed that a vortex line has a fixed probability of branching in an elementary cube as it propagates through spacetime~\cite{Spengler:2018dxt}. If this holds, the distribution of vortex chain lengths between consecutive branching points on the lattice would be expected to follow a geometric distribution,
\begin{equation} \label{eq:geodist}
	\Pr\,(k) = p\,(1-p)^{k-1} \,.
\end{equation}
In a sequence of independent trials, each with fixed probability of success $p$, this gives the probability of first success (in our case, a branching point) after $k$ trials.

This has been tested at zero temperature in both pure gauge and full QCD~\cite{Biddle:2023lod}, where the conjecture was found to be supported for large chain lengths but breaks down at short distances due to a tendency for branching points to cluster near each other. The clustering is significantly greater with dynamical fermions than without. Recent work has performed this analysis at finite temperature in pure gauge, where this clustering was observed to weaken across the phase transition and vanish entirely at very high temperatures~\cite{Mickley:2024zyg}. Motivated by these findings, we repeat this investigation with dynamical fermions to explore the extent to which they translate to full QCD.

We continue to scrutinise temporal and spatial slices independently. This is decidedly important here as we seek to understand the impact of anisotropy on branching point geometry. For instance, it seems probable that the chain lengths between branching points will not ``see" the anisotropy, with the same probability of branching irrespective of the physical distance between each elementary ``cube". This would provide a more fundamental explanation of the larger density of branching points in spatial slices. An alternative possibility is that the branching \textit{rates} (i.e. probability per unit length) are equal along the spatial and temporal dimensions. This would imply that the corresponding probabilities differ, and would make the analysis in spatial slices especially challenging.

\begin{figure*}
	\centering
	\includegraphics[width=0.48\linewidth]{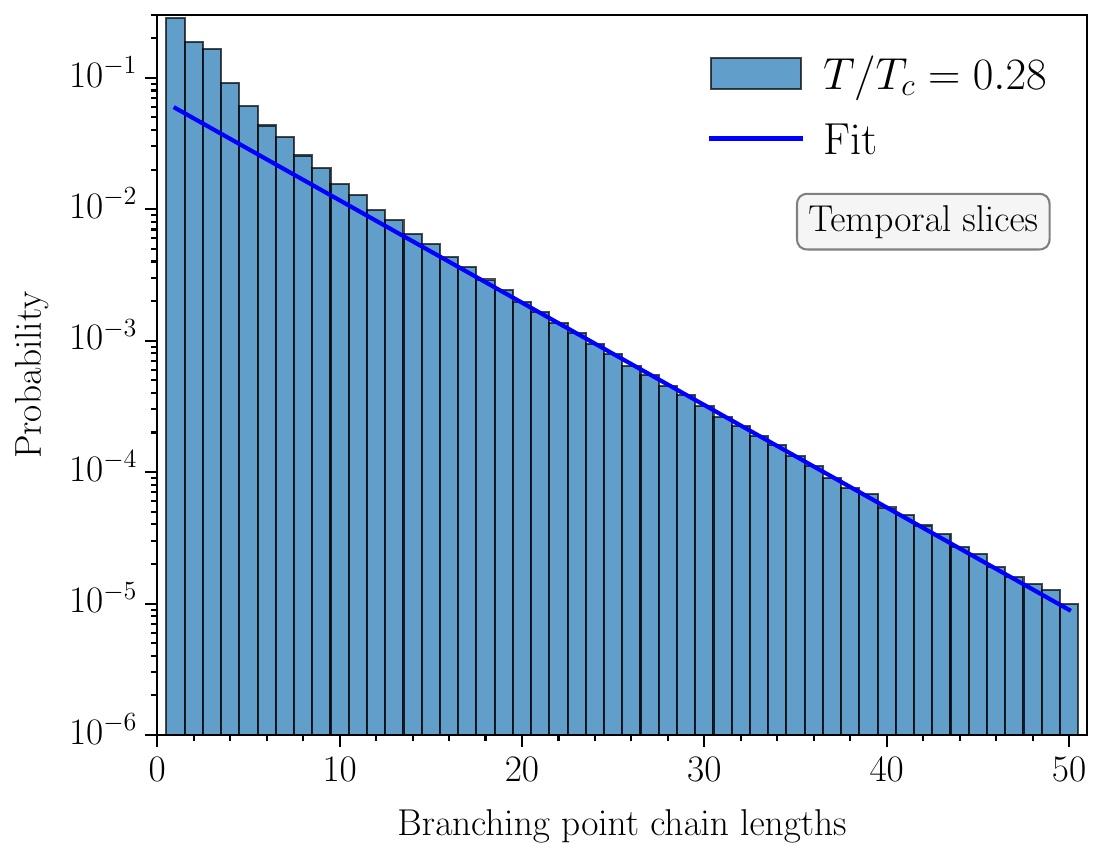}
	\hfill
	\includegraphics[width=0.48\linewidth]{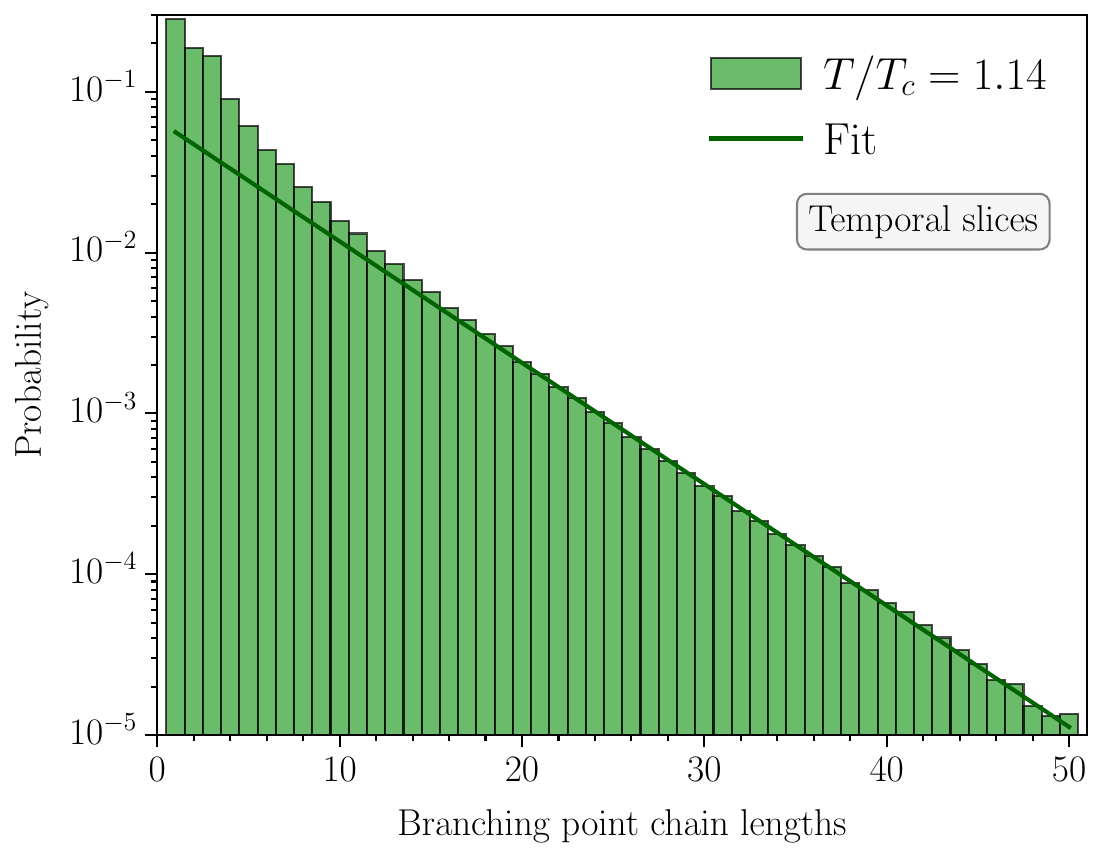}
	
	\vspace{0.9em}
	
	\includegraphics[width=0.48\linewidth]{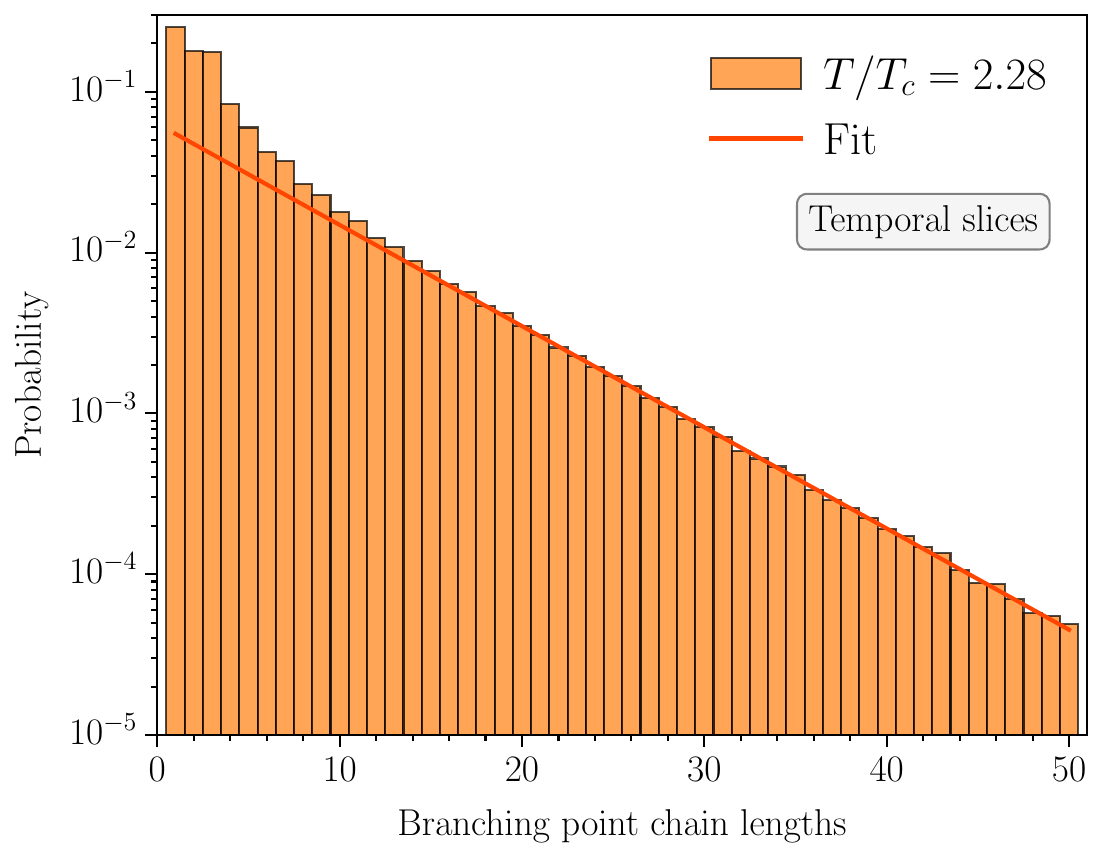}
	\hfill
	\includegraphics[width=0.48\linewidth]{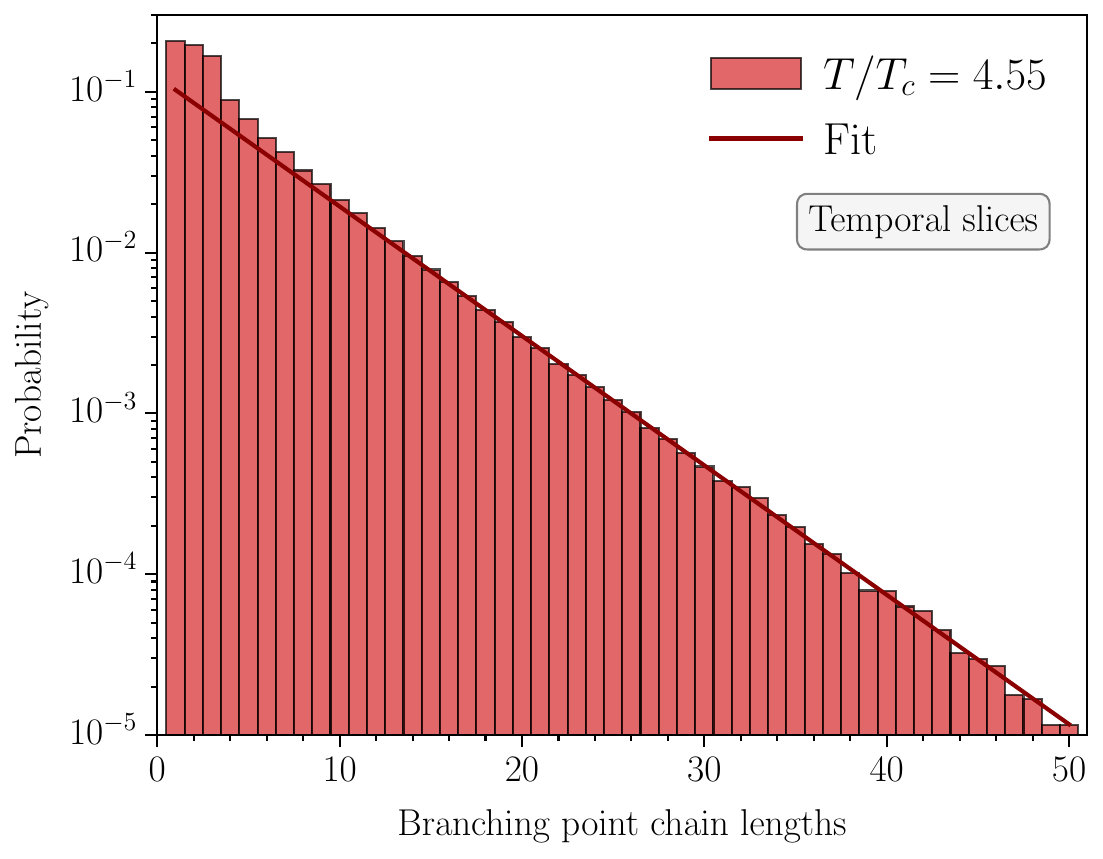}
	\caption{\label{fig:temporalbpsep} The distribution of branching point chain lengths in temporal slices of the lattice for $T/T_c \simeq 0.28$ (\textbf{upper left}), $1.14$ (\textbf{upper right}), $2.28$ (\textbf{lower left}) and $4.55$ (\textbf{lower right}). The distributions are all similar, exhibiting a clear clustering at short lengths with a smooth exponential falloff. The fits for separations $k > 20$ described in text are overlaid on the distributions and are observed to accurately capture the long-range behaviour of vortex branching.}
\end{figure*}

\begin{figure*}
	\centering
	\includegraphics[width=0.48\linewidth]{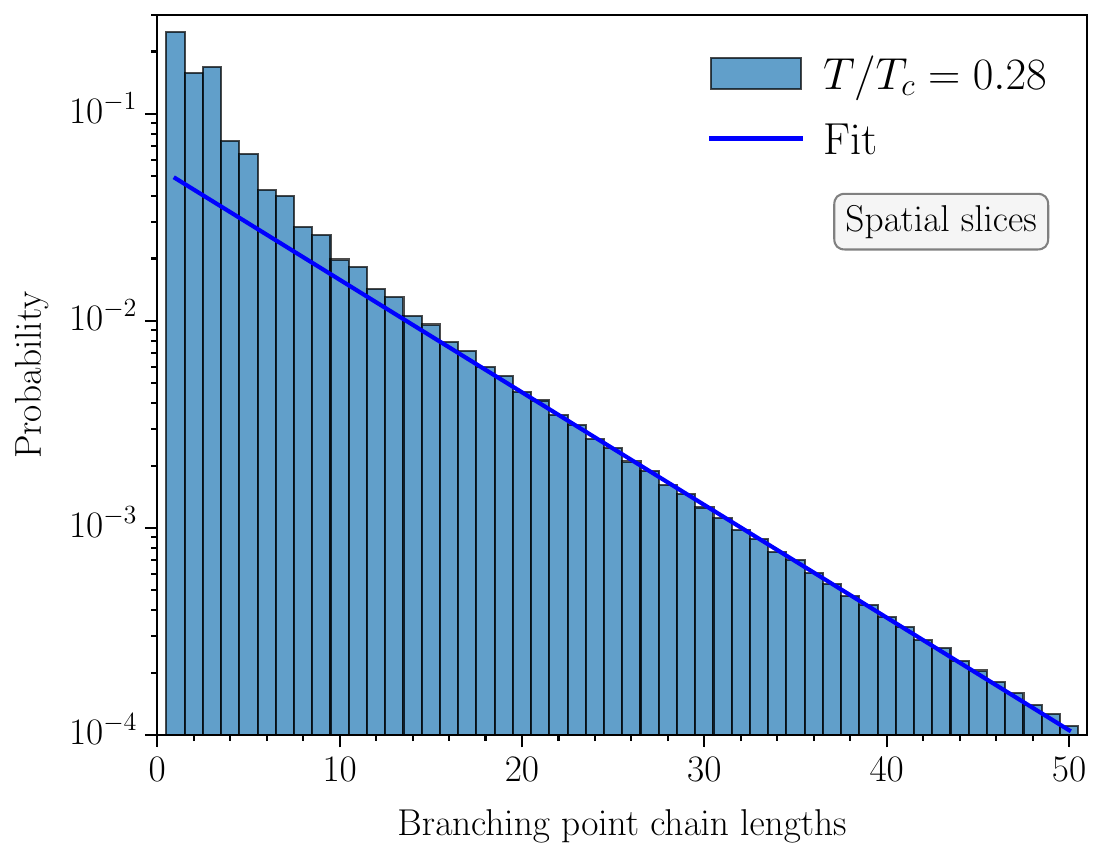}
	\hfill
	\includegraphics[width=0.48\linewidth]{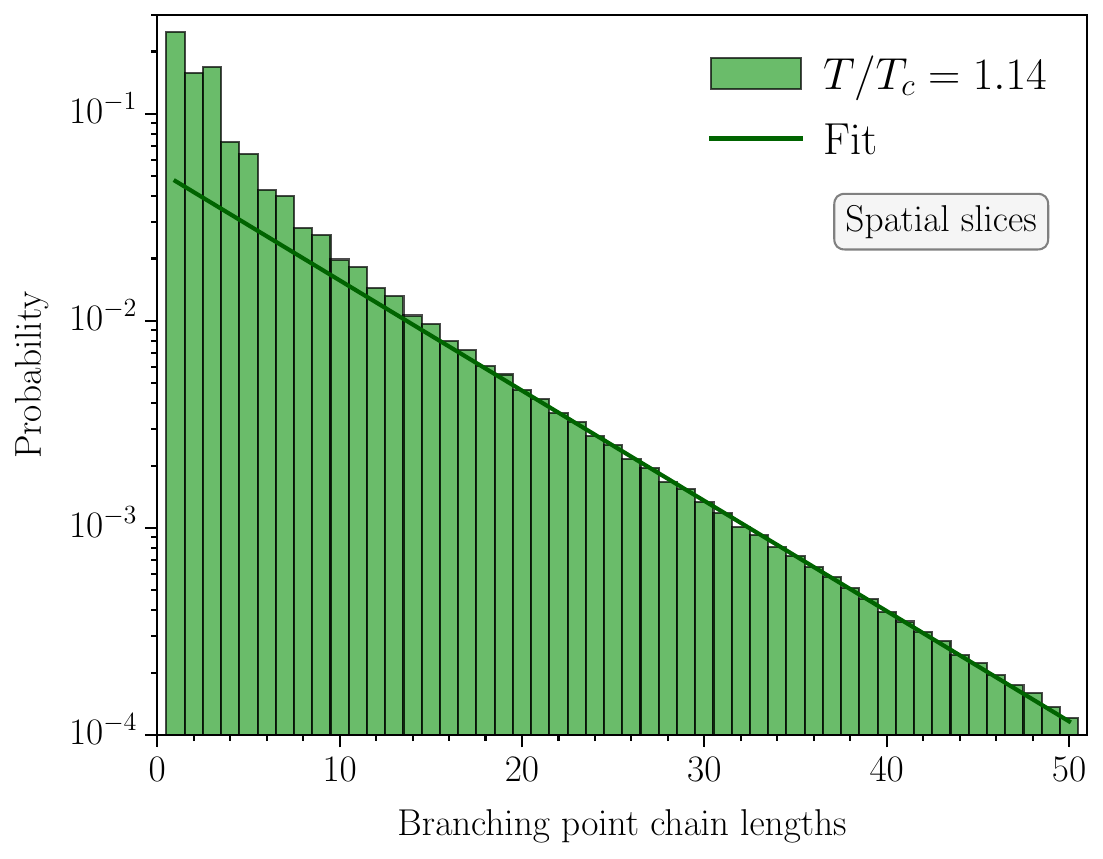}
	
	\vspace{0.9em}
	
	\includegraphics[width=0.48\linewidth]{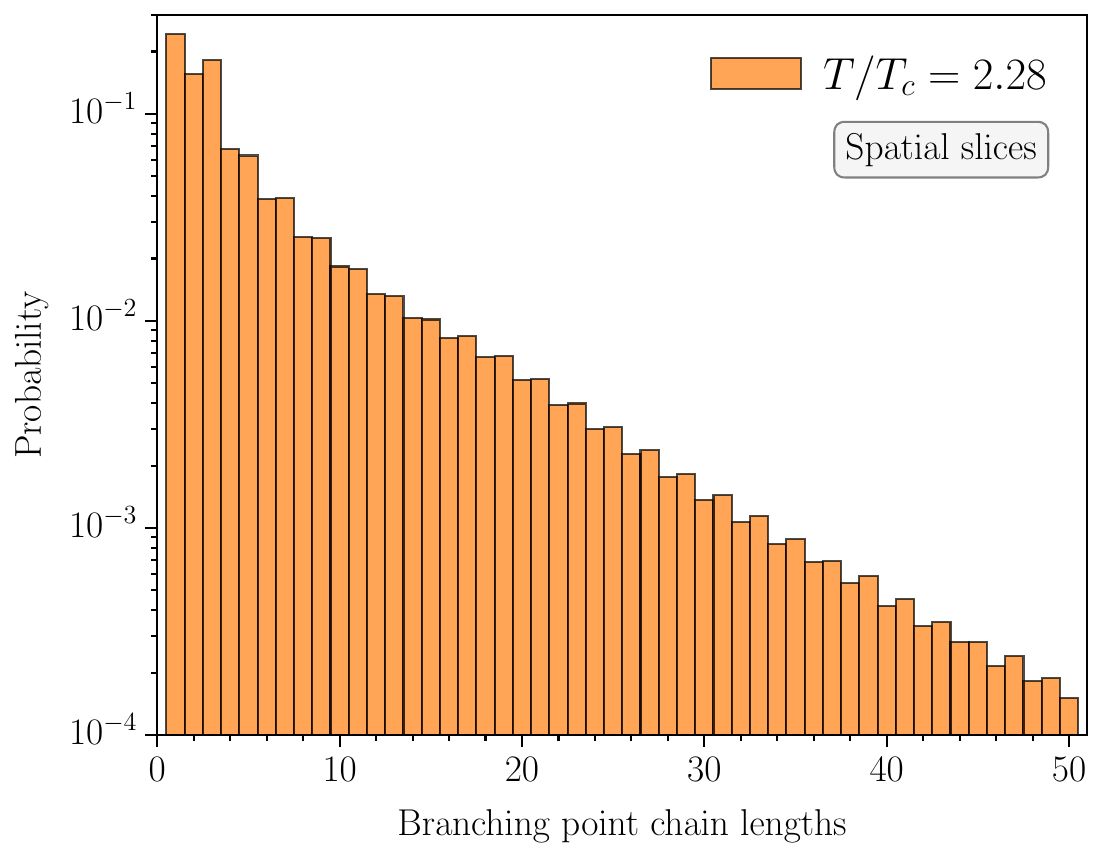}
	\hfill
	\includegraphics[width=0.48\linewidth]{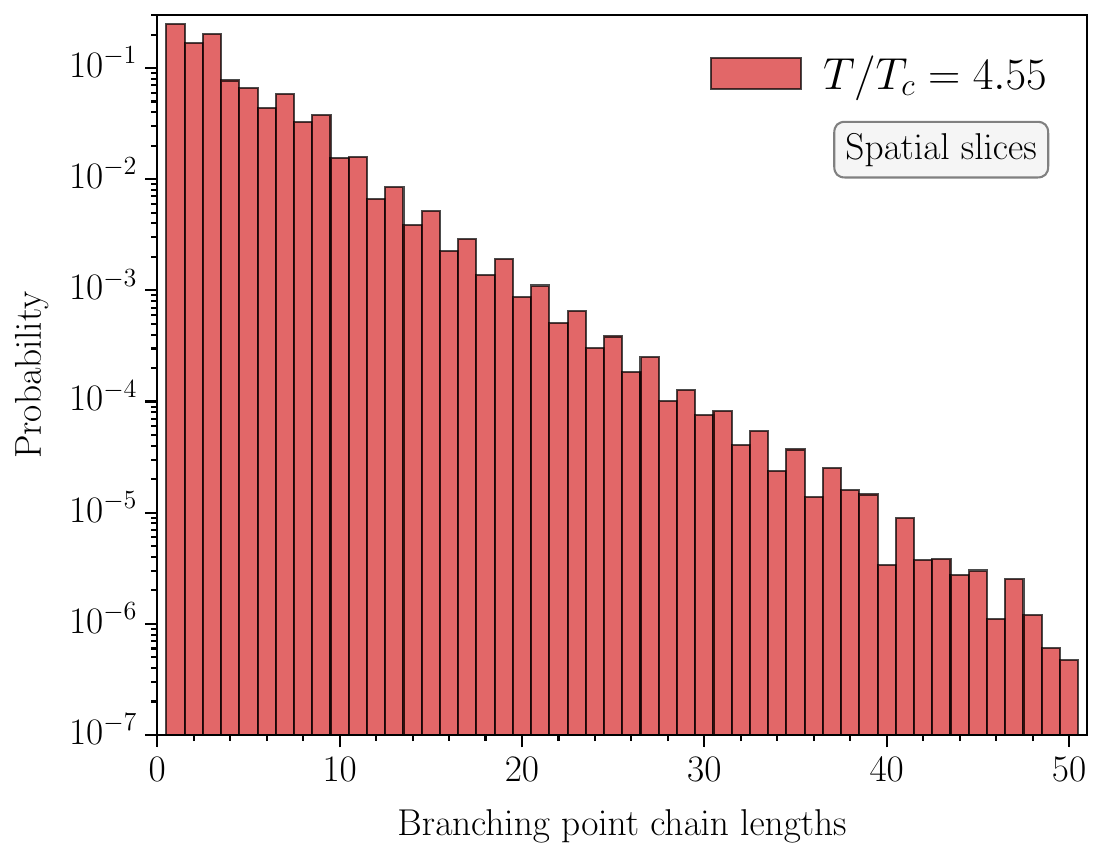}
	\caption{\label{fig:spatialbpsep} The distribution of branching point chain lengths in spatial slices of the lattice for $T/T_c \simeq 0.28$ (\textbf{upper left}), $1.14$ (\textbf{upper right}), $2.28$ (\textbf{lower left}) and $4.55$ (\textbf{lower right}). The distributions at low and intermediate temperatures coincide with those seen in the temporal slices, and the analogous fits are displayed. There is curious behaviour at very high temperatures where odd separations are favoured over their adjacent even separations. This spoils the geometric distribution interpretation.}
\end{figure*}

We produce the distributions of branching point chain lengths for all temperatures and show a selection corresponding to $N_\tau = 128$, $32$, $16$ and $8$ (four of the five ensembles visualised earlier in this paper). These are shown in Figs.~\ref{fig:temporalbpsep}
and \ref{fig:spatialbpsep}
for temporal and spatial slices, respectively. They are given on a logarithmic scale where the distribution should be linear if it follows Eq.~(\ref{eq:geodist}).

All distributions demonstrate the clustering at short separations. Although this is most pronounced for $k\leq 3$, the histograms do not truly settle into a linear trend until $k\gtrsim 20$. In temporal slices, there is little variation across the entire temperature range. Case in point, the clustering persists at all temperatures with dynamical fermions, yet it does appear less severe at our highest temperature, $T/T_c \simeq 4.55$, compared to low temperatures. This is another marked distinction from the pure gauge theory. That said, with the gradual changes in vortex geometry observed with dynamical fermions compared to the first-order phase transition in the pure gauge theory, it seems plausible that the clustering will continue to soften and possibly vanish at even higher temperatures.

Turning our focus to spatial slices, it appears that the anisotropy has not affected the distributions at low-to-intermediate temperatures. Indeed, the upper histograms in Fig.~\ref{fig:spatialbpsep} are similar to those seen in the temporal slices of Fig.~\ref{fig:temporalbpsep}. This confirms our suspicion that the probability of branching, with the current vortex-identification procedure, is insensitive to the anisotropy. As a result, with the increase in space-space-time elementary ``cubes" arising from the anisotropy, there is necessarily a greater branching point density in spatial slices over temporal slices.

Continuing to the two highest temperatures, there is strange behaviour in spatial slices where the distributions of chain lengths are no longer smooth. Instead, the probability for a given length alternates between increasing and decreasing. To be precise, the probability of an \textit{odd} separation is greater than the previous \textit{even} separation. As a by-product, the notion of a branching probability is ill defined in spatial slices at high temperatures. Peculiarly, the onset of this behaviour coincides with the loss of percolation and the vortex sheet developing an alignment with the temporal dimension. We believe there is a connection between the increase in vortex lines forced to wind around the short temporal axis and the branching point chain lengths.

\begin{figure}
	\centering
	\includegraphics[width=0.95\linewidth]{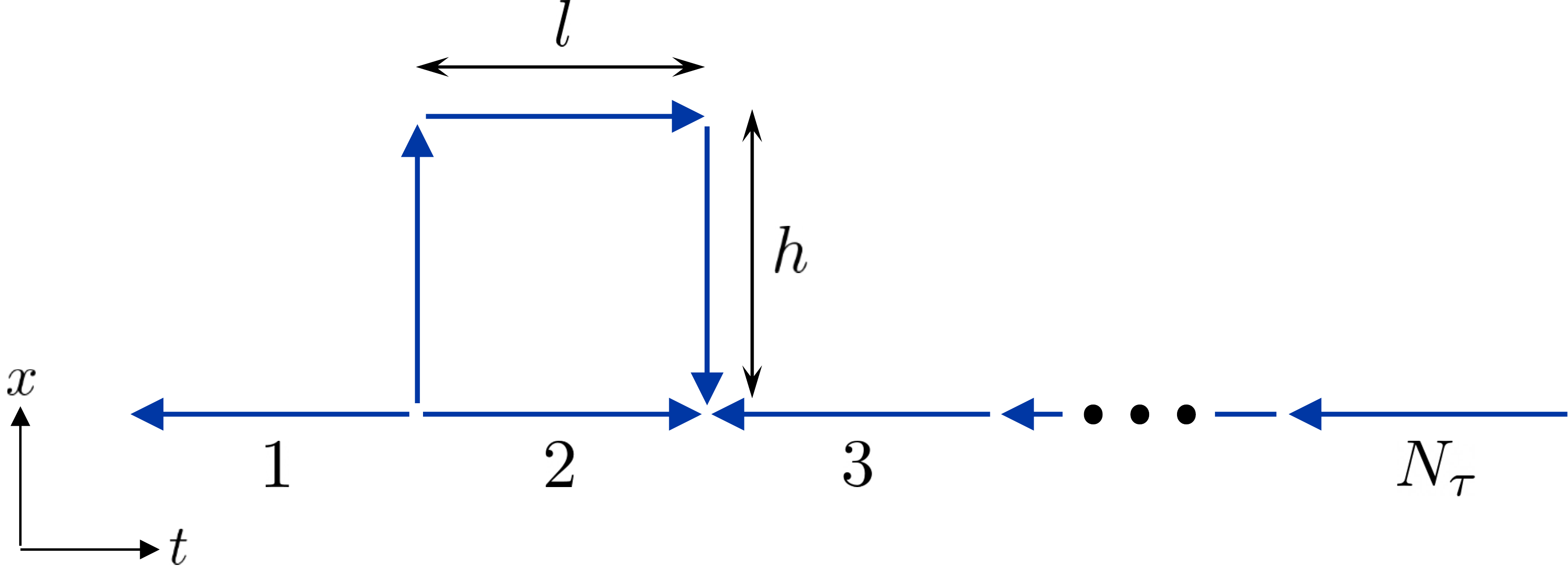}
	\caption{\label{fig:branchingcorrection} One possible branching scenario for vortex lines that predominantly wind around the temporal axis at high temperatures. The three distinct paths that connect the monopole and antimonopole have chain lengths of $l$, $l+2h$ and $N_\tau-l$ (from winding around the temporal dimension), here shown for $l=1$ and $h=1$. For even $N_\tau$, these chain lengths all have the same parity as $l$. Since $l$ is constrained to $1, \, \hdots, \, N_\tau-1$, such that there is one additional odd length allowed than even, the net effect with an even temporal extent is a bias toward odd chain lengths.}
\end{figure}

To make this explicit, we imagine ``leading-order" branching corrections to a vortex line parallel to the temporal axis in spatial slices. Noting a preference for branching point clustering, the simplest such occurrence is illustrated in Fig.~\ref{fig:branchingcorrection},
consisting of two adjacent branching points (monopole and antimonopole) connected by a vortex path of length three that temporarily deviates from the temporal axis. In this instance, following the three possible paths between the two branching points would give chain length contributions of $1$, $3$ and $N_\tau-1$ (due to periodic boundary conditions) to the histograms. If $N_\tau$ is even, these are all odd lengths.

As an extension, the length labelled $l$ in Fig.~\ref{fig:branchingcorrection} could encompass any number of vortices from $1$ through $N_\tau - 1$. The three chain lengths are then $l$, $l+2h$ and $N_\tau-l$; these all have the same parity as $l$ for even $N_\tau$. However, $l$ cannot have a length of $N_\tau$ as this would dictate there is only one branching point, whereas branching points necessarily come in pairs. Specifically, a monopole must terminate at an antimonopole. This dictates that there is one extra opportunity for $l$ to be odd compared to even. Thus we see that if $N_\tau$ is even, we incur a bias toward odd chain lengths between branching points.

These corrections are the ``least expensive", in the sense that only two space-time plaquettes need to be pierced. That said, the favouritism survives also in more complicated clusters as any chain that connects the two branching points, regardless of its deviation from the temporal axis, automatically has the same parity as $l$. In essence, this ``locks" the branching point chains to prefer odd over even lengths at all distances. The effect becomes large for small $N_\tau$ as the ratio of odd to even lengths for $l$ is $N_\tau/2 : N_\tau/2-1$. For example, for $N_\tau = 8$ the ratio is $4:3$.

\begin{figure}
	\centering
	\includegraphics[width=\linewidth]{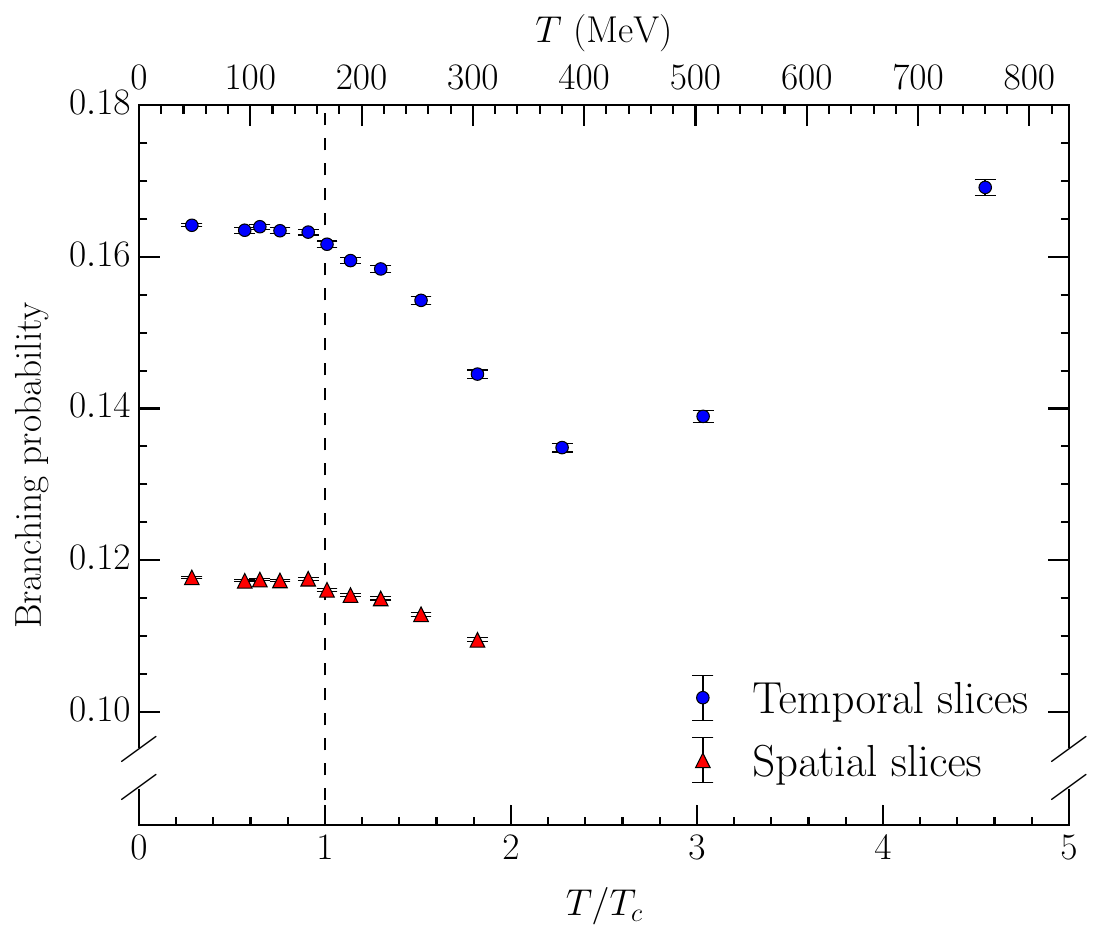}
	
	\vspace{-1em}
	
	\caption{\label{fig:branchingprob} The long-distance probability for a vortex line to branch in temporal and spatial slices, estimated by performing an exponential fit to the distribution of branching point chain lengths for $k > 20$. As in the vortex and branching point densities, it decreases for temperatures $1 \leq T/T_c \lesssim 2$, though distinctly turns to increase from $T/T_c \approx 2$ in temporal slices. This suggests the branching probability is also sensitive to the deconfinement transition.}
\end{figure}

We now proceed to obtain an estimate of the branching probability for each ensemble to observe its detailed temperature dependence. This is carried out in temporal slices for all temperatures and in spatial slices for $T/T_c < 2$ owing to the peculiar behaviour at high temperatures. Even though the detailed connection between anisotropy and vortex branching remains an open question, we still consider it interesting to examine the constant branching probability in spatial slices observed here over intermediate temperatures. An exponential provides the continuum analogue of the geometric distribution; therefore, exponential fits of the form
\begin{equation} \label{eq:exp}
	f(x) = \zeta \,e^{-\lambda x}
\end{equation}
are performed to the histograms for $k > 20$, over which the geometric distribution holds. These are overlaid on the histograms in Figs.~\ref{fig:temporalbpsep} and \ref{fig:spatialbpsep} and are seen to accurately describe the long-range branching behaviour. As per the discussion in Refs.~\cite{Biddle:2023lod, Mickley:2024zyg}, the probability is estimated by
\begin{equation} \label{eq:probexp}
	p_\mathrm{branch} = 1 - e^{-\lambda} \,.
\end{equation}
The resulting temperature evolution of the branching probability is presented in Fig.~\ref{fig:branchingprob}.

It is interesting to note that the probabilities sit above the linear trend up to chain lengths as long as $k\simeq 20$. This differs in comparison to the histograms in Ref.~\cite{Biddle:2023lod}, for which the linear fit is accurate from $k\simeq 10$. We have carried out calculations in the pure gauge theory that suggest this is a by-product of the action used to generate the ensembles. Both the pure gauge and dynamical ensembles of Ref.~\cite{Biddle:2023lod} use an Iwasaki renormalisation-group improved action. Generating a pure gauge ensemble with the Wilson action reproduces these enhanced probabilities over chain lengths up to $k \approx 10$.

We next address the lower branching probability found in spatial slices compared to temporal slices. At first glance, this might appear contradictory to the greater branching point densities in spatial slices, as seen in Figs.~\ref{fig:branchingpointdensity} and \ref{fig:linearbranchingdensity}. This apparent conundrum is a simple artefact of the two densities incorporating the anisotropy into their definitions, while here it has been ignored in counting chain lengths. The inclusion or otherwise of the anisotropy results in a scaling relative to the value in temporal slices, so it is not unexpected that the \textit{physical} probability per unit length is larger (as seen with the linear branching point density), while simultaneously the raw (dimensionless) probability is lower.

With that understood, we now turn to the temperature dependence of the branching probabilities. We can again discern a small but statistically significant drop in their values at the chiral transition, before continuing to decline over intermediate temperatures. In temporal slices, for which we have access to the full temperature range, there is noteworthy behaviour at high temperatures. The probability at our highest temperature exceeds the values attained below $T_c$, indicating that branching is enhanced at very high temperatures. Additionally, unlike the vortex and branching point densities, which only increased in moving to our highest temperature, here we find a turning point closer to the proposed deconfinement transition $T_d/T_c \approx 2$. This qualitatively aligns with the vortex structure correlation (Fig.~\ref{fig:temporalcorrelation}) and number of secondary clusters (Fig.~\ref{fig:secondaryclusters}), suggesting the branching probability is sensitive to the deconfinement transition in the same manner. One could imagine performing separate fits to the probability for $1 \leq T/T_c \leq 2$ and $T/T_c \geq 2$, and the intersection point of the two curves would provide another means of estimating the second transition. Due to our rather coarse temperature grid for $T/T_c \geq 2$, we defer this possibility to future analyses.

\section{Discussion} \label{sec:discussion}
In this paper, we have identified thin centre vortices on dynamical QCD ensembles via the standard procedure of fixing to maximal centre gauge and projecting out the centre elements. The vortex flux is mapped out through the centre-projected plaquettes with nontrivial values. These centre vortices have a rich clusterlike structure which we have dissected by studying
\begin{itemize}
	\item temporal correlation functions,
	\item the linear extent of the clusters,
	\item the physical density of vortices,
	\item the prevalence of secondary clusters,
	\item the physical branching point density,
	\item the linear branching point density, and
	\item the branching probability.
\end{itemize}
In every case, there is quantitative evidence of both the conventional transition at $T_c$ {\em and} a second transition at $T_d \approx 1.9\, T_c$. The former is usually linked to the restoration of chiral symmetry. We have illustrated that the second transition is associated with deconfinement.

Indeed, the correspondence between the behaviour discovered herein at $T_d$ and that observed in the quenched theory \cite{Mickley:2024zyg} at the critical phase transition temperature $T_\mathrm{quenched}$ is remarkable. In both cases, the temporal correlation functions grow with temperature above the transition as the vortex sheet aligns with the temporal axis. This alignment acts to freeze the temporal dependence of the three-dimensional spatial percolating cluster.

Similarly, the normalised cluster extent in spatial slices of the lattice undergoes a sharp drop as the transition temperature is crossed and percolation in all four spacetime dimensions is lost. 

Moreover, the vortex, branching, and linear branching densities all take values much smaller than their low-temperature counterparts as the deconfinement temperature is crossed. Above the deconfinement transition, they grow with higher temperatures. Both the bulk linear branching density and the local branching probability display this behaviour. 

The prevalence of secondary clusters also shows similar behaviour in quenched and full QCD across the deconfinement transition. The secondary cluster density increases just below the deconfinement temperature and then drops to lower values as the vortex sheet aligns with the temporal axis, suppressing the curvature associated with large secondary clusters.

However, the transitions disclosed herein at $T_d$ have a smoother property than in the quenched theory, which shows the characteristic features of a first-order phase transition. This suggests the possibility of a second-order phase transition or perhaps a crossover in the dynamical case considered here.

We note that in the quenched approximation, which corresponds to dynamical quark masses of $m_q=\infty$, the two transition temperatures, $T_c$ and $T_d$, coincide at $T_\text{quenched} \simeq 290\,$MeV. Furthermore, $T_c$ has been measured at a variety of $m_q$ values from the chiral value $m_q=0$ up to values corresponding to pion masses of $m_\pi \approx 400\,$MeV or more.

\begin{figure}
	\centering
	\includegraphics[width=0.95\linewidth]{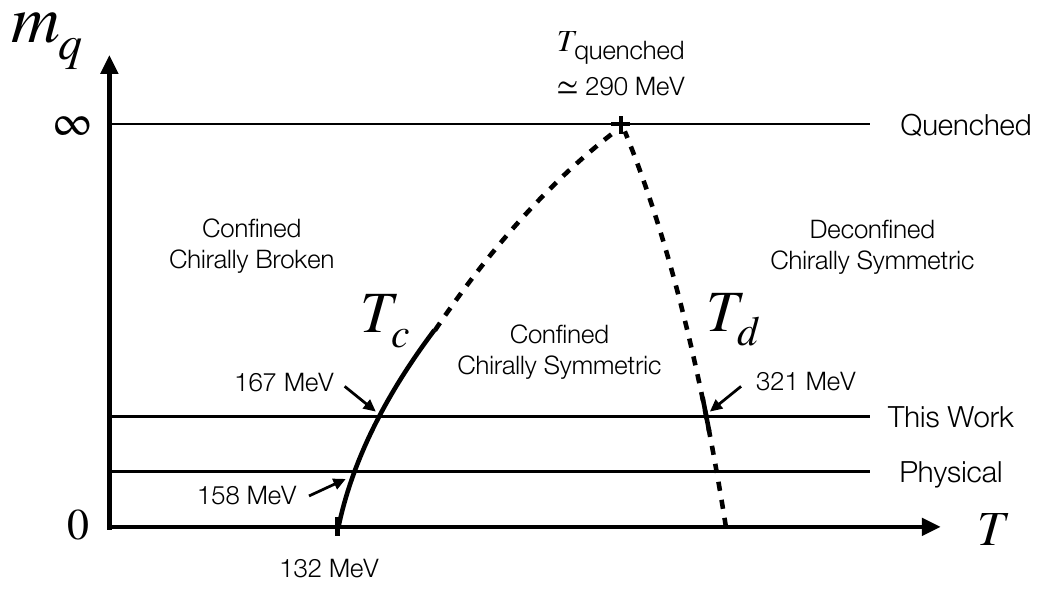}
	
	\vspace{-0.5em}
	
	\caption{\label{fig:qcd_phase_diagram} The proposed phase structure of QCD highlighting the two transition temperatures analysed in this work. Here, $m_q$ represents the masses of the light and strange dynamical quarks. The known dependence of the conventional (chiral) transition temperature $T_c$ on $m_q$ is plotted as a solid curve, together with the proposed variation up to the quenched value $T_\text{quenched} \simeq 290\,$MeV \cite{Borsanyi:2022xml} as a dashed curve. The second transition temperature $T_d = 321(6)\,$MeV defined in this work [see Eq.~(\ref{eq:finalTd})] is displayed accompanied by its hypothesised variation with $m_q$ such that $T_d \rightarrow T_\text{quenched}$ in the quenched limit. These transition temperatures define three phases of QCD depending on the chiral symmetry and confinement properties. The values of $T_c = 132(+6)(-3)\,$MeV at the chiral point \cite{HotQCD:2019xnw}, $T_c \simeq 158\,$MeV at the physical point \cite{HotQCD:2018pds, Borsanyi:2020fev, Gavai:2024mcj}, and $T_c = 167(3)\,$MeV at the quark masses considered in this work \cite{Aarts:2022krz} are shown.}
	
	\vspace{-0.2em}
\end{figure}

This can be schematically summarised in Fig.~\ref{fig:qcd_phase_diagram} where the variation of $T_c$ with $m_q$ is plotted.
Here, $m_q$ represents the masses of the light and strange quarks. We use solid lines to represent established trends of the transition temperature. In this work, we propose a second transition at $T_d$, which is also plotted. We propose that both $T_c$ and $T_d$ merge in the quenched limit, where we expect the chiral and deconfinement transition to occur at the same temperature $T_\text{quenched}$. Dashed lines tentatively illustrate how these two transition temperatures vary with $m_q$. Figure~\ref{fig:qcd_phase_diagram} thus displays three phases of QCD: (i) confined and chirally broken, (ii) confined and chirally symmetric, and (iii) deconfined and chirally symmetric. It is the extra feature of quark loops in the vacuum that drives a rich phase structure.

\section{Conclusion} \label{sec:conclusion}
Through the consideration of the centre vortex structure of QCD ground-state fields, we have obtained lattice evidence for a second finite-temperature transition in QCD associated with the deconfinement of quarks.

Our calculations are founded on the anisotropic \textsc{Fastsum} Collaboration~\cite{Aarts:2020vyb} gauge fields with $2+1$ flavours of $\order{a}$-improved Wilson fermions on a Symanzik-improved gauge action at $m_\pi = 239(1)\,$MeV. The spatial lattice spacing is $a_s = 0.11208(31)$\,fm with anisotropy $\xi \equiv a_s/a_\tau = 3.453(6)$. The renormalised chiral condensate was used to determine the pseudocritical chiral transition temperature $T_c = 167(3)\,$MeV \cite{Aarts:2020vyb, Aarts:2022krz}.

A comprehensive range of temperatures are examined including five temperatures below $T_c$, five temperatures in the intermediate regime $T_c \lesssim T \lesssim T_d$ between the chiral and deconfinement transitions, and another three temperatures above $T_d$. Both transition temperatures are observed and quantified via the sigmoid functional form of Eq.~(\ref{eq:crossover_ansatz}) to precisely determine the inflection points of the transitions.

While the prevailing wisdom assumes the presence of only one transition temperature, we now understand the first transition at $T_c$ is a chiral transition, with confinement persisting through this transition.  The centre vortex structure continues to percolate through all four spacetime dimensions in this intermediate temperature range, $T_c < T < T_d$, such that a space-time Wilson loop will acquire an area law signaling confinement.

The second transition at $T_d$ is a deconfinement transition, associated with the loss of centre vortex percolation in spatial slices of the four-dimensional lattice.  Through the consideration of six quantitative measures of the vortex structure from Table~\ref{tab:inflectionpoints}, we find the deconfinement temperature $T_d = 321.1(1.0)(5.6)\,$MeV as described in detail in the following.  In comparison to the review of other indicators of a second transition listed in Sec.~\ref{sec:rev}, this is a precise determination.

Critical to the calculation is the creation and development of an anisotropic extension of maximal centre gauge (MCG). A large temporal link weighting of $\xi^2$ in the gauge-fixing functional was derived.  In the current investigation $\xi^2 \simeq 12$, a factor so large that direct fixing to the anisotropic-improved functional left spatial links far from the centre gauge group elements. An isotropic MCG preconditioning was introduced to ensure all links are optimised and the algorithms were tested by ensuring the physical space-space and space-time vortex densities match at our lowest temperature well below the chiral transition.  Both mesonic and baryonic gauge-fixing functionals were considered with the baryonic measure preferred.

Visualisations of the vortex matter revealed a transition from a single percolating cluster in all four dimensions of the anisotropic lattice to percolation only in the three spatial dimensions at the highest temperatures, a signature of deconfinement.  However, percolation in all four dimensions of the lattice was observed to persist through the chiral transition, indicating the persistence of confinement there.  Other features of the deconfinement transition were noted including a reduction in vortex density and a loss of secondary vortex loops associated with the vortex sheet curving back in the time dimension.

To quantify the two transitions, a collection of vortex measures were defined and evaluated over the ensembles, providing a robust bootstrap statistical analysis.  Measures include	the alignment of the vortex sheet with the temporal dimension via a correlation function in time, the vortex cluster extent, the secondary cluster density, the physical vortex density, the physical branching point density, and a linear branching point density associated with the rate of branching. The latter three measures are carried out on both spatial and temporal slices of the four-dimensional lattice, providing ten quantitative measures of centre vortex structure.

We observe increasing disorder in the vortex alignment correlation function of Eq.~(\ref{eq:correlation}) from temperatures commencing at the chiral transition and persisting up to $T \simeq 304\,$MeV.  This behaviour contrasts that observed above the deconfinement temperature $T_d$ where the vortex sheet does indeed align with the temporal direction as in the pure gauge theory.  It is this alignment that ultimately leads to a drop in secondary clusters at the highest two temperatures considered.  With the vortex sheet aligning, curvature of the sheet in the temporal direction is suppressed.  In spatial slices of the lattice exposing vortex behaviour in space-time dimensions, the secondary cluster density also signals changes in vortex structure at $T \simeq 304\,$MeV.

The cluster extent provides a direct quantitative measure of percolation in space-time dimensions.  The loss of percolation in the temporal dimension at the deconfinement temperature is decisive and further highlights the $T \simeq 304\,$MeV ensemble as very close to $T_d$.

The vortex, branching and linear branching densities provide robust signatures at both transitions, admitting an inflection point analysis to quantify the exact transition temperatures. The results for these six measures are summarised in Table~\ref{tab:inflectionpoints}.  Taking a statistically weighted average of our results and estimating the systematic error from the standard deviation of the distribution about the weighted average, we conclude
\begin{equation} \label{eq:finalTc}
	\begin{gathered}
		\frac{T_c}{T_c^{\rm Fastsum}} = 0.983(8)(0)(18)\, , \\
		{T_c} = 164.2(1.4)(0.0)\,{\rm MeV}
	\end{gathered}
\end{equation}
for the chiral transition, and
\begin{equation} \label{eq:finalTd}
	\begin{gathered}
		\frac{T_d}{T_c^{\rm Fastsum}} = 1.921(6)(33)(35)\, , \\
		{T_d} = 321.1(1.0)(5.6)\,{\rm MeV}
	\end{gathered}
\end{equation}
for the deconfinement transition. The indicated uncertainties are statistical, systematic and the contribution from $T_c^\mathrm{Fastsum}$ (where relevant), respectively.

The value of $T_c$ determined here is in agreement with the value obtained from the renormalised chiral condensate, $T_c^{\rm Fastsum} = 167(3)\,$MeV  \cite{Aarts:2020vyb,Aarts:2022krz}. It is interesting to note that the main source of uncertainty in $T_d$ follows from systematic differences among the different vortex measures.

The changes to vortex geometry that occur at $T_d$ broadly match with the transition in the pure gauge theory. That said, whereas the quenched theory exhibits the attributes of a critical first-order transition, the smoother behaviour observed herein at $T_d$ suggests that the deconfinement transition in full QCD is either second order in nature or a crossover.

In addition to examining the bulk properties of vortices and their branching points, the latter's intrinsic distribution within the lattice has also been examined.  By counting the number of vortices between successive branching points along the vortex path in a three-dimensional slice, we are able to show the presence of a branching probability that is independent of the path length for paths greater than $\sim\! 10$ spatial steps.  The branching probability undergoes a step change through $T_c$, diminishing further through the intermediate regime before turning around at the highest two temperatures as the vortex-sheet alignment becomes manifest.

We emphasise that evidence for a second transition at $T_d$ follows from both local and bulk quantities. Local measures such at the correlation function of Eq.~(\ref{eq:correlation}) describing the alignment of the vortex sheet with the temporal axis, or the probability of branching at each step along the vortex chain, show a qualitative change as one moves through $T_d$ to higher temperatures.  These are complemented by bulk measures such as the vortex extent and the various densities based on large-scale features across the lattice volume.  Indeed the vortex, branching and linear branching densities in temporal and spatial slices show quantitative inflections in behaviour that have enabled the precise determination of the deconfinement temperature $T_d$.

Finally we have outlined the quark-mass dependence of the QCD transitions in Fig.~\ref{fig:qcd_phase_diagram}, introducing the new deconfinement transition at $T_d$ and illustrating its convergence with the chiral transition, $T_c$, at the infinite-quark-mass limit of quenched QCD.  Here it is clear that extensive research remains to be done.

For example, to make contact with the physical quark mass point relevant to experiment, it will be important to determine the quark-mass trend of $T_d$.  Figure~\ref{fig:qcd_phase_diagram} indicates a trend opposite that of $T_c$, with $T_d$ growing somewhat larger toward the physical quark-mass point.  Ultimately, direct simulations at the physical point will precisely determine the physical value of $T_d$.

With the presence of a second transition in QCD now established and knowledge of its position ascertained, significant resources can be focused on this region to obtain the precise deconfinement transition temperature at the physical point relevant to experiment.

\begin{acknowledgments}
It is a pleasure to thank Dr.\ Waseem Kamleh for discussions regarding vortex identification on anisotropic lattices, and Jeff Greensite and Massimo D’Elia for other useful conversations. This work was supported with supercomputing resources provided by the Phoenix HPC service at the University of Adelaide. This research was undertaken with the assistance of resources and services from the National Computational Infrastructure (NCI), which is supported by the Australian Government. We acknowledge EuroHPC Joint Undertaking for awarding the project EHPC-EXT-2023E01-010 access to LUMI-C, Finland. This work used the DiRAC Data Intensive service (DIaL2 and DIaL) at the University of Leicester, managed by the University of Leicester Research Computing Service on behalf of the STFC DiRAC HPC Facility (www.dirac.ac.uk). The DiRAC service at Leicester was funded by BEIS, UKRI and STFC capital funding and STFC operations grants. This work used the DiRAC Extreme Scaling service (Tesseract) at the University of Edinburgh, managed by the Edinburgh Parallel Computing Centre on behalf of the STFC DiRAC HPC Facility (www.dirac.ac.uk). The DiRAC service at Edinburgh was funded by BEIS, UKRI and STFC capital funding and STFC operations grants. DiRAC is part of the UKRI Digital Research Infrastructure. This research was supported by the Australian Research Council through Grant No.\ DP210103706. C.A. is grateful for support via STFC Grant No.\ ST/X000648/1 and the award of a Southgate Fellowship from the University of Adelaide. R.B. acknowledges support from a Science Foundation Ireland Frontiers for the Future Project award with Grant No.\ SFI-21/FFP-P/10186. We are grateful to the Hadron Spectrum Collaboration for the use of their zero-temperature ensemble. For the purpose of open access, the authors have applied a Creative Commons Attribution (CC BY) licence to any Author Accepted Manuscript version arising.
\end{acknowledgments}


\bibliography{main}

\end{document}